\documentclass[reprint, amsmath, amssymb, showpacs, floatfix, longbibliography, aps, twocolumn, superscriptaddress, nofootinbib]{revtex4-2}
\usepackage{CJKutf8}
\usepackage{amsthm}
\usepackage{amsmath}
\usepackage{graphicx}
\usepackage{dcolumn}
\usepackage{bm}
\usepackage{amsfonts,stmaryrd}
\usepackage{physics}
\usepackage{dsfont}
\usepackage{graphicx}
\usepackage{xcolor}
\usepackage{cancel}
\usepackage{tikz-cd}
\usepackage{natbib}
\usepackage[section]{placeins}
\usepackage{amssymb}
\usepackage{adjustbox}
\usepackage{color}
\usepackage{tikz}
\usepackage{mathrsfs}
\usepackage{relsize}
\usepackage{caption}
\usepackage{subcaption}
\usepackage[all]{xy}

\usepackage[draft]{changes}

\usepackage{multirow}
\usepackage{environ}

\usepackage[unicode,linktocpage,colorlinks=true,
linkcolor=blue,citecolor=blue,urlcolor=blue]{hyperref}
\NewEnviron{eqs}{%
\begin{equation}\begin{split}
    \BODY
\end{split}\end{equation}}

\usepackage{geometry}
\newcommand{\NN}{{\mathbb N}}
\newcommand{\ZZ}{{\mathbb Z}}
\newcommand{\RR}{{\mathbb R}}

\newcommand{\tZ}{{\widetilde{\mathbb Z}}}

\newcommand{\supp}{\operatorname{supp}}

\newcommand{\id}{{\operatorname{id}}}
\newcommand{\inv}{{\operatorname{inv}}}

\theoremstyle{definition}

\newtheorem{conjecture}{Conjecture}[section]
\newtheorem{theorem}{Theorem}[section]
\newtheorem{lemma}{Lemma}[section]
\newtheorem{corollary}{Corollary}[section]

\newtheorem{definition}{Definition}[section]

\newtheorem{example}{Example}[section]
\newtheorem{remark}{Remark}[section]

\usetikzlibrary{decorations.markings}

\graphicspath{{./Figures/}}

\begin{document}

\title{Statistics of Abelian topological excitations}

\author{Hanyu Xue}
\email{xhy2002@mit.edu}
\affiliation{International Center for Quantum Materials, School of Physics, Peking University, Beijing 100871, China}
\affiliation{Department of Physics, Massachusetts Institute of Technology, Cambridge, Massachusetts 02139, USA}

\begin{abstract}
	In this paper, we develop a novel theory that generalizes the concept of anyon statistics to Abelian topological excitations of any dimension. We axiomatize excitations as a selected collection of states and operators satisfying the configuration axiom and the locality axiom, purely based on many-body quantum mechanics. Upon these axioms, we define a rigorous and self-contained theory of statistics using only basic algebra and can be implemented on a computer. While our theory is developed independently, the results align with existing physical theories.
\end{abstract}

\date{\today}

\maketitle

\tableofcontents

\section{Introduction}

Condensed-matter systems exhibit remarkable phenomena arising from their collective behavior -- a direct consequence of the large number of interacting particles and emergent degrees of freedom. Under a tight-binding approximation, the physics of interacting electrons in a solid is usually captured by a lattice model, consisting of
\begin{itemize}
	\item a $d$-dimensional spatial lattice with a local Hilbert space $\mathcal{H}_x$ at each site $x$;
	\item a total Hilbert space $\mathcal{H}=\otimes_x \mathcal{H}_x$;
	\item a local Hamiltonian $H=\sum_x H_x$, where each $H_x$ acts only near $x$.
\end{itemize}

Although the microscopic interactions are complicated, different systems in the same phase share many universal macroscopic properties. For gapped Hamiltonians, it is common to say that two systems are in the same phase if their ground states are related by a finite–depth local quantum circuit \cite{zeng-2019}. This notion, however, is strictly well–defined only in the thermodynamic limit.

Landau’s theory of spontaneous symmetry breaking asserts that different phases are classified by a pair of groups $H\subset G$, where $G$ is the symmetry group of the Hamiltonian and $H$ is the unbroken symmetry of states. It was later realized that many quantum phases cannot be described within this framework, such as spin liquids, fractional quantum Hall states, and a large family of exactly solvable lattice models \cite{Wen_2015}. A modern guiding principle proposes that gapped topological phases of matter (topological orders) are classified by certain higher categories built from their topological excitations \cite{Kong:2014qka,kong2024highercondensationtheory}. This program has seen substantial progress, but is still far from complete: traditional tools, both analytical and numerical, are not effective for numerous tensor products. 

A particular puzzle is how to interpret the thermodynamic limit. This limit is inevitable in the definition of phase, and its mathematical description must involve analytical tools, such as operator algebras. In contrast, higher categories are purely algebraic objects. Some works do study phases using operator-algebraic methods \cite{Kapustin_2022,NAAIJKENS_2011}, but most of the literature on topological order formulates classification in purely algebraic terms, and then applies it to finite lattice systems. Conceptually, it is unsatisfactory that phases are defined only in infinite systems, while almost all practical calculations are carried out in finite ones.

Our answer to this puzzle is that in the study of phases and phase transitions, many notions at different logical levels are mixed together. This notion mixing already appears in Landau’s symmetry-breaking theory, where one can distinguish three levels:

\begin{enumerate}
	\item \textbf{Mathematics level}: A symmetry-breaking pattern specified by the total symmetry $G$ and an unbroken subgroup $H\subset G$. These are purely algebraic data with no physical realization yet.
	\item \textbf{Kinematics level}: A set of states carrying a transitive action of $G$ such that $H$ is the stabilizer of a chosen element. In the language of quantum mechanics, this is a Hilbert space $\mathcal{H}$ spanned by vectors 
	$\{|gH\rangle\}$ with the obvious $G$-action. Here we already have states in a Hilbert space, and this space may arise as the ground-state subspace of a Hamiltonian, but there is still no dynamics, and no reference to system size or spatial dimension.
	\item \textbf{	Dynamics level}: One now talks about free energy, spatial dimensions, mean-field theory, perturbations, finite temperature, quantum fluctuations, Landau–Ginzburg field theory, renormalization, and so on. All of these concern the stability of a given realization of a symmetry-breaking pattern under various perturbations and physical conditions.
\end{enumerate}

At the first two levels, the thermodynamic limit and the dimension of space are completely irrelevant; from the viewpoint of generalized symmetries, Landau’s symmetry is effectively zero-dimensional. Only at the third level—roughly, the theory of stability under perturbations—does the thermodynamic limit become central.

The transverse-field Ising model provides a familiar example. At the mathematics level, there are two symmetry-breaking patterns: paramagnetic or ferromagnetic. At the kinematics level, the ferromagnetic pattern is realized by $|\uparrow\cdots\uparrow\rangle$ and $| \downarrow\cdots\downarrow\rangle$ for any system size $L$. At the dynamics level, we perturb the ferromagnetic interaction by a small transverse field and ask whether this realization is stable. For any finite $L$ the degeneracy is lifted by a splitting that decays exponentially in $L$, and a genuine phase transition only appears in the thermodynamic limit, with its existence and critical behavior depending on the spatial dimension.

In the symmetry-breaking theory, the kinematical level is almost trivial, and Landau’s mean-field free energy $F(\phi)=a\phi^2+b\phi^4$ is already at the dynamics level. For topological phases of matter, however, the situation is more complicated, and keeping these levels conceptually separate is useful. In the same three-layer scheme \textit{for lattice models}, higher categories live at the mathematical level; continuum topological quantum field theories (TQFTs) sit between the mathematical and kinematical levels, as they are not directly built from quantum mechanics; lattice gauge theories and exactly solvable lattice Hamiltonians belong to the kinematical level; and questions involving thermodynamic limits, functional analysis, or perturbation theory are part of the dynamical level.

Topological orders are expected to be classified by higher categories built from all topological excitations in a given system. This provides a coherent picture at the mathematical level, but it is not yet sufficient at the kinematical level, because the notion of a “topological excitation” in a concrete lattice model remains somewhat vague. Researchers have systematically constructed many exactly solvable lattice models from categorical data, such as Kitaev’s toric-code model and quantum double models \cite{Kitaev:1997wr},
string-net models \cite{Levin_2005,generalizedStringNet},  Walker-Wang models \cite{walker201131tqftstopologicalinsulators}, and lattice realizations of Dijkgraaf–Witten gauge theories. This approach provides a rich supply of examples and strong evidence for the categorical picture, but not yet a general microscopic theory.
We can easily be persuaded that the $e$ anyon in the toric-code model is a topological excitation, but to really understand what a topological excitation \textit{is}, we must be able to define and detect it in generic lattice models. 

This is more difficult in practice. Even if a generic lattice Hamiltonian is in the same phase as some exactly solvable string-net model, we do not have the underlying fusion category \textit{a priori}. Such information must be extracted from the many-body Hilbert space, states, and operators in a systematic way. In this direction, Haah’s theory of Pauli stabilizer models goes beyond the construction of individual examples \cite{HaahThesis}. Under the assumption that the local Hamiltonian is built from translation-invariant commuting Pauli operators, Haah’s method converts any such system to a problem in commutative algebra; the anyon data are encoded in this algebraic structure and can even be computed automatically on a computer \cite{Zijian1,Zijian2}. This formalism has also led to the discovery of fracton phases and other exotic orders \cite{Fractonphasesofmatter,Haah_2011}. Nevertheless, the class of Hamiltonians it covers is still very special: translation-invariant, Pauli, and exactly solvable.

In more generic situations we neither want to assume spatial homogeneity\footnote{We say a lattice model is spatially homogeneous in two situations: (i) the lattice and the Hamiltonian are translation invariant; (ii) the Hamiltonian or ground state is defined uniformly on any triangulation of a fixed manifold, as in lattice gauge theories.} nor expect to solve the Hamiltonian exactly. In such cases the only robust starting point is to \textit{axiomatize lattice models} in terms of families of states and operators satisfying appropriate locality and compatibility conditions. We do not attempt to diagonalize the Hamiltonian. Rather, in a Platonic sense, a concrete lattice system exhibits anyonic statistics and topological order insofar as it realizes such an axiom system. On top of these axioms, we develop a theory of statistics for general Abelian topological excitations in arbitrary dimension, whose structure mathematically corresponds to the coherence data of certain higher categories. We will explain more background in the next section.

 In short, for $p$-dimensional Abelian excitations with fusion group $G$ in $d$-dimensional space, we define its statistics and then prove that it is naturally isomorphic to $H^{d+2}(K(G,d-p),\RR/\ZZ)$, where $K(G,d-p)$ is the Eilenberg-MacLane space. This is in agreement with the cohomological classification of higher-form SPT phases, discrete gauge theories, and anomalies in the continuum literature \cite{Chen2013Symmetry,Kapustin:2013uxa,feng2025higherformanomalieslattices}. Our definition of statistics is purely from two axioms, basic quantum mechanics, and linear algebra. They are logically independent of categorical and field-theoretical arguments, so this agreement therefore provides nontrivial consistency checks for both sides.

Our theory is completely at the kinematics level. All our invariants are defined directly in a \textit{single and finite} lattice system, without invoking families of systems or taking any thermodynamic limit. The spatial dimension enters only through combinatorial topology, which makes the framework both rigorous and computationally tractable. Conceptually, our approach is a rebellion against TQFT approaches, which start from a functor on a category of manifolds and cobordisms, and thus inherently deal with \textit{families }of systems on different spatial manifolds. Although two theories agree in $H^{d+2}(K(G,d-p),\RR/\ZZ)$, they can produce different (but explainable) invariants in some other situations, reflecting a sharp conflict between their philosophical position of spacetime. This conflict is not about correct and wrong, but is about which mathematical definition captures the real physics. Then, the theory introduced in this paper at least provides a possibility to think outside the traditional way.

\section{Background of Statistics}\label{secBackground}

In this section, we review the background of statistics. Initially, statistics refers to the Bose-Einstein statistics of bosons and the Fermi-Dirac statistics of fermions,  corresponding to symmetric and antisymmetric wave functions, respectively. When the number of particles is not conserved, one uses second quantization, and then whether a particle is bosonic or fermionic depends on whether creation operators corresponding to different modes commute or anti-commute.

But emergent quasi-particles in many-body systems are beyond this classical scope. In (2+1)-dimensional systems, particles can obey anyonic statistics, neither bosonic nor fermionic. For example, the quasi-hole in the fractional quantum Hall phase is anyonic \cite{Laughlin1983Anomalous, RevModPhys.71.875}, but one cannot argue that from the two-anyon wave function. 
 Anyons also appear and have applications in topological quantum computing \cite{freedman2003topological, Nayak2008NonAbelian, Kitaev:2005hzj}. In the toric-code model \cite{Kitaev_2003}, anyons are always created in pairs at the end of a string, and the creation operator of a single anyon does not exist. Beyond anyons, extended excitations, such as flux loops in gauge theories and domain walls, also exhibit statistics-like behavior. Typical examples include fermionic loops \cite{CH21,FHH21,Previous,Thorngren:2014pza,johnsonfreyd2021minimal,Johnson-Freyd:2020twl} and three-loop braiding \cite{Wang2014Braiding}. In all, statistics is meaningful within a broader framework, though it remains to be understood.

In the modern view, all topological excitations in a topological order form a higher category \cite{gaiotto2019condensationshighercategories,kong2024highercondensationtheory}; then, topological excitations correspond to morphisms of various levels, and statistics corresponds to coherence data that determine the category. We call this the macroscopic theory of statistics, and it is closely related to the classification of topological phases and has similar problems. Given the types and the fusion rules of some topological excitations, researchers can predict the classification of statistics by classifying certain higher categories and construct corresponding lattice models. However, to extract these categorical data from given lattice models is a completely different story, because these data do not have direct quantum-mechanical meanings. To fill this logical gap, a microscopic theory of statistics based only on many-body Hilbert spaces, states, and operators is highly in need.

Since the macroscopic theory is well-developed and has already given classifications, the major attempt of researchers is to translate these categorical data into lattice models. Roughly speaking, these categorical data can be viewed as deformation of geometric shapes in the spacetime, and then such deformation is translated into the action of a sequence of excitation operators.
However, the choices of excitation operators are not unique, which will lead to phase ambiguities. This ambiguity causes a severe problem, making the current understanding quite limited. 

This problem already appears in the simplest notion of statistics, the topological spin, which is the phase factor $\Theta$ assigned to the exchange of two identical anyons of the type $c$ in $2$ dimensions. We should have $\Theta=0$ if $c$ is a boson and $\Theta=\pi$ if $c$ is a fermion. However, to rigorously define $\Theta$ on lattice already involves many subtleties. In particular, it is not obvious why the anyon $f$ in the toric-code model is a fermion.

The classical way to measure the topological spin is as follows. Instead of considering the whole plane, we only focus on four points labeled by $\{\overline{0},\overline{1},\overline{2},\overline{3}\}$, and then we link every pair of vertices by an edge. Anyons can be located at these points, and for every $i$ and $j$, we assign a string operator $U_{ij}$ which moves the anyon from $\overline{i}$ to $\overline{j}$. Now, the topological spin is measured by the following T-junction process\cite{Previous,Levin2003Fermions, Kawagoe2020Microscopic, FHH21}, see Fig.~\ref{figTJunction}:

\begin{figure*}
	\centering
	\includegraphics[width=0.94\linewidth]{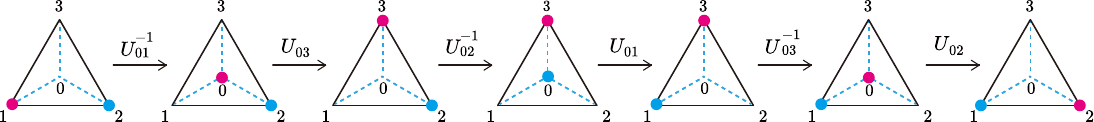}
	\caption{(Eq. (4) in \cite{Previous}) The T-junction process exchange two anyons at $\overline{1}$ and $\overline{2}$. Although the two anyons are colored differently, they must be identical, so the initial state and the final state are the same.}
	\label{figTJunction}
\end{figure*}

\begin{enumerate}
	\item Initially, assume two anyons $c$ are at  $\overline{1}$ and $\overline{2}$.
	\item move these anyons by applying string operators $U_{01}^{-1}, U_{03} , U_{02}^{-1}, U_{01}, U_{03}^{-1},  U_{02}$ one-by-one\footnote{In this paper, the term \textit{process} always refers to a multiplication sequence of excitation operators and their inverses.}.
	\item the system returns to the initial configuration, and we measure the total phase difference $e^{i\Theta}$ between the initial state and the final state.
\end{enumerate}

In other words, the topological spin is determined by
\begin{equation}\label{eqTjunction}	e^{i\Theta}=\left\langle \hbox{\raisebox{-2ex}{\includegraphics[width=1.2cm]{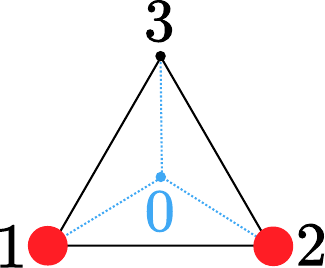}}}\right| U_{02} U_{03}^{-1} U_{01} U_{02}^{-1} U_{03} U_{01}^{-1}
	\left|\hbox{\raisebox{-2ex}{\includegraphics[width=1.2cm]{T_junction_state_12_detailed.pdf}}}\right\rangle
\end{equation}

This formula measures the topological spin for two reasons. Basically, the anyon trajectories in Fig.~\ref{figTJunction} exhibit an exchange, which aligns with the physical intuition of topological spin. Moreover, Eq.~\eqref{eqTjunction} is \textbf{operator-independent}: 
\begin{itemize}
	\item During this measurement of topological spin, \textbf{the choice string operators $U_{jk}$ is completely arbitrary}; only if the right hand side of the equation does not depend on the choice of any $U_{jk}$, the phase factor $\Theta$ can be a well-defined invariant.
\end{itemize}
 For a simple counterexample, the phase factor

\begin{equation}
	e^{i\Theta'}=\left\langle \hbox{\raisebox{-2ex}{\includegraphics[width=1.2cm]{T_junction_state_12_detailed.pdf}}}\right| U_{32}U_{21}U_{13}
	\left|\hbox{\raisebox{-2ex}{\includegraphics[width=1.2cm]{T_junction_state_12_detailed.pdf}}}\right\rangle
\end{equation}
fails to measure the topological spin: although it still exchanges two anyons geometrically, $\Theta$ is not stable when we change the choice of $U_{jk}$. For example, if $U_{21}$ is a valid string operator, then $U_{21}'=U_{21}e^{i\alpha}$ will also be a valid choice; when we replace $U_{21}$ by $U_{21}'$, $\Theta'$ becomes $\Theta'+\alpha$. To eliminate this dependence, any $U_{jk}$ and $U_{jk}^{-1}$ should always be paired, which is satisfied by the T-junction process. More generally, one can consider a modification $U_{jk}\mapsto U_{jk}V_x$, where $V_x$ is a local operator at $x\in\overline{jk}$ satisfying $V_x|a\rangle=e^{i\alpha(a|_x)}|a\rangle$. Here, $a$ labels the locations of anyons and is called a configuration; the operator $V_x$  introduces a phase factor $\alpha(a|_x)$, which depends on the local configuration at $x$. 
The cancellation of these phase factors for all vertices $x$ and phase factors $\alpha$ imposes a stronger constraint on the form of the process, We call this \textit{the local-perturbation invariance}. We show in Section~\ref{subsecSymmetry} that the T-junction process indeed satisfy this condition (also shown in \cite{Kawagoe2020Microscopic, FHH21,Levin2003Fermions}).

Intuitively, $\Theta'$ and $\Theta$ are both Berry phases of some anyon movements, but the latter is more special.
\begin{itemize}
	\item $\Theta'$ is a geometric phase: it only depends on the trajectory (string operators) but not dynamics (time).
	\item $\Theta$ is a statistical phase: it does not even depend on the trajectory (string operators)!
\end{itemize}

\begin{remark}
	One may raise a question that whether the local-perturbation argument is strong enough: in principle, we allow completely arbitrary choices of $U_{jk}$, and there is no reason to claim any two choices are connected by local perturbations. The answer to this question is very interesting: we do propose some stronger statements and call them the \textit{strong operator independence}. Both the weaker one and strong one are important: the local-perturbation argument is very good to define statistics on any simplicial complex $X$, however, the statistics may have bad behavior because the failure of the strong operator independence. If, however, we additionally assume that $X$ is a combinatorial manifold (Definition~\ref{defCombinatorialManifold}), then the strong operator independence is a consequence of the local-perturbation argument, discussed in Section~\ref{secUniversal}. This indicates that we should always work on manifolds.
\end{remark}

\begin{figure*}[t]
	\centering
	\includegraphics[width=0.85\textwidth]{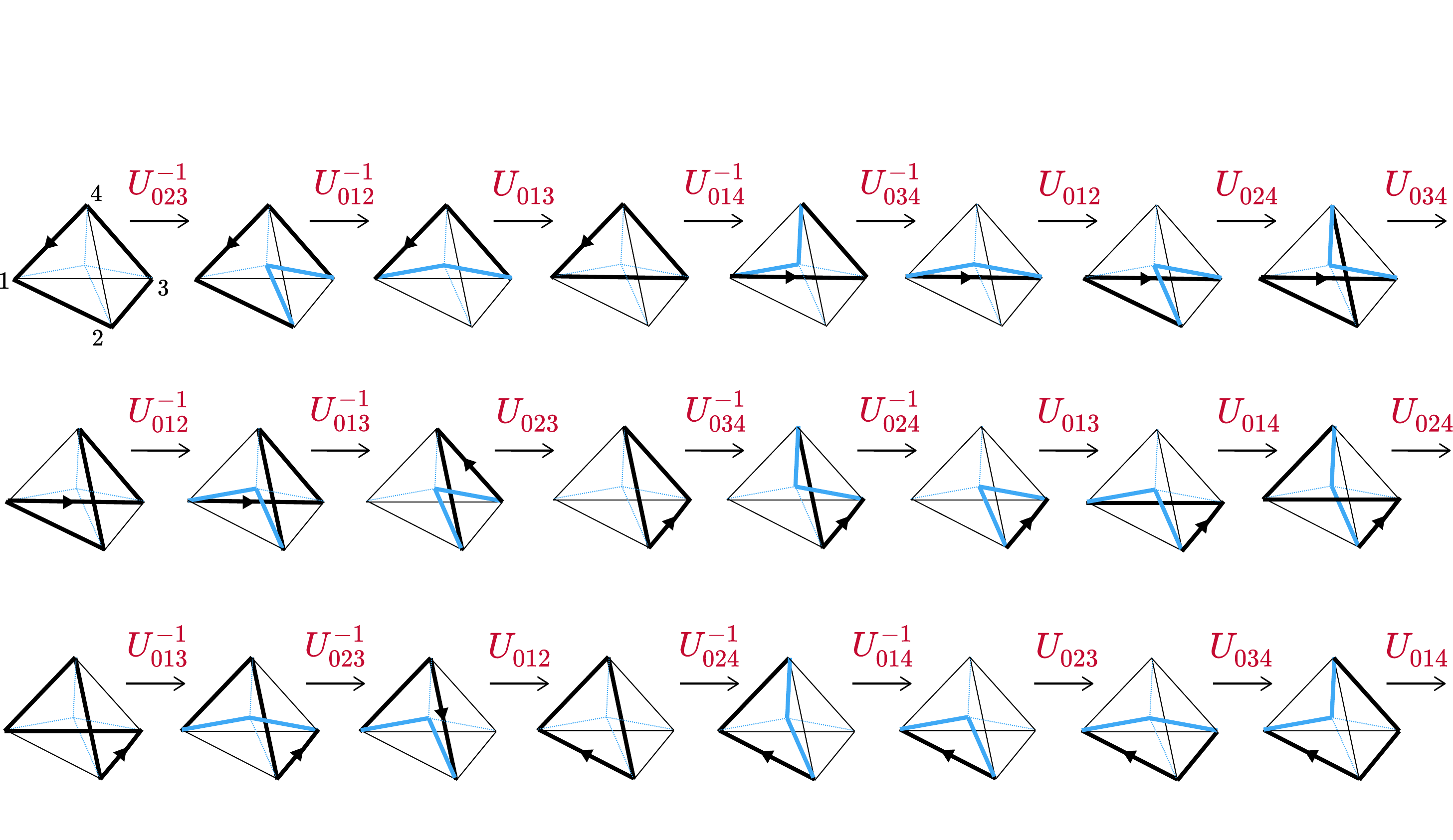}
	\caption{(Fig.~4 of \cite{Previous}) The 24-step process for detecting the statistics of loops with \( G = \mathbb{Z}_2 \) fusion in (3+1)D, which is an optimization of the $36$-step process found in \cite{FHH21}. For \( \mathbb{Z}_2 \) loops, different orientations correspond to the same configuration state, indicating that the initial and final configurations are reversed and illustrating the loop-flipping process. }
	\label{fig: 24 step process}
\end{figure*}

Since the T-junction process was found by Levin and Wen in 2003, there was essentially no higher-dimensional generalization for a long time. It is predicted for a long time from field theories that $\ZZ_2$-loops in $3$ dimensions can be "bosonic" or "fermionic", but only recently, Fidkowski, Haah, and Hastings \cite{FHH21} construct a $36$-step process to detect this statistics on lattice. In their setting, loop configurations are described by $Z_1(M,\ZZ_2)$ (1-cycles with $\ZZ_2$ coefficient on a $3$-dimensional triangulated manifold $M$), and 
excitation operators are membrane-like, creating loop excitations at their boundaries. In Fig.~\ref{fig: 24 step process}, we exhibit an equivalent but shorter process consisting $24$ steps. Using our notation, their original $36$-step process is
\begin{equation}\label{eq36Step}
	\begin{aligned}
		&U_{034}^{-1} U_{012}^{-1} U_{023}^{-1} U_{012} U_{023} U_{014} U_{013}^{-1} U_{024}^{-1} U_{012}^{-1} U_{024} U_{034} U_{012} \\
		&U_{023}^{-1} U_{014}^{-1} U_{024}^{-1} U_{014} U_{013} U_{024} U_{034}^{-1} U_{012}^{-1} U_{014}^{-1} U_{034} U_{023} U_{014} \\
		&U_{013}^{-1} U_{024}^{-1} U_{034}^{-1} U_{013} U_{034} U_{012} U_{023}^{-1} U_{014}^{-1} U_{013}^{-1} U_{023} U_{013} U_{024}
	\end{aligned}
\end{equation}

The geometric meaning of these (36 or 24 steps) processes is that they rotate a loop around a diameter for an angle of \( \pi \), so its orientation is reversed. This is similar to the T-junction process, where we rotate $S^0$ for an angle of $\pi$. However, the operator independence condition for loops becomes much more restrictive. The local-perturbation argument produces many complicated constraints, and that is why flipping a loop should use $24$ or $36$ steps. Indeed, these processes are asymmetric, and the fact that they satisfy these constraints looks very magical. Without doubt, generalizing this method to higher dimensions will be much more difficult, and geometric pictures also become less intuitive. 

From our point of view, one should not translate the macroscopic theory into the microscopic theory; instead, one should give up categories and geometric pictures and establish the microscopic theory directly. In our systematic generalization of the T-junction process and the 36-step process, the operator independence is no longer a difficulty but as the definition of a statistical process. Using this theory, we have successfully found many new statistical processes, such as the self-statistics of membranes \cite{feng2025anyonicmembranespontryaginstatistics}. As a side product, we find an equivalent and shorter statistical process consists of 24 steps (Fig.~\ref{fig: 24 step process}), which we believe to be the shortest.

The core of our paper is Section~\ref{sec: Framework}, where we establish our theory of excitations and statistics. In Section~\ref{secComputation}, we show how to use a computer to compute statistics and discuss the result. In some cases the statistics is isomorphic to $H^{d+2}(K(G,d-p),\RR/\ZZ)$, and we will give the proof for a special case in Appendix \ref{appendixSimplicialSet}. In Section~\ref{secFunctorial}, we will discuss the relationship of statistics in different dimensions and having different fusion groups. In Section~\ref{secUniversal}, we will study the strong operator independence and discuss its relationship with the manifold structure.

Since our definitions are novel, it is natural to ask why our definitions are "correct". In fact, we have designed our definitions very carefully so that they will lead to both good theoretical properties and computational results, and we have aimed to make this work self-contained, unambiguous, rigorous, and how it is against the common intuitions. Unfortunately, this style may also make the presentation feel too abstract. In such a situation, we recommend that the reader compare this paper with our previous paper \cite{Previous}, whose definitions are less rigorous but more intuitive. Also, reading \cite{FHH21} is helpful to obtain more motivation for this work.

\section{Framework for Statistics}\label{sec: Framework}

Our theory is all about a given family of \textit{configuration states $\{|a\rangle|a\in A\}$} and \textit{excitation operators $\{U(s)|s\in S\}$} satisfying the configuration axiom and the locality axiom. Here, $A$ and $S$ are at the mathematical level and together form an excitation pattern. The corresponding configuration states and excitation operators live on the kinematical level and will be called realizations of this pattern. This follows the three–layer philosophy of the introduction, but with a small difference. When discussing the three layers of Landau’s symmetry-breaking theory, we attribute different phases to different symmetry-breaking patterns.
 But in our framework, we usually fix an excitation pattern, corresponding to a specific dimension and fusion group; different statistics correspond to different classes of realizations of this single pattern. A better analogy between our theory and Landau's symmetry breaking theory is as follows: 

\begin{itemize}
	\item an excitation pattern $m$ $\longleftrightarrow$ a symmetry $G$;
	\item a realization of $m$ $\longleftrightarrow$ a set of states with a transitive $G$-action;
	\item the statistics of a realization $\longleftrightarrow$ the stabilizer subgroup of some state;
	\item the statistics of $m$ $\longleftrightarrow$ all symmetry-breaking patterns of $G$.
\end{itemize}

To illustrate the meaning of $|a\rangle$ and $U(s)$ more concretely, let us consider the $\ZZ_2$-Toric code on a triangulated sphere $S^2$ \cite{Kitaev_2003}; we denote this triangulation by $X$ as before. Every edge $l$ is assigned a qubit degree of freedom, and the Hamiltonian is $H=-\sum \mathcal{A}_v-\sum \mathcal{B}_p$. The operator $\mathcal{A}_v=\prod_{v\subset l}X_l$ detects whether an $e$ anyon is at the vertex $v$, and the operator $\mathcal{B}_p=\prod _{l\subset p}Z_l$ detects whether an $m$-anyon is at the plaquette $p$. To study the statistics of the $e$ anyon, we define that a configuration state $|a\rangle$ is a common eigenstate of all $\mathcal{A}_v,\mathcal{B}_p$ operators such that $\mathcal{B}_p|a\rangle=|a\rangle$. This means that we restrict to states without $m$-anyons, and the numbers and locations of $e$-anyons should be definite. Different configuration states are distinguished by the number and locations of $e$-anyons, characterized by the eigenvalues $\mathcal{A}_v|a\rangle=\pm|a\rangle$ for all vertices $v$. Therefore, we can identify the label $a$ as the $0$-chain (with $\ZZ_2$ coefficients) encoding the locations of $e$-anyons: $a=\sum_{v:\mathcal{A}_v|a\rangle=-|a\rangle}v$. Because the total number of $e$-anyons must be even, configurations are always $0$-boundaries: $A=B_0(X,\ZZ_2)$.

Next, excitation operators are string operators that create two $e$-anyons at two ends. We here denote the string by $L$ and the two ends by $x$ and $y$. The choice of string operators is not unique, but the simplest way is to take the product of Pauli $Z$ along $L$: such an operator commutes with all $\mathcal{B}_p$ and $\mathcal{A}_v$ except that it anti-commutes with $\mathcal{A}_x$ and $\mathcal{A}_y$. This string operator can be interpreted in multiple ways: it can move an anyon from $x$ to $y$, create a pair consisting of an anyon and an anti-anyon (a hole), or annihilate a pair of anyon and anti-anyon at $x$ and $y$, respectively\footnote{In the toric-code model, $e$ is the anti-anyon of itself, so there is no distinguish between them.}. We do not distinguish these pictures: in any case, we assume that a string operator creates a pairs of anyon and anti-anyon at two ends, and anyons at the same location fuse automatically.

A long string operator can always be decomposed into shorter ones. So, we only need to include some "elementary" strings for our set $\{U(s)\}$ of excitation operators. For $e$ anyon in the toric-code model, it is enough to take $S$ to be the set of all edges and to take $U(s)$ as the Pauli $Z$ at the edge $s$.

These data satisfy the two axioms below; here we give a sketch, and the rigorous version is in Definition~\ref{defRealization}.
\begin{enumerate}
	\item The configuration axiom: the action of $U(s)$ brings any configuration state $|a\rangle$ to another configuration state $|a+\partial s\rangle$ up to a phase factor, where $\partial s\in B_0(X,\ZZ_2)$ is the chain-level boundary of $s$.
	\item The locality axiom: any $U(s)$ is supported near $s$, in the sense of that they satisfy some commutativity relations (Eq.~\eqref{axiomLocalityIdentity}) related to the support.
\end{enumerate}

Both axioms are physically reasonable and very basic. The configuration axiom is implicitly assumed in the literature and is necessary to defining statistics. If the final state after the T-junction process was not proportional to the initial state, the phase $e^{i\Theta}$ as in Eq.~(\ref{eqTjunction}) would become ill-defined.
  The locality axiom is our characterization of many-body structure. This axiom holds trivially in this example of $e$-anyons because all Pauli $Z$ commute. In general, $U(s)$ for edge $s$ may involve some nearby degrees of freedom, and $U(s),U(s')$ for disjoint $s,s'$ may not commute; however, we can always repair the locality axiom by using longer (but still finite) string operators. In other words, we consider excitations on a coarser triangulation than the original lattice.

Whenever these two axioms are satisfied, our theory will produce an Abelian group $T^*$ and an element $\sigma\in T^*$ as the output. In this example, we will get $T^*\simeq \ZZ_4$ and $\sigma=0$. The former means that an anyon of fusion group $\ZZ_2$ can be boson, semion, fermion, or anti-semion; the latter means that the $e$ anyon in the toric-code model is, without doubt, a boson. The Abelian group $T^*$ only depends on the excitation pattern, while $\sigma\in T^*$ is different for different realizations, dividing realizations into several equivalence classes.

This picture naturally extends to general cases. There are three parameters: the spatial dimension $d$, the dimension of excitation $p$, and the fusion group $G$. For the $e$ anyon in the toric-code model, we have $d=2$, $p=0$, and $G=\ZZ_2$. In general, we can take $X$ as a triangulation of the $d$-sphere $S^d$, and the configuration group is $A=B_p(X,G)$. Excitation operators should correspond to the set $(p+1)$-simplexes of $X$ (denoted by $X_{p+1}$) and create excitations at their boundaries. We can take $S=X_{p+1}$ when $G\simeq \ZZ_n$; when $G$ requires more than one generators, we have to take $S=G_0\times X_{p+1}$, where $G_0$ is a subset generating $G$. For example, the four anyons $1,e,m,f$ in the toric-code model correspond to $G=\ZZ_2\times \ZZ_2$, and then we need to take $G_0=\{e,m\}$. For any $X,p$, and $G$, we define and compute the statistics $T^*$ in the same way. Whenever our computer program can compute $T^*$ in a reasonable time, we find $T^*\simeq H^{d+2}(K(G,d-p),\RR/\ZZ)$. Note that this does not depend on the choice of triangulation $X$ and the generating set $G_0$.

With these intuitions, we can begin our formal definition.

\subsubsection*{(Optional) Historical remarks and reading suggestions}

Readers can go directly to the next section; but according to our experience, many readers struggle about the intuition of our definitions, especially for \textit{localization}. In this case, knowing how we establish this theory may help organize the logic. 

At the beginning of the project, we were struggling to understand the complicated $36$-step process Eq.~\eqref{eq36Step}, and a particular difficulty was that excitation operators did not have to commute. However, we find that \cite{FHH21} implicitly assumes a weak version of commutativity, which we write as the configuration axiom Eq.~\eqref{eqChangeConfig}. Under this assumption, non-commutative processes are translated into the  \textit{expression group} $E$, a free Abelian group generated by all "steps" (applying an operator on a configuration). Then, the fact that the phase factor of $36$-step process is $0$ or $\pi$ quantized becomes a puzzle. If we use $e\in E$ to denote the expression of the $36$-step process, then $0\ne 2e\in E$, but the quantization of the phase factor suggests that $2e$ is always trivial! There is an argument in \cite{FHH21} using locality, but it is not completely correct. Regardless of the correct proof, there must be some mechanism making some expressions "trivial", which should form a subgroup $E_{\operatorname{id}}\subset E$. We tried many definitions of $E_\id$, and in some cases, $E/E_\id$ contains exactly one torsion-$2$ element, exactly corresponding to the $36$-step process. This explains the quantization of statistics, and then we propose that the statistics $T_f$ is the torsion part of $E/E_\id$ (Definition~\ref{defTf}). This definition is simpler than the local-perturbation argument in the literature. 

The next thing is what the correct definition of $E_\id$ in general dimensions is. Although locality is usually argued in the literature in an intuitive way, we believe we must give up all intuitions and build everything from basic facts of quantum mechanics. For example, how to understand the notion "support"? In a many-body Hilbert space, it is natural to talk about the support of an operator---the degrees of freedom it acts on. \textbf{However, one should never talk about the support of a state, which has no mathematical meaning at all}. People usually talk about the support of an anyon; \textbf{in contrast, we only treat it as an illusion or a derived notion from the support of operators.} This principle is the main reason we introduce the locality axiom Eq.~\eqref{axiomLocalityIdentity} and define $E_\id$ using Eq.~\eqref{eqBasicLocalityIdentityInEid}, which is different from traditional arguments \cite{FHH21,Previous}. The correctness of this definition is confirmed by the computation of $T_f$ on a computer, which produce $T_f\simeq H_{d+2}(K(G,d-p),\ZZ)$.\footnote{ Many readers may see $H^{d+2}(K(G,d-p),\RR/\ZZ)$ more often; actually, $T_f$ is the dual concept of classification, so it is not $H^{d+2}(K(G,d-p),\RR/\ZZ)$ but its Pontryagin dual $H_{d+2}(K(G,d-p),\ZZ)$. These are discussed in Section~\ref{subsecPhaseData}. } Following this logic line, the quickest way is to read Section~\ref{subsecDefinition}, Section~\ref{subsecExpression}, and Definition~\ref{defTf}; then, one can go to Section~\ref{secComputation} to see the computation. 

After that, we try to answer whether taking the torsion part is equivalent to the criterion used in \cite{FHH21}; we rewrite it in Theorem~\ref{thmCheckByHand}. At that time, we felt this criterion questionable because we can propose many different criteria to interpret operator independence, and we do not know which one is better. We have develop many technical tools to study their relationships. Two of them, which we called quantum cellular automata and condensation, are put together with the discussion of operator independence in Section~\ref{secUniversal}. Another concept named \textit{localization}, turns out to be the best way to define statistics (Definition~\ref{defStatisticsMain}), so we introduce it earlier.

\subsection{Excitation patterns and their realizations}\label{subsecModel}

\begin{definition}\label{defexcitationModel}
	An \textit{excitation pattern} \(m\) (in the topological space $M$) consists of the following data $(A,S,\partial,\supp)$:
	\begin{enumerate}
		\item A (finite) \footnote{One may assume $A,S$ are finite for simplicity, while many constructions still work when they are infinite.} Abelian group \(A\), referred to as the configuration group.
		\item A finite set \(S\), referred to as formal excitation operators, and a map \(\partial: S \longrightarrow A\) such that $\{\partial s|s\in S\}$ generates \(A\).
		\item A topological space $M$ and a closed subspace \(\operatorname{supp}(s) \subset M\) for every $s\in S$.
	\end{enumerate}

\end{definition}

Our primary example involves \(p\)-dimensional invertible topological excitations with Abelian fusion group \(G\) in a simplicial complex $X$ (Appendix~\ref{appendixSimplicialComplex}). Before proceeding, we mention a technical detail. Let $\alpha,\beta\in G$ be two excitation types, and then the excitation type $\alpha+\beta\in G$ corresponds to the fusion of $\alpha$ and $\beta$. When considering the corresponding excitation operators, we can either take the product of excitation operators of $\alpha$ and $\beta$ or introduce new ones. In general, we only need to introduce excitation operators for several types, which form a generating subset $G_0\subset G$ (i.e., elements in $G_0$ generate $G$). For example, one can take $G_0=G$ or $G_0=G-\{0\}$. Different choices of $G_0$ essentially mean different choices of excitation operators. This should influence the statistics $T^*$ because the statistics should be operator independent.  In fact, the statistics defined later indeed depends on $G_0$ for some simplicial complexes (see Example~\ref{expNotManifold}), while it is independent of $G_0$ when $X$ is a combinatorial manifold (see Theorem~\ref{thmG0independence},~\ref{thmTrivialStatisticsInduction} and Definition~\ref{defCombinatorialManifold}). This fact indicates that we should always work on a manifold. In that case, we want to choose the smallest $G_0$ during computation. Writing $G\simeq \ZZ_{n_1}\oplus\cdots\oplus \ZZ_{n_k}$, we will take the generator $1\in \ZZ_{n_i}$ for every $\ZZ_{n_i}$ component; when $G=\ZZ_n$, taking $G_0=\{1\}$ is enough.

\begin{definition}\label{expSimplicialComplex}
	The excitation pattern $m_p(X,G)$ is constructed using the following data:
	\begin{enumerate}
		\item \(A\) is the group of \(p\)-dimensional simplicial boundaries, \(A = B_p(X, G)\).
		\item \(S = G_0\times X_{p+1}\), where \(X_{p+1}\) is the set of all \((p+1)\)-simplexes of $X$, and \(G_0\) is a generating subset of \(G\). 
		\item Viewing $X$ as a topological space, \(\operatorname{supp}(s)\) gives the geometric image of the simplex itself, while \(\partial\) is the usual homological boundary map.
	\end{enumerate}

\end{definition}

\begin{remark}
	Only the $(p+1)$-skeleton of $X$ contributes to the excitation pattern, so we do not care about simplexes in higher dimensions.
\end{remark}

This definition covers the T-junction process in $2$ dimensions and $24$-step process in $3$ dimensions, where $X$ is chosen to be \hbox{\raisebox{-2ex}{\includegraphics[width=1.2cm]{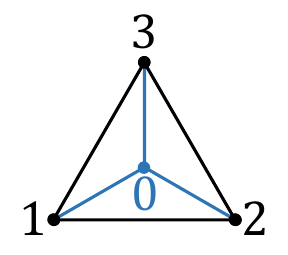}}} and \hbox{\raisebox{-2ex}{\includegraphics[width=1.2cm]{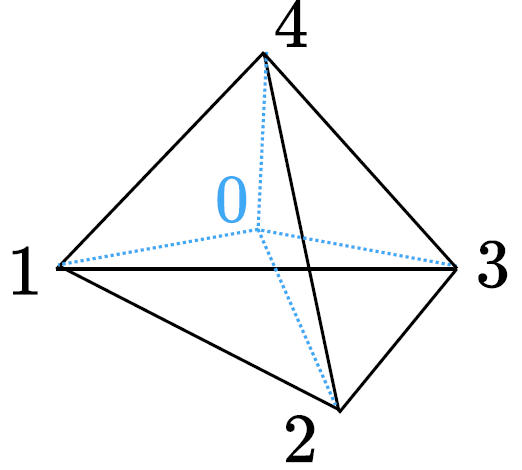}}}, respectively. These simplicial complexes are studied in the literature because they are the smallest simplicial complex embedded in the spatial manifold that exhibit anyon statistics and fermionic loop statistics, respectively. We find a better explanation: the two simplicial complex correspond to the simplex $\Delta^{d+1}$, or equivalently, its boundary $\partial \Delta^{d+1}\simeq S^d$, where $d$ is the spatial dimension. Computation suggests that different triangulations of $S^d$ all give $T^*\simeq H^{d+2}(K(G,d-p),\RR/\ZZ$, and $\partial \Delta^{d+1}$ is the smallest triangulation.

An excitation pattern $m=(A,S,\partial,\supp)$ is at the mathematics level. It obtains physical meaning only when it is realized by a lattice model, where $a\in A$ corresponds to a \textit{configuration state} $|a\rangle$ and $s\in S$ corresponds to a \textit{excitation operator} $U(s)$, additionally satisfying two axioms.

\begin{definition}\label{defRealization}
	A realization of the excitation pattern \(m=(A,S,\partial,\supp)\) consists of a Hilbert space $\mathcal{H}$, a collection of \textit{configuration states} $\{|a\rangle|a\in A\}$ such that $|a\rangle$ and $|a'\rangle$ are either orthogonal or collinear\footnote{One can always assume that $\{|a\rangle\}$ is a basis of \(\mathcal{H}\); this is equivalent to considering general cases in the sense of Definition~\ref{defEqualRealization}.}, and a collection of \textit{excitation operators} $\{U(s)|s\in S\}$, satisfying the following two axioms.
	\begin{itemize}
			\item \textbf{The configuration axiom (state version):} For any $s\in S$ and $a\in A$, the equation
			\begin{equation}\label{eqChangeConfig}
				U(s)|a\rangle=e^{i\theta(s,a)}|a+\partial s\rangle
			\end{equation}
			holds for some $\theta(s,a)\in \RR/2\pi\ZZ$.
			\item \textbf{The locality axiom:} \(\forall s_1, s_2, \dots, s_k \in S\) satisfying \(\operatorname{supp}(s_1) \cap \operatorname{supp}(s_2) \cap \cdots \cap \operatorname{supp}(s_k) = \emptyset\), the equation
			\begin{align}\label{axiomLocalityIdentity}
				[ U(s_k), [\cdots, [ U(s_2),  U(s_1)]]]=1 \in U(\mathcal{H}),
			\end{align}
			holds, where \([a,b] = a^{-1}b^{-1}ab\).
	\end{itemize}
	We use the symbol $U$ to label a realization, and we denote all realizations of an excitation pattern $m$ by $R(m)$.
\end{definition}

A process is a multiplicative sequence of excitation operators; for example, the T-junction process is
\begin{equation}\label{eqO}
	O=U(s_{02})U(s_{03})^{-1}U(s_{01})U(s_{02})^{-1}U(s_{03})U(s_{01})^{-1}.
\end{equation}
 It is convenient to write $O=U(g)$ and $g=s_{02} s_{03}^{-1} s_{01} s_{02}^{-1} s_{03} s_{01}^{-1}$, where $g$ is nothing but a formal product of formal excitation operators and their inverse. Mathematically, $g$ is an element of the non-Abelian free group $\operatorname{F}(S)$ generated by the set $S$. The set-theoretical maps $\partial: S\rightarrow A$ and $U: S\rightarrow U(\mathcal{H})$ naturally extend to group homomorphisms $\partial: \operatorname{F}(S)\rightarrow A$ and $U: \operatorname{F}(S)\rightarrow U(\mathcal{H})$. Under this notation, Eq.~\eqref{axiomLocalityIdentity} should be written as
 \begin{equation}
 	U([s_l,[\cdots,[s_2,s_1]]])=1.
 \end{equation}
 
 Now we explain more about the two axioms. Eq.~\eqref{eqChangeConfig} is necessary to define statistics: in previous studies, statistical phases are all measured in the form of $\langle a|U(g)|a\rangle$, where $\partial g=0$; Eq.~\eqref{eqChangeConfig} is a natural condition to ensure $U(g)|a\rangle\propto |a\rangle$. 
 
 \begin{remark}
 	 Eq.~\eqref{eqChangeConfig} does impose a strong constraint on excitation operators. In fact, we can rewrite it in an purely "operator version". 
 	\begin{itemize}
 		\item \textbf{The configuration axiom (operator version)}: Any two operators in the set $\{U(g)|g\in \operatorname{F}(S),\partial g=0\}$ commute.
 	\end{itemize}
 	The state version and the operator version of the configuration axiom are almost equivalent. Assuming the state version, operators $\{U(g)|g\in \operatorname{F}(S),\partial g=0\}$ share all $|a\rangle$ as their common eigenstates, so they must commute, at least in the subspace of $\mathcal{H}$ generated by $\{|a\rangle\}$. For example, we have
 	\begin{equation}\label{eqCommute}
 		U([[s_4,s_3],[s_2,s_1]])=1.
 	\end{equation}
 	If the fusion group is $\ZZ_2$, then we also have equations like
 	\begin{equation}
 		U([s^2,[s_2,s_1]])=1.
 	\end{equation}
 	Conversely, if we assume the operator version of configuration axiom, we can recover the configuration states as their common eigenstates. There is a freedom of choosing the vacuum state, but this is not important in our analysis. 
 	One can view $\{U(g)|\partial g=0\}$ as symmetry operators and $\{U(s)\}$ as symmetry patch operators in the sense of \cite{chatterjee-2023}. Our axioms are different from them, so we leave further comparison to future works.
 	
 	The problem of the operator version is that the set $\{U(g)|g\in \operatorname{F}(S),\partial g=0\}$ contains infinite elements and inconvenient to use. Thus, we only use the state version in the rest of this paper.
 \end{remark}

The locality axiom Eq.~(\ref{axiomLocalityIdentity}) characterizes that $\mathcal{H}$ is the tensor product of local degrees of freedom. If the supports of two excitation operators \(s_1\) and \(s_2\) are disjoint, i.e., \(\operatorname{supp}(s_1) \cap \operatorname{supp}(s_2) = \emptyset\), then for every realization \(U\), the operators \( U(s_1)\) and \( U(s_2)\) act on disjoint regions and therefore commute. We have
 \begin{equation}\label{eqCommutator}
 	 U([s_2, s_1]) = 1,
 \end{equation}

which corresponds to Eq.~(\ref{axiomLocalityIdentity}) for the case of \(k=2\).

It can be argued that excitation operators \( U(s)\) are always finite-depth local circuits. This implies that $ U([s_2,s_1])$ is a finite-depth local circuit supported in a small neighborhood of  \(\operatorname{supp}(s_1) \cap \operatorname{supp}(s_2)\). We ignore this neighborhood, and then we obtain Eq.~(\ref{axiomLocalityIdentity}) for $k=3$. Other cases are argued similarly. 

\begin{remark}\label{remarkCombinationData}
Ignoring this small neighborhood does not affect the rigor of our framework because we take Eq.~\eqref{axiomLocalityIdentity} rather than "finite-depth local quantum circuits" as our axiom. In fact, the notions of support and tensor-product Hilbert space are only involved combinatorially through the locality axiom, so the exact geometric shape of $\supp(s)$ is not important.
\end{remark}

We give some example of realizations in the end of this subsection, while skipping them will not affect the main logic of our theory. Example~\ref{expTrivialFSymbol} and \ref{expTwisted} show that on the same $1$-dimensional ferromagnetic ground state, different domain-wall excitations can have different statistics in our framework, which suggests that statistics should be interpreted from the view of symmetry anomaly and symmetry defect.
 Example~\ref{expH2(G)} and \ref{exampleGroupCohomology} provides some intuitions about how our constructions are relevant but different to the traditional construction of group cohomology.

\begin{example}\label{expTrivialFSymbol}
	Let $X$ be a polygon with $n$ edges, and then $m_0(X,\ZZ_2)$ is the excitation pattern of point excitation on this 1-dimensional chain. We give a realization describing the domain-wall excitation in the ferromagnetic phase. We assign a qubit on every edge and the Hilbert space $\mathcal{H}$ is their tensor product. In other words, $\mathcal{H}$ is spanned by $|b\rangle\;(b\in C_1(X,\ZZ_2))$. We define the ground state  $\frac{1}{\sqrt{2}}(|0\cdots0\rangle+|1\cdots1\rangle)$ and  excitation operators $ U(s_i)=X_i$, where $s_i$ is the $i$th edge, and $X_i$ is the Pauli $X$ operator at that site. All such excitation operators commute with the global symmetry operator $P=\prod_{i}X_i$. Since the ground state satisfies $P|0\rangle=|0\rangle$, configuration states do not span the whole Hilbert space $\mathcal{H}$ but only the $P=1$ subspace. Specifically, a configuration $a\in B_0(X,\ZZ_2)$ describes the location of the domain walls, and the corresponding state is $\frac{1+P}{\sqrt{2}}|b\rangle$, where $\partial b=a$.
\end{example}

\begin{example}\label{expTwisted}
	We give another realization of $m_0(X,\ZZ_2)$ which is very similar to the previous example but has different statistics. The ground state is still $\frac{1}{\sqrt{2}}(|0\cdots0\rangle+|1\cdots1\rangle)$, but we define excitation operators $ U(s_i)=CZ_{i-1,i}X_iCZ_{i,i+1}$, where $CZ_{ij}|b\rangle=(-1)^{b_ib_j}|b\rangle\;(b\in C_1(X,\ZZ_2))$. There is an anomalous global symmetry operator $P'=(\prod_{i}X_i)(\prod_{i}Z_iCZ_{i,i+1})$ that commutes with all $ U(s_i)$, so all configuration states span the subspace $P'=1$. In fact, for any configuration $(a\in B_0(X,\ZZ_2))$, the corresponding state is
	
	\begin{equation}
		|a\rangle=\frac{1+P'}{\sqrt{2}}|b\rangle,\; b\in C_1(X,\ZZ_2), 
	\end{equation}
	where $\partial b=a$.
	
	These string operators do not commute; for example, $ U(s_{i+1}) U(s_i)=- U(s_i) U(s_{i+1})Z_iZ_{i+1}$ (note that all configuration states are eigenstates of $Z_iZ_{i+1}$). Specifically, we have $[ U(s_{i+1})^2, U(s_i)]=-1$. This $-1$ is statistical in the sense that it is invariant under local perturbations.
	
	One may find this example puzzling, since by multiplying these $ U(s_i)$ we obtain a long string operator $ U(s_i) U(s_{i+1})\cdots U(s_{i+k})=CZ_{i-1,i}X_iX_{i+1}\cdots X_{i+k}CZ_{i+k,i+k+1}$ (all intermediate $CZ$ cancel). Compared to the previous example, the only difference is the two $CZ$ operators decorated at the ends of the string. This is of course a local perturbation, which seems to contradict the local-perturbation invariance of statistics. The answer to this puzzle is that we only allow local perturbations that do not change configuration states. In this example, the decoration of $CZ$ operators changes the subspace spanned by the configuration states from $P=1$ to $P'=1$, and then the statistical phase is not protected.
\end{example}

\begin{example}\label{expH2(G)}
The statistics can be extended to $(-1)$-dimensional excitations through the following definition of $m_{-1}(X,G)$.
	\begin{enumerate}
		\item \(A=G\).
		\item \(S = G\times X_0\), where \(X_0\) is the set of all vertices of $X$. We use $gx$ where $ g\in G,x\in X_0$ to denote $(g,x)\in S$.
		\item $\partial(gx)=g$. 
	\end{enumerate}
	
	In this simplest example, only the number of vertices of $X$ influences the excitation pattern. In fact, the statistics is nontrivial only when $X$ has exactly two vertices $x_1,x_2$, i.e., when $X=S^0$. We now construct realizations of $m_{-1}(S^0,G)$ using group cohomology. Recall that $2$-cocycle is a function $\omega:G\times G\rightarrow \operatorname{U}(1)$ satisfying $\omega(k,h)\omega(g,hk)=\omega(gk,h)\omega(g,k)$. Consider a Hilbert space $\mathcal{H}$ spanned by $|g\rangle\;(g\in G)$; given a cocycle $\omega$, we define $ U(gx_1)|h\rangle=\omega(g,h)|gh\rangle$ and $ U(gx_2)|h\rangle=\omega(h,g)|gh\rangle$. The cocycle condition ensures Eq.~\eqref{axiomLocalityIdentity} that $[ U(gx_1), U(hx_2)]|k\rangle=\omega(k,h)\omega(g,hk)\omega(gk,h)^{-1}\omega(g,k)^{-1}|h\rangle=|h\rangle$, so we have defined a realization of $m_{-1}(S^0,G)$ for any 2-cocycle. As to be defined in the next section, coboundaries correspond to "local realizations" and should be considered trivial, so the equivalence classes are classified by $H^2(G,\RR/\ZZ)$. 
	
\end{example}

\begin{example}\label{exampleGroupCohomology}
	We now construct a realization of $m_{d-1}(\partial\Delta^{d+1},G)$ related to the group cohomology $H^{d+2}(G,\operatorname{U}(1))$. By passing to the dual cell complex, we set $A=B^1(\partial\Delta^{d+1},G)$ and the excitation operators are labeled by $(d+2)$ vertices of $\Delta^{d+1}$ \footnote{By definition, we say their support has empty intersection is the vertices do not share a common $d$-simplex.} and the group element $g\in G$.  Let $\nu(g_0,\cdots,g_{d+2})$ be a cocycle, i.e., it satisfies $\nu(gg_0,\cdots,gg_{d+2})=\nu(g_0,\cdots,g_{d+2}),\forall g\in G$ and $\sum (-1)^i \nu(g_0,\cdots,\hat{g_i},\cdots,g_{d+3})=0,\forall g_0,\cdots,g_3\in G$. We construct a realization as follows:
	\begin{widetext}
	\begin{enumerate}
		\item the Hilbert space is spanned by $|g_0,\cdots,g_{d+1}\rangle$ and the ground state is $|0\rangle=\sum_{g\in G}|g,g\cdots ,g\rangle$.
		\item Let $s_i(g)$ be the operator labeled by vertex $i$ and $g\in G$, $ U(s_i(g))$ is defined by
		
				\begin{equation}
				 U(s_i(g))|g_0,\cdots ,g_{i},\cdots,g_{d+1}\rangle=(-1)^i\nu(g_0,\cdots,g_{i-1},g_i,gg_i,g_{i+1},\cdots,g_{d+1})|g_0,\cdots ,gg_{i},\cdots,g_{d+1}\rangle
			\end{equation}

	\end{enumerate}
	
	Why do Eq.~\eqref{axiomLocalityIdentity} hold in this case is very interesting. First, we notice that every vertex should appear in the higher commutator, so without loss of generality, we only need to check
	\begin{equation}
		 U([s_{d+1}(g_{d+1}),[\cdots,[s_1(g_1),s_0(g_0)]]])|h_0,\cdots,h_{d+1}\rangle=|h_0,\cdots,h_{d+1}\rangle.
	\end{equation}
\end{widetext}
	We sketch the proof using double-cone construction. To begin with, we draw a tetrahedron (as a triangulation of $S^2$) whose vertices are labeled by $h_0,g_0h_0,h_1,g_1h_1$. Next, we introduce two new vertices labeled by $h_2,g_2h_2$ and join each of them with every $2$-simplex. After this process, we get a triangulation of $S^3$ with eight $3$-simplexes. Repeating the process, we get a triangulation of $S^{d+2}$ with $2^{d+2}$ of $(d+2)$-simplex. Choosing the orientation carefully, one will find the total phase of the higher commutator exactly coincide with the sum of $\nu$-values on this triangulation of $S^{d+2}$. This phase being zero is strictly weaker than the cocycle condition. 
	
	More generally, one can construct a realization of $m_p(\partial\Delta^{d+1},G)$ using a cocycle in $Z^{d+2}(K(G,d-p),\RR/\ZZ)$; see Theorem~\ref{thmEvaluation} and \cite{feng2025anyonicmembranespontryaginstatistics}.
\end{example}

\subsection{Phase data, localization, and statistics}\label{subsecPhaseData}

Let $U$ be a realization of an excitation pattern $m$. Then, for any $s\in S$ and $a\in A$, the phase factor \( \theta(s,a)\in \RR/2\pi\ZZ \) is determined by

\begin{align}\label{eqPhaseData}
	 U(s) |a\rangle = e^{i\theta(s,a)} |a + \partial s\rangle.
\end{align}

More generally, for any process $g\in\operatorname{F}(S)$, we have a phase factor $\theta(g,a)\in \RR/2\pi\ZZ$ determined by
\begin{align}\label{eqDefTheta(g,a)}
	U(g) |a\rangle = e^{i\theta(g,a)} |a + \partial s\rangle.
\end{align}

Using
\begin{equation}
	\left\{
	\begin{aligned}
		&U(g_2)U(g_1)|a\rangle=e^{i\theta(g_1,a)+i\theta(g_2,a+\partial g_1)}|a+\partial g_1+\partial g_2\rangle\\&U(g)^{-1}|a\rangle=e^{-i\theta(g,a-\partial g)}|a-\partial g\rangle,
	\end{aligned}
	\right.
\end{equation}
we get that
\begin{equation}\label{eqThetaP(g,a)}
	\left\{
	\begin{aligned}
		\theta(g_2g_1,a)&=\theta(g_1,a)+\theta(g_2,a+\partial g_1)\\
		\theta(g^{-1},a)&=-\theta(g,a-\partial g);
	\end{aligned}
	\right.
\end{equation}
applying this formula inductively, we can expand any $\theta(g,a)$ as a linear combination of $\{\theta(s,a)|s\in S,a\in A\}$. For example, the statistical phase of the T-junction process (Eq.~\eqref{eqTjunction}) can be written as:
	\begin{align}\label{eq3.2.2}
		\Theta =& -\theta(s_{01}, \hbox{\raisebox{-1.5ex}{\includegraphics[width=0.8cm]{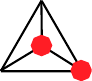}}}) + \theta(s_{03}, \hbox{\raisebox{-1.5ex}{\includegraphics[width=0.8cm]{T_junction_state_02.pdf}}}) - \theta(s_{02}, \hbox{\raisebox{-1.5ex}{\includegraphics[width=0.8cm]{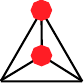}}})  \notag\\&+ \theta(s_{01}, \hbox{\raisebox{-1.5ex}{\includegraphics[width=0.8cm]{T_junction_state_03.pdf}}}) - \theta(s_{03}, \hbox{\raisebox{-1.5ex}{\includegraphics[width=0.8cm]{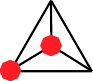}}}) + \theta(s_{02}, \hbox{\raisebox{-1.5ex}{\includegraphics[width=0.8cm]{T_junction_state_01.pdf}}})
	\end{align}
	
Actually, we assume that all information relevant to statistics can be extract from these phase factors. This indicates that rather study states and operators, we should study these phase factors directly. 
\begin{definition}\label{defEqualRealization}
Let $m=(A,S,\partial,\supp)$ be an excitation pattern. For any realization $U$ of $m$, the corresponding phase factors $\{\theta(s,a)|s\in S,a\in A\}$ are called the phase data of $U$. We say that two realizations $U,U'$ are equal if and only if their phase data are equal.
\end{definition}

This shift in perspective is important because linear operations of phases are much easier than non-commutative multiplications of operators. In the new perspective, Eq.~\eqref{axiomLocalityIdentity} converts to

\begin{equation}\label{eqLinearLocalityIdentityEquation}
	\theta([s_k, [\cdots, [s_2, s_1]]],a)=0,
\end{equation}
for $\cap_{i=1}^k\supp(s_i)=\emptyset$ and $\forall a\in A$. Using Eq.~\eqref{eqThetaP(g,a)}, Eq.~\eqref{eqLinearLocalityIdentityEquation} can be expanded to linear equations of $\{\theta(s,a)|s\in S,a\in A\}$. Conversely, given any solution $\{\theta(s,a)|s\in S,a\in A\}$, by taking $\{|a\rangle\}$ as the basis of $\mathcal{H}$ and defining $U(s)$ using Eq.~\eqref{eqDefTheta(g,a)}, we reconstruct a realization $\in R(m)$. Thus, $R(m)$ is identified with solution space of Eq.~\eqref{eqLinearLocalityIdentityEquation}, which is an Abelian group. Physically, the addition can be interpreted as stacking two systems together:

\begin{example}
	Let $m=(A,S,\partial,\supp)$ be an excitation pattern and $P_i=(\mathcal{H}_i,|\cdot\rangle_i, U_i)\in R(m)\; (i=1,2)$ are two realizations. Then,  $P_1+ P_2=(\mathcal{H},|\cdot\rangle, U)\in R(m)$ is defined by:
	\begin{enumerate}
		\item $\mathcal{H}=\mathcal{H}_1\otimes \mathcal{H}_2$.
		\item $|a\rangle=|a\rangle_1\otimes|a\rangle_2,\forall a\in A$.
		\item $ U(s)= U_1(s)\otimes  U_2(s),\forall s\in S$.
	\end{enumerate}
\end{example}

The solution space of Eq.~\eqref{eqLinearLocalityIdentityEquation} is very large; to get meaningful invariants, we must quotient out some trivial solutions. To do that, we introduce the concept of localization. This is the simplest way to derive new excitation patterns from an existing excitation pattern $m=(A,S,\partial,\supp)$: configurations and formal excitation operators are completely the same, while we only modify the function $\supp$. A common confusion is that the notion of support only applies to a formal excitation operator $s\in S$ or a unitary excitation operator $U(s)$; \textbf{we never introduce the notion of support for a configuration $a\in A$ or a configuration state $|a\rangle$.}

\begin{definition}\label{defLocalization}
		Let $m=(A,S,\partial,\supp)$ be an excitation pattern in the topological space $M$, and let $N\subset M$ be a subspace. The \textit{localization} of $m$ at $N$ is a new excitation pattern $m|_N=(A,S,\partial,\supp_N)$ defined by  $\supp_N(s)=\supp(s)\cap N$. For notation simplicity, we also refer to $m|_s$ as $m|_{\supp(s)}$ for $s\in S$ and $m|_x$ to $m|_{\{x\}}$ for $x\in M$.
\end{definition}

The only difference between $m|_N$ and $m$ is the support of formal excitation operators, and the only effect on realizations in through the locality axiom Eq.~(\ref{axiomLocalityIdentity}). Suppose there are $s,s'\in S$ such that $\supp(s)\cap\supp(s')\ne \emptyset$ but $\supp(s)\cap\supp(s')\cap N= \emptyset$. Then, a realization $U$ of $m|_N$ must satisfy $[U(s),U(s')]=1$, but a realization of $m$ need not. More extremely, if $\supp(s)\cap N=\emptyset$ for some $s\in S$, then in any $U\in R(m|_N)$, $U(s)$ must commute with any other operators. In short, realizations of $m|_N$ are special realizations of $m$ that are more localized. From the view of phase data, and $R(m|_N)$ is a subspace of $R(m)$ because it satisfies the stronger version of Eq.~\eqref{eqLinearLocalityIdentityEquation}.

An extreme case is the localization at a point $x$, where all operators of $U\in R(m|_x)$ are supported at $x$. Because we only care about invariants stable under local perturbations, realizations in $R(m|_x)$ for any point $x$ should be considered trivial. Therefore, we should quotient out all solutions of local realizations.

\begin{definition}\label{defEquivalentClass}
	For an excitation pattern $m$ in topological space $M$, the statistics of $m$, denoted by $T^*(m)$\footnote{We reserve the notation $T(m)$ for its Pontryagin dual, defined in Definition~\ref{defStatisticsMain}.}, is defined by
	\begin{equation}
		T^*(m)=R(m)/(\sum_{x\in M}R(m|_x))\footnote{$\sum R(m|_x)$ means the subgroup of $R(m)$ spanned by these $R(m|_x)$. For $m=m_p(X,G)$, taking all vertices $x\in X_0$ is enough.}
	\end{equation}
\end{definition}

When $A$ and $S$ are finite, the computation of $T^*(m)$ is purely finite linear algebra. Moreover, we will prove in Appendix~\ref{appendixSimplicialSet} that
\begin{theorem}\label{conjectureCohomology}
	\begin{equation}
		T^*(m_p(\partial\Delta^{d+1},G))\simeq H^{d+2}(K(G,d-p),\RR/\ZZ).
	\end{equation}
\end{theorem}

This result is actually very surprising. The same classification also appears in higher SPT phases and gauge theories, where they use algebraic topology in the very beginning; however, our theory is nothing but several axioms about many-body Hilbert spaces, and the presence of Eilenberg-MacLane spaces is not obvious.  Although the locality axiom Eq.~(\ref{axiomLocalityIdentity}) and cocycles have some relations, as shown in Example~\ref{exampleGroupCohomology}, they are not exactly the same. Our locality axiom is strictly weaker than the traditional cocycle condition;  unlike field-theoretical approaches, we do not assume spatial homogeneity and allow arbitrary local perturbations. The group $R(m)$ is actually larger than the group of cocycles, and $\sum_{x\in M}R(m|_x)$ is larger than the group of coboundaries; on the other hand, their quotients are exactly the same. 

\begin{remark}\label{remarkMixDimension}
	 This theorem also applies to excitations of mixed dimensions. For example, if we want to consider excitations of dimensions $p_1,p_2$ and fusion groups $G_1,G_2$ at the same time, we can construct an excitation pattern "$m_{p_1}(X,G_1)\times m_{p_2}(X,G_2)$", defined by $A=B_{p_1}(X,G_1)\oplus B_{p_2}(X,G_2)$ and $(G_1\times X_{p_1})\cup (G_2\times X_{p_2})$. When choosing $X=\partial\Delta^{d+1}$, we have
	\begin{equation}
		T^*=H^{d+2}(K(G_1,d-p_1)\times K(G_2,d-p_2),\RR/\ZZ).
	\end{equation}
	Taking $G_1=G_2=\ZZ_2$, $d=3$, $p_1=1$, and $p_2=0$, we get $T^*\simeq \ZZ_2^3$, corresponding to fermion, fermionic loop, and particle-loop braiding.
	
	In simplicial homotopy theory, products of Eilenberg-MacLane spaces are one-to-one correspond to the homotopy types of simplicial Abelian groups. In this sense, our current theory is able to describe all Abelian excitations. However, this picture of statistics is still too narrow: a more general concept is simplicial group, which includes $K(G,1)$ for $G$ non-Abelian and twisted products of $K(G,n)$ of different dimensions; higher-group excitations belong to this type. Moreover, non-invertible excitations are beyond simplicial groups. We have no idea how to describe them yet.
\end{remark}

\begin{remark}\label{remarkEmptySet}
	In Definition~\ref{defLocalization}, one may take $N$ to be the empty set.
	 Expanding the definition, we find that $U\in R(m_\emptyset)$ if and only if all $U(s)$ commute. In this case, phase factors $\theta(s,a)$ are only some gauge transformations. More precisely, we have the following lemma.
\end{remark}
\begin{lemma}\label{lemmaEmptySet}
	Let $U\in R(m)$ be a realization. If all excitation operators $U(s)$ commute, then there exists a "gauge transformation" $|a\rangle\mapsto e^{i\phi(a)}|a\rangle$, $U(s)\mapsto e^{i\varphi(s)}U(s)$, making $U(s)|a\rangle=|a+\partial s\rangle$ for any $s\in S,a\in A$. In other words, all $\theta(s,a)=0$.
\end{lemma}
\begin{proof}
	The map $\partial: \operatorname{F}(S)\rightarrow A$ naturally decomposes to $\operatorname{F}(S)\xrightarrow{f} \ZZ[S]\xrightarrow{\partial'}A$, where $\ZZ[S]$ is the free Abelian group generated by $S$ and $f$ is the Abelianization. Because all $U(s)$ commute, $U(g)$ only depends on $f(g)\in \ZZ[S]$. As a subgroup of $\ZZ[S]$, a classical group-theoretical result says that $\ker\partial'$ is generated by not more than $|S|$ elements. Modifying $U(s)$ by phase factors, we can impose $U(g)|0\rangle=|0\rangle$ for these generators. This further implies that for any $g\in \operatorname{F}(S)$, $U(g)$ only depends on $\partial g$. Finally, we are able to redefine $|a\rangle$ as $U(g)|0\rangle$ by choosing any $\partial g=a$. Note that this proof also works when $A$ is infinite.
\end{proof}

\subsection{Expressions: a dual point of view}\label{subsecExpression}

Up to now, we have already established a self-contained theory about the solution of Eq.~\eqref{eqLinearLocalityIdentityEquation} and Theorem~\ref{conjectureCohomology}. However, our initial motivation is to understand the structure of statistical processes, such as Fig.~\ref{figTJunction} and \ref{fig: 24 step process}. The two concepts are very different, but they are dual to each other:
\begin{equation}
	\text{realizations }\times \text{ statistical processes }\rightarrow \RR/\ZZ.
\end{equation}
More concretely, a realization $U$ is roughly a set of unitary operators $\{U(s)\}$, and a statistical process is a carefully designed formal product, such as $g=s_{02} s_{03}^{-1} s_{01} s_{02}^{-1} s_{03} s_{01}^{-1}$. When a realization $U$ and a process $g$ are paired together, we get a unitary operator $U(g)$. After we define statistical process, we will prove that $U(g)$ is always a pure phase, and then we say the phase factor $U(g)$ detects the statistics of $U$.
In the case of $\ZZ_n$-anyons, this map is as follows:
\begin{equation}
	\begin{aligned}
		&U\in R(m_0(\hbox{\raisebox{-2ex}{\includegraphics[width=1.2cm]{triangle0123}}},\ZZ_n))\times \text{ the T-junction processes }  g\\&\mapsto \text{the topological spin }U(g)\in \operatorname{U}(1).
	\end{aligned}
\end{equation}
In the remaining part of this paper, the term process will always refer to a formal product of excitation operators: \textbf{a process is an element in the free group $\operatorname{F}(S)$}.

The product of process is non-commutative and inconvenient to use, so we are going to translate them into some linear objects, referred to as \textit{expressions}. More concretely, given a process $g\in \operatorname{F}(S)$ and a configuration $a$, we can define a map that maps any realization $U$ to a phase factor $\theta(g,a)$, defined by 

\begin{equation}
	U(g)|a\rangle=e^{i\theta(g,a)}|a+\partial g\rangle.
\end{equation}

Using Eq.~\eqref{eqThetaP(g,a)}, we can always expand $\theta(g,a)$ as the sum of the phase data $\{\theta(s,a)|s\in S,a\in A\}$.

\begin{equation}
	\sum_{s\in S,a\in A}c(s,a)\theta(s,a),\;\;c(s,a)\in \ZZ.
\end{equation}

So in generally, an expression is nothing but a linear map $e:R(m)\rightarrow \RR/2\pi\ZZ$, determined by the coefficients $\{c(s,a)\in \ZZ\}$. These integer coefficients form an Abelian group $E(m)\simeq \ZZ[S\times A]$ called the \textit{expression group} of $m$. We will denote the corresponding expression by
\begin{equation}
	\begin{aligned}
		e=\sum_{s\in S,a\in A}c(s,a)\left(s,a\right),
	\end{aligned}
\end{equation}
which maps $\{\theta(s,a)\}$ to $\sum_{s\in S,a\in A}c(s,a)\theta(s,a)$. Here, $(s,a)$ are not real values but only symbols which represent the basis of $E(m)$. In this language, the expression of the T-junction process is

\begin{align}\label{eqExpressionTjunction}
	e =&(s_{02} s_{03}^{-1} s_{01} s_{02}^{-1} s_{03} s_{01}^{-1},\hbox{\raisebox{-1.5ex}{\includegraphics[width=0.8cm]{T_junction_state_02.pdf}}})\\
	&= -(s_{01}, \hbox{\raisebox{-1.5ex}{\includegraphics[width=0.8cm]{T_junction_state_02.pdf}}}) + (s_{03}, \hbox{\raisebox{-1.5ex}{\includegraphics[width=0.8cm]{T_junction_state_02.pdf}}}) - (s_{02}, \hbox{\raisebox{-1.5ex}{\includegraphics[width=0.8cm]{T_junction_state_03.pdf}}})\notag \\& + (s_{01}, \hbox{\raisebox{-1.5ex}{\includegraphics[width=0.8cm]{T_junction_state_03.pdf}}}) - (s_{03}, \hbox{\raisebox{-1.5ex}{\includegraphics[width=0.8cm]{T_junction_state_01.pdf}}}) + (s_{02}, \hbox{\raisebox{-1.5ex}{\includegraphics[width=0.8cm]{T_junction_state_01.pdf}}})\in E(m),
\end{align}
which maps any realization to the phase factor Eq.~\eqref{eq3.2.2}.

In Eq.~\eqref{eqExpressionTjunction}, we actually \textit{define} $(g,a)$ using the expansion formula similar to Eq.~\eqref{eqThetaP(g,a)}:
\begin{equation}\label{eqTheta(g,a)}
	\begin{aligned}
		(g_2g_1,a)&=(g_1,a)+(g_2,a+\partial g_1),\\
		(g^{-1},a)&=-(g,a-\partial g).
	\end{aligned}
\end{equation}

There is an important structure on expressions that we have not discussed yet. If phase factors $\{\theta(s,a)|s\in S,a\in A\}$ could take arbitrary values, then the expression group $E(m)\simeq \ZZ[S\times A]$ would be the Pontryagin dual of phases $(\RR/2\pi\ZZ)[S\times A]$. However, $R(m)$ is not $(\RR/2\pi\ZZ)[S\times A]$ but only the solution space of Eq.~\eqref{eqLinearLocalityIdentityEquation}, so the dual group of $R(m)$ is actually the quotient group $E(m)/E_{\operatorname{id}}(m)$, where $E_{\operatorname{id}}(m)=\{e\in E(m)|e(U)=0,\forall U\in R(m)\}$ is the annihilator group of $R(m)$\footnote{ When taking duals, sub-objects and quotient-objects are always reversed.}. An expression $e\in E_{\operatorname{id}}$ is ineffective in distinguishing realizations because it assigns zero to any realizations.
If two expression $e_1,e_2$ has the same image in $E/E_\id$, they should be considered identical; because these are derived from the locality axiom, we call $E_{\operatorname{id}}(m)$  \textit{locality identities}.

The introduction of $E_\id$ explains many puzzles in the literature. It is strange why the phase of the $36$-step process is always $0$ or $\pi$; the real reason is that the expression of the $36$-step process is an order-2 element in $E/E_\id$. Also, although our 24-step process (Fig.~\ref{fig: 24 step process}) and the 36-step process in \cite{FHH21} looks very different, their images in $E/E_\id$ are the same, and that is why we say our processes are equivalent. This also explains why these processes look so asymmetric: the equivalence class in $E/E_\id$ is symmetric, but its symmetry is lost when one represent it by a statistical process. We will analyze the T-junction process as an example in Section~\ref{subsecSymmetry}.

Working in $E/E_{\operatorname{id}}$ also provides us with a new view of the operator independence. Let two formal excitation operators $s,s'$ share the same support and anyon creation effects. When we construct a statistical process, we can use either $s$ or $s'$,producing different expressions $e_s$ and $e_{s'}$ respectively. Although $e_s\ne e_{s'}$, we often have $[e_s]=[e_{s'}]\in E/E_{\operatorname{id}}$, so $e_s(U)=e_{s'}(U)$ for any realization $U$. This view of operator independence is slightly stronger than the argument of local perturbation, which we explain in Section~\ref{secUniversal}. The concrete analysis for the T-junction process is in Section~\ref{subsecSymmetry}.
\begin{widetext}
	We close this subsection by discussing the explicit generators of $E_\id$. If $\supp(s_1)\cap \supp(s_2)=\emptyset$, we have
	\begin{align}
		 U([s_2, s_1]) |a\rangle = e^{i\theta([s_2, s_1], a)} |a\rangle = |a\rangle,
	\end{align}
	which implies
	\begin{align}
		\theta([s_2, s_1], a) = \theta(s_1, a) + \theta(s_2, a + \partial s_1) 
		- \theta(s_1, a + \partial s_2) - \theta(s_2, a) = 0.
	\end{align}
	From the dual space \( E \), we see that the expression,
	\begin{align}\label{eqSimpleCommutator}
		([s_2, s_1], a) = (s_1, a) + (s_2, a + \partial s_1)- (s_1, a + \partial s_2) - (s_2, a),
	\end{align}
	vanishes for all realization $U$. Generally, $E_\id$ is generated by expressions of the form
	\begin{equation}\label{eqBasicLocalityIdentityInEid}
		\begin{aligned}
			&([s_k, [\cdots, [s_2, s_1]]], a)\\=&\sum_{c_3,\cdots,c_k\in\{0,1\}}(-1)^{c_3+\cdots+c_k}([s_2, s_1],a+\sum_{i=3}^kc_i\partial s_i)\\=&\sum_{c_2,\cdots,c_k\in\{0,1\}}(-1)^{c_2+\cdots+c_k}(s_1,a+\sum_{i=2}^kc_i\partial s_i)-\sum_{c_1,c_3,\cdots,c_k\in\{0,1\}}(-1)^{c_1+c_3+\cdots+c_k}(s_2,a+\sum_{i\in\{1,3,\cdots,k\}}c_i\partial s_i)
		\end{aligned}	
	\end{equation}
	where \( \operatorname{supp}(s_1) \cap \operatorname{supp}(s_2) \cap \cdots \cap \operatorname{supp}(s_k) = \emptyset \). 
	
	These generators are highly redundant, and several observations help to reduce the size of generators.
	
			\begin{enumerate}
			\item We have the equation
			\begin{equation}
				([s_k, [\cdots, [s_2, s_1]]], a)=([s_{k-1}, [\cdots, [s_2, s_1]]], a)-([s_{k-1}, [\cdots, [s_2, s_1]]], a+\partial s_k).
			\end{equation}	
			Therefore, if we already have \( \operatorname{supp}(s_1) \cap \operatorname{supp}(s_2) \cap \cdots \cap \operatorname{supp}(s_{k-1}) = \emptyset \) and have used $([s_{k-1}, [\cdots, [s_2, s_1]]], a),\;\forall a\in A$ as generators, $([s_k, [\cdots, [s_2, s_1]]], a)$ becomes redundant. In particular, we may only include generators such that $s_1,\cdots,s_k$ are all different.
			
			\item $([s_k, [\cdots, [s_2, s_1]]], a)$ only involves $(s_1,\cdot)$ and $(s_2,\cdot)$ and is anti-symmetric in $s_1$ and $s_2$ So, we may assume $s_1<s_2$.
			
			\item $s_3,\cdots,s_k$ are anti-symmetric. So, we may assume $s_3<\cdots<s_k$.
			
			\item The equation
			\begin{equation}
				([s_3,[s_2,s_1]],a)+([s_1,[s_3,s_2]],a)+([s_2,[s_1,s_3]],a)=0
			\end{equation}
			always holds.
			\item Because of the equation
			\begin{equation}\label{eqBilinearCommutator}
				([g_3,g_2g_1],a)=([g_3,g_1],a)+([g_3,g_2],a+\partial g_1),
			\end{equation} 
			we can always expand a (higher-)commutator of processes $g\in \operatorname{F}(S)$ to some (higher-) commutator of excitation operators $s\in S$.
		
		\end{enumerate}

These observations enable us to write down a finite set of generators of $E_\id$. We call these generators \textit{basic locality identities}.

\end{widetext}

\subsection{Classification of statistical expressions}\label{subsecDefinition}

Recall that previous studies \cite{Kawagoe2020Microscopic,FHH21} characterize statistical processes by the local-perturbation invariance argument. Our first definition is equivalent to their arguments, while we use our language of expressions and localizations. Statistical processes should correspond to "statistical expressions", which should form a subgroup $E_{\operatorname{inv}}(m)\subset E(m)$. Our definition of $E_{\operatorname{inv}}(m)$ is essentially dual to Definition~\ref{defEquivalentClass}, where we construct equivalence classes of realizations modulo realizations supported at a single point.
To be specific, while $E(m)=E(m|_N)\simeq \ZZ[S\times A]$,  $E_{\operatorname{id}}(m)\subset E_{\operatorname{id}}(m|_N)$ because $m|_N$ has more locality identities. A statistical expression $e$ should only be sensitive to long-range physics but not to local perturbations, so it must satisfy $e(U)=0 $ for any realization $ U\in R(m|_x)$ and any point $x\in M$. This means $e\in \bigcap_{x\in M}E_{\operatorname{id}}(m|_x)$.

\begin{definition}\label{defStatisticsMain}
	The statistics $T(m)$ of an excitation pattern $m$ is defined as
	\begin{equation}
		T(m)=E_{\operatorname{inv}}(m)/E_{\operatorname{id}}(m),
	\end{equation}
 where
	\begin{equation}
		E_{\operatorname{inv}}(m)=\bigcap_{x\in M}E_{\operatorname{id}}(m|_x).
	\end{equation}
\end{definition}

Here we give a table for some concepts we have introduced. The left hand side and the right hand side are roughly dual to each other.

\begin{tabular}{l l}
	\hline
	$(\RR/\ZZ)[S\times A]$ &  processes, expressions, $E$ \\
	Eq.~\eqref{eqLinearLocalityIdentityEquation}, realizations & locality identities, $E_\id$ \\
	$\sum_x R(m|_x)$ & statistical processes, $E_\inv$ \\
	$T^*(m)$ & $T(m)$\\
	$H^{d+2}(K(G,d-p),\RR/\ZZ)$ & $H_{d+2}(K(G,d-p),\ZZ)$
	 \\ \hline
\end{tabular}

In particular,  $T(m)$ and $T^*(m)$ are Pontryagin duals: $T^*(m)\simeq \operatorname{Hom}(T(m),\RR/\ZZ)$. Thus, Theorem~\ref{conjectureCohomology} is equivalent to 
\begin{theorem}\label{thmHomology}
	$T(m_p(\partial\Delta^{d+1},G))\simeq H_{d+2}(K(G,d-p),\ZZ)$.
\end{theorem}
 Theorem~\ref{thmFiniteOrder} says that $G$ is finite implies that $T$ is also finite, and then $T(m)$ and $T^*(m)$ are non-canonically isomorphic. This claim is false when $G$ is infinite: for example, the anyon ($p=0,d=2$) with fusion group $G=\ZZ$ has $T^*=\RR/\ZZ$ and $T\simeq \ZZ$.

Actually, the finite order of statistics is highly nontrivial. We take this as another definition of statistics and denote it by $T_f(m)$. When $G$ is finite, we have $T_f(m)\simeq T(m)$ and Theorem~\ref{thmHomology} also applies.

\begin{definition}\label{defTf}
	For an excitation pattern $m$, its \textit{finite-order statistics}, denoted by $T_f$ or $T_f(m)$, is the torsion subgroup of $E/E_{\operatorname{id}}$: 
	\begin{equation}
		T_f = \{ [e] \in E/E_{\operatorname{id}} \mid e \in E, \exists n > 0, ne \in E_{\operatorname{id}} \}.
	\end{equation}
\end{definition}

For any $[e] \in T_f$, let $n$ be the minimum positive integer satisfying $ne \in E_{\operatorname{id}}$. This implies $ne(U)=0,\forall U\in R(m)$, and hence $e(U) = k \frac{2\pi}{n}$ with $k \in \mathbb{Z}$. The discreteness of $e(U)$ reflects the quantization nature of statistics, as seen in fermionic loop statistics and anyon self-statistics (when the fusion group $G$ is finite), contrasting sharply with continuous varying geometric Berry phases.

As the first definition means that statistics is stable under local perturbations, the second definition means that the statistics is stable under continuous perturbations. The adiabatic deformation of excitations and excitation operators corresponds to a parameterized family of realizations $U_t$ evolving continuously with $t$. Since $e(U_t)$ should remain quantized under continuous variation, it must remain constant.

Fully characterizing $E_\id(m)$ and $E_\inv(m)$ is generally difficult, but at least, a statistical expression $e$ should be invariant under the gauge transformation $|a\rangle\mapsto e^{i\phi(a)}|a\rangle$ and $U(s)\mapsto e^{i\varphi(s)}U(s)$. The first condition means that $e$ can be written as $(g,0)$ where $\partial g=0$, while the second condition means that for any $s\in S$, the total sum of the coefficients of $(s,a),\forall a\in A$ is zero. We will call these two conditions \textit{closed} and \textit{zero-weight}, respectively.
These conditions characterize $e\in E_\id(m|_\emptyset)$; see Remark~\ref{remarkEmptySet} and Lemma~\ref{lemmaEmptySet}. Here is an obvious lemma used for later references.

\begin{lemma}\label{lemmaClosedLoop}
	For $e\in E$,  $e$ is closed and zero-weight, i.e., $e\in E_\id(m|_\emptyset)$, if and only if there exists $g\in \operatorname{F}(S)$ such that $e=(g,0)$, satisfying that the total numbers of $s$ and $s^{-1}$ in $g$ are the same for any $s\in S$.
\end{lemma}

A geometric pictures about $E(m)$ may also be helpful.

\begin{definition}\label{constructionG(m)}
	For an excitation pattern $m$, we construct the \textit{configuration graph} $\operatorname{G}(m)$ as follows: vertices of $\operatorname{G}(m)$ correspond to $a \in A$; for every equation $a + \partial s = a'$, we draw a directed edge labeled $s$ from $a$ to $a'$\footnote{When $\partial s$ has order 2, and then we have two distinct edges for $\theta(s,a)$ and $\theta(s,a')$.}. These edges one-to-one correspond to $(s,a)$. Thus, the group of 1-chains $C_1(\operatorname{G}(m))$ is identified with the expression group $E$.
	A process $g \in \operatorname{F}(S)$ acting on $a\in A$ is identified with the corresponding path in $\operatorname{G}(m)$; its "Abelianization" as a $1$-chain is exactly $(g,a)$. If $\partial g=0$, the path becomes a loop, and $(g,a)$ becomes a $1$-cycle. 
\end{definition}

Using this picture, closed expressions are identified with $1$-cycles in $\operatorname{G}(m)$. For example, we represent the expression Eq.~\eqref{eqExpressionTjunction} of the T-junction process (Fig.~\ref{figTJunction}) as follows. 
\begin{equation}
	\includegraphics[width=0.99\linewidth]{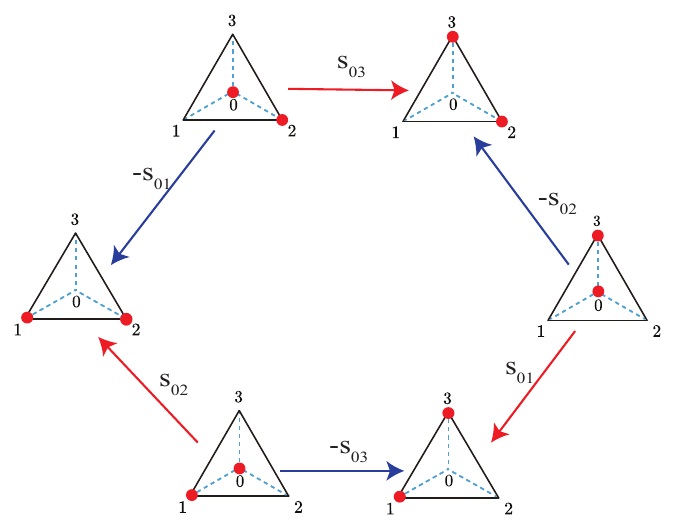}
\end{equation}
An arrow labeled $s$ and pointing from $a$ to $a+\partial s$ corresponds to a term $c(s,a)(s,a)$; red arrows means $c(s,a)>0$, and blue arrows means $c(s,a)<0$. We also draw a statistical expression for fermionic-loop statistics in Fig.~\ref{figFermionicloop20step}.

The quickest way to test whether an expression $e$ is in $E_\inv(m)$ is to use the following theorem. This is essentially the same as the criterion in \cite{FHH21}. We will give the proof after Theorem~\ref{thmQuotientEmptySupport}.

\begin{theorem}\label{thmCheckByHand}
	Let $m=(A,S,\partial,\supp)$ be an excitation pattern in $M$, and $e\in E_\id(m|_\emptyset)$ (see Lemma~\ref{lemmaClosedLoop}). The following procedure tests if $e\in E_{\operatorname{id}}(m|_x)$.
	\begin{enumerate}
		\item Choose a point $x\in M$ and pick all operators whose support contain $x$ which form a subset $V(x)\subset S$.
		\item All operators \textit{not} in $V(x)$ generate a subgroup of $A$, denoted by $A_{V(x)^c}$.
		\item Note that $e\in E(m)\simeq \ZZ[S\times A]$. Now, we define a map $\tilde{q}_{x*}: \ZZ[S\times A]\rightarrow \ZZ[V(x),A/A_{V(x)^c}]$ by deleting all $(s,a)$ for $s\notin V(x)$ and applying the quotient map on $a$. Then, 
		\begin{equation}
			e\in E_{\operatorname{id}}(m|_x)\iff q_{x*}(e)=0.
		\end{equation}

	\end{enumerate}
	Thus,
	\begin{equation}\label{eqEinv}
		e\in E_{\operatorname{inv}}(m)\iff q_{x*}(e)=0,\forall x.
	\end{equation}
\end{theorem}

\subsection{Example: hidden permutation symmetry and properties of the T-junction process}\label{subsecSymmetry}

The self-statistics of a $\ZZ_n$-anyon corresponds to the excitation pattern $m=m_0(\hbox{\raisebox{-2ex}{\includegraphics[width=1.2cm]{triangle0123}}},\ZZ_n)$. Computer computation shows that $T(m)\simeq\ZZ_{n(2,n)}$ is generated by the T-junction process $s_{02} s_{03}^{-1} s_{01} s_{02}^{-1} s_{03} s_{01}^{-1}$. This is consistent with the fact that the topological spin of an $\ZZ_n$-anyon has order $n$ and $2n$ for $n$ odd and even, respectively. The expression of the T-junction process actually has many interesting properties, and we will discuss them in detail. These properties are universal and are exhibited by all kinds of statistical expression; we will give formal proofs in Section~\ref{secUniversal}.

Let $s_{ij}$ denote a formal excitation operator moving the anyon from point $\bar{i}$ to the point $\bar{j}$, and we identify $s_{ji}$ with $s_{ij}^{-1}$\footnote{This is only for notation convenience and is not required in general.}. States involved in the process are labeled by vertex pairs (e.g., $|ij\rangle$ for anyons at $\bar{i}$ and $\bar{j}$)\footnote{Strictly speaking, $|ij\rangle$ and the vacuum may not be in the same super-selection sector; this does not affect our analysis because the choice of vacuum is not important.  }.

In this notation, we rewrite Eq.~\eqref{eqExpressionTjunction} as follows:
\begin{equation}\label{eqThetaInSecFive}
	\begin{aligned}
		e =&\;\Bigl(s_{02} s_{03}^{-1} s_{01} s_{02}^{-1} s_{03} s_{01}^{-1}, 12\Bigr) \\
		=& -(s_{01}, 02) + (s_{03}, 02) - (s_{02}, 03) \\&+ (s_{01}, 03) - (s_{03}, 01) + (s_{02}, 01).
	\end{aligned}
\end{equation}

When applying the procedure of Theorem~\ref{thmCheckByHand}, by symmetry, we only need to check the points $\bar{0}$ and $\bar{1}$.

\begin{itemize}
\item $\overline{0}$: Since $s_{12},s_{13},s_{23}\notin V(\overline{0})$, configurations $01,02,03$ are in the same equivalence class modulo $A_{V(\overline{0})^c}=\langle \partial s_{12},\partial s_{13},\partial s_{23}\rangle$. Terms in Eq.~\eqref{eqThetaInSecFive} cancel pairwise.
	\item $\overline{1}$: Because $s_{02},s_{03}\notin V(\overline{1})$, we only care about $s_{01}$. $-(s_{01},02)$ and $(s_{01},03)$ cancel because $\partial{s_{23}}=03-02\in A_{V(\overline{1})^c}$.
\end{itemize}

\begin{remark}\label{remarkOuterEdges}
	While the T-junction process only involves $s_{01},s_{02},s_{03}$ explicitly, the outer edges and operators $s_{12},s_{13},s_{23}$ also play a crucial implicit role. Their existence enables us to write down locality identities and to define statistics correctly. 
	
	This phenomenon generalizes: In the study of statistics of $m_p(\partial\Delta^{d+1},G)$, all $(p+1)$-simplexes are important. On the other hand, it is possible to represent
	any statistical process (for example, the 24-step loop process) using operators containing $\bar{0}$ (Theorem~\ref{thmStar}).
\end{remark}

\subsubsection{Symmetry restoration in the T-junction process}

From Eq.~(\ref{eqThetaInSecFive}), we see that $e$ exhibits clear $S_3$ symmetry under permutations of vertices $\{ \bar{1}, \bar{2}, \bar{3}\}$; even permutations ($A_3$) preserve $e$, while odd permutations flip its sign. However, the graph \hbox{\raisebox{-2ex}{\includegraphics[width=1.2cm]{triangle0123}}} has $S_4$ permutation symmetry. Actually, the equivalence class $[e]\in T(m)$ has the full symmetry, but it is hidden in $e\in E_\inv(m)$.

\begin{widetext}
 To show the full symmetry of $[e]$, we consider two locality identities
	\begin{equation}
		([s_{13}, s_{02}], 01) = (s_{02}, 01) + (s_{13}, 12) - (s_{02}, 03) - (s_{13}, 01) \in E_{\operatorname{id}}
	\end{equation}
	and
	\begin{equation}
		([s_{12}, s_{03}], 01) = (s_{03}, 01) + (s_{12}, 13) - (s_{03}, 02) - (s_{12}, 01) \in E_{\operatorname{id}}.
	\end{equation}
	
	Adding $([s_{12}, s_{03}], 01)-([s_{13}, s_{02}], 01)$ to Eq.~(\ref{eqThetaInSecFive}), we get
	\begin{equation}\label{eqPermutedTjunction}
		\begin{aligned}
			e &\sim -(s_{01}, 02) + (s_{12}, 13) - (s_{13}, 12) + (s_{01}, 03) - (s_{12}, 01) + (s_{13}, 01) \\
			&= \theta\Bigl(s_{13} s_{12}^{-1} s_{10} s_{13}^{-1} s_{12} s_{10}^{-1}, 03\Bigr).
		\end{aligned}
	\end{equation}

The process $s_{13} s_{12}^{-1} s_{10} s_{13}^{-1} s_{12} s_{10}^{-1}$ is analogous to the original process, except that the vertices are permuted as $(\bar{0}, \bar{1}, \bar{2}, \bar{3}) \mapsto (\bar{1}, \bar{0}, \bar{3}, \bar{2})$. Similarly, one can show that the permutation group $S_4$ acts on $\{[e], -[e]\}$, with kernel $A_4$.

In Remark~\ref{remarkOuterEdges}, we have said that the topological spin $e(U)$ is independent with $U(s_{12}),U(s_{13}),U(s_{23})$ because these edges do not even appear in $e$. But in Eq.~\eqref{eqPermutedTjunction}, we construct an equivalent expression such that $s_{02}$ and $s_{03}$ do not appear, which means that $e(U)$ is also independent with $U(s_{02})$ and $U(s_{03})$. Actually, $e(U)$ is independent with any excitation operators! This amazing phenomenon happens generally; see the Fig.~\ref{figFermionicloop20step} and Theorem~\ref{thmReplacable2}.

\subsubsection{Initial state independence}

	In the previous discussion, we sometimes do not distinguish a statistical process $g$ and a statistical expression $e\in E_\inv$. On one hand, any statistical expression can be written as $(g,a)$ for any $a\in A$; on the other hand, if $g$ is a statistical process, then $(g,a)$ is always a statistical expression. The interesting is that different choices of $a\in A$ give the same $[(g,a)]\in T$ under some conditions, which we will prove in Theorem~\ref{thmInitialStateIndependence}. Here, we give a direct proof for the T-junction process.
	 
	 By the $S_4$ permutation symmetry, it suffices to prove that:
	\begin{equation}\label{eqInitialStateIndependenceSecFive}
		\Bigl(s_{02} s_{03}^{-1} s_{01} s_{02}^{-1} s_{03} s_{01}^{-1}, a\Bigr) - \Bigl(s_{02} s_{03}^{-1} s_{01} s_{02}^{-1} s_{03} s_{01}^{-1}, a + \partial s_{13}\Bigr) \in E_{\operatorname{id}}, \forall a \in A.
	\end{equation}
	
	This amounts to proving:
	\begin{equation}
		\Bigl(\Bigl[s_{13}, s_{02} s_{03}^{-1} s_{01} s_{02}^{-1} s_{03} s_{01}^{-1}\Bigr], a\Bigr) \in E_{\operatorname{id}}.
	\end{equation}
	
	We have 
	\begin{equation}
		\begin{aligned}
			&\Bigl(\Bigl[s_{13}, s_{02} s_{03}^{-1} s_{01} s_{02}^{-1} s_{03} s_{01}^{-1}\Bigr], a\Bigr)\\=&+([s_{13}, s_{01}^{-1}],a)+([s_{13},s_{03}],a-\partial s_{01})+([s_{13},s_{02}^{-1}],a+\partial s_{03}-\partial s_{01})\\&+([s_{13},s_{01}],a-\partial s_{02}+\partial s_{03}-\partial s_{01})+([s_{13},s_{03}^{-1}],a-\partial s_{02}+\partial s_{03})+([s_{13},s_{02}],a-\partial s_{02})\\=\;&-([s_{13}, s_{01}],a-\partial s_{01})+([s_{13},s_{03}],a-\partial s_{01})-([s_{13},s_{02}],a+\partial s_{03}-\partial s_{01}-\partial s_{02})\\
		&+([s_{13},s_{01}],a-\partial s_{02}+\partial s_{03}-\partial s_{01})-([s_{13},s_{03}],a-\partial s_{02})+([s_{13},s_{02}],a-\partial s_{02}).
		\end{aligned}
	\end{equation}
	
	In the first step, we use Eq.~(\ref{eqBilinearCommutator}) to expand the multiplication inside the commutator; in the second step, we use Eq.~(\ref{eqTheta(g,a)}) to remove their minus signs. All $[s_{13},s_{02}]$ terms are in $E_{\operatorname{id}}$, and the remain part is 
	
	\begin{equation}
		\begin{aligned}
			\Bigl([s_{32},[s_{13},s_{01}]],a-\partial s_{02}+\partial s_{03}-\partial s_{01}\Bigr)+\Bigl([s_{21},[s_{13},s_{03}]],a-\partial s_{01}\Bigr),
		\end{aligned}
	\end{equation}
	also in $E_{\operatorname{id}}$, thereby Eq.~(\ref{eqInitialStateIndependenceSecFive}) is proved. 
\end{widetext}

\subsubsection{Finite order of $[e]$}

Finally, we prove that $[e]$ has finite order. Let 

\begin{equation}
	f=\sum_{a\in A} \Bigl(s_{02} s_{03}^{-1} s_{01} s_{02}^{-1} s_{03} s_{01}^{-1}, a\Bigr).
\end{equation}

Every $(s,a_1)$ appears together with another $-(s,a_2)$ in $\Bigl(s_{02} s_{03}^{-1} s_{01} s_{02}^{-1} s_{03} s_{01}^{-1}, a\Bigr)$, so after the sum we find $f=0$. On the other hand, the initial state independence implies $[f]=n^3[e]$, where $n^3$ is the size of $A$. Therefore, we have $n^3 e \in E_{\operatorname{id}}$.

\section{Computational Methods and Conjectures}\label{secComputation}

\subsection{Computing $T(m)$}

A major advantage of our theory is that everything can be finite and computable. Thus, the best way to use our framework is to construct various excitation patterns $m$, compute $T(m)$ or $T^*(m)$, and think about their physical interpretations. This is much easier than doing mathematical proofs: although we have found some theorems like Theorem~\ref{conjectureCohomology}, there are still many computational results that we cannot explain yet.

We focus on $T(m_p(X,G))$ when $X$ and $G$ are both finite, in which case $T\simeq T_f$ is the torsion part of $E/E_{\operatorname{id}}$. By the structure theorem of finitely generated Abelian groups, there is an isomorphism $E/E_{\operatorname{id}} \simeq \mathbb{Z}/\lambda_1\mathbb{Z} \oplus \dots \oplus \mathbb{Z}/\lambda_N\mathbb{Z}$ such that $\lambda_i|\lambda_{i+1}$ ($\lambda_k=0$ when $k>\dim E_{\operatorname{id}}$); then, $T$ is isomorphic to $\oplus_{i:\lambda_i>1} \mathbb{Z}/\lambda_i\mathbb{Z}$.

This computation can generally be done using Smith decomposition. We write generators of $E_{\operatorname{id}}$ as an $(N'\times \dim E)$-dimensional integer matrix $M$, where each row is a basic locality identity; see the discussion at the end of Section~\ref{subsecExpression}. Usually, $N'$ is much larger than $\dim E_{\operatorname{id}}$. The output of the Smith decomposition contains three integer matrices $U\in \operatorname{GL}(N',\ZZ)$, $V\in \operatorname{GL}(\dim E,\ZZ)$, and an $N'\times \dim E$ diagonal matrix $D = UMV$. $D=\operatorname{diag}\{\lambda_1, \dots, \lambda_N\}$ satisfies $\lambda_i|\lambda_{i+1}$ and $\lambda_k=0$ for $k>\dim E_{\operatorname{id}}$. The first $\dim E_{\operatorname{id}}$ rows of $DV^{-1}$, $\{e_1,\cdots e_{\dim E_{\operatorname{id}}}\}$, are a basis of $E_{\operatorname{id}}$ satisfying $\lambda_i|e_i$. Typically, there are very few of $\lambda_i>1$, and the corresponding $e_i/\lambda_i$ generate the statistics $T$.

General algorithms of Smith decomposition are very consuming in time and memory. However, in our case, where most of the entries during computation are $0$ or $\pm1$, the naive row reduction algorithm works much better. We realize it from the package Sheafhom \cite{sheafhom} who has already used this strategy and works well in algebraic-topology computations. To begin with, we select an entry $M_{ij}=\pm1$; choose the $i$-th row as a basis element $e_1$ of $E_{\operatorname{id}}$, and perform row transformations to eliminate other entries in the $j$-th column; remaining rows then form a smaller matrix $M'$. This reduce our question: if $e$ represents an element of $T$, let the $j$-th entry of $e$ be $c$, and then $e'=e\pm c e_1$ is an equivalent representative that its $j$-th entry is zero. Repeating this process, we eventually get a sequence of linear independent expressions $e_1,\cdots,e_k\in E_{\operatorname{id}}$ and a residue matrix without any $\pm1$ entries. In practice, this matrix is very small, and whether some $n>1$ can divide sum combination of the rows can be read out directly. We implement this algorithm in \cite{GitHubProject} with some optimization. 
 For example, we fix $s_1,s_2\in S$, use this algorithm to get a basis of $\operatorname{span}\{([s_k, [\cdots, [s_2, s_1]]], a)\}\subset E_{\operatorname{id}}$, and then gather the result for all $s_1,s_2\in S$. This helps to maintain the matrix sparsity during the computation because $([s_k, [\cdots, [s_2, s_1]]], a)$ only involves $(s_1,\cdot)$ and $(s_2,\cdot)$. For the same reason, when choosing $M_{ij}=\pm1$, we require that (elements in the $i$-th row) $\times$ (elements in the $j$-th column) is the smallest. The largest example we have computed is the statistics of $\ZZ_2$ membranes in $4$ dimensions, that is, $T_2(\partial\Delta^5,\ZZ_2)\simeq \ZZ_2$; we get $\dim E=15360,\dim E_{\operatorname{id}}=11405$, and the computation takes about one day.

 \begin{figure}
 	\centering
 	\begin{tikzpicture}
 		\draw[thick] (0,0) -- (2,0) -- (1,1.732) -- cycle; 
 		\draw[thick] (3,0) -- (5,0) -- (5,2) -- (3,2) -- cycle; 
 		
 		\draw[thick] (6.5,0) -- (7.5,0) -- (8,0.877)--(7.5,1.732)-- (6.5,1.732) -- (6,0.877) -- cycle;
 		\foreach \x in {(0,0), (2,0), (1,1.732), (3,0), (5,0), (5,2), (3,2),(6.5,0) , (7.5,0) , (8,0.877),(7.5,1.732), (6.5,1.732) , (6,0.877)} \fill[red] \x circle (2pt);
 	\end{tikzpicture}
 	\caption{When $X$ is a triangle, square, or other polygons, computations suggest $T_0(X,G)\simeq H_3(G,\ZZ)$.}
 	\label{figFSymbol}
 \end{figure}
 
 \begin{figure*}
 	\centering
 	\begin{tikzpicture}
 		\coordinate (A) at (0,0,0);
 		\coordinate (B) at (1,1.732,0);
 		\coordinate (C) at (2,0,0);
 		\coordinate (D) at (1,-1,-1);
 		\draw[thick, fill=gray!50,fill opacity=0.5] (A) -- (B) -- (D) -- cycle;
 		\draw[thick, fill=gray!30,fill opacity=0.5] (A) -- (C) -- (D) -- cycle;
 		\draw[thick] (B) -- (C) -- (D) -- cycle;
 		\draw[thick] (A) -- (B) -- (C) -- cycle;
 		
 		\fill[red] (D) circle (2pt);
 		\foreach \x in {(A), (B), (C),(D)} \fill[red] \x circle (2pt);
 		
 		\coordinate (E) at (3,0.5,0);
 		\coordinate (F) at (4,2,0);
 		\coordinate (G) at (5,0.5,0);
 		\coordinate (H) at (4,-0.5,-1);
 		\coordinate (I) at (4,-1,0);
 		
 		\draw[thick, fill=gray!50,fill opacity=0.5] (E) -- (F) -- (H) -- cycle;
 		\draw[thick, fill=gray!30,fill opacity=0.5] (E) -- (F) -- (G) -- cycle;
 		\draw[thick] (F) -- (G) -- (H) -- cycle;
 		\draw[thick, fill=red!20,fill opacity=0.5] (E) -- (I) -- (H) -- cycle;
 		\draw[thick, fill=green!10,fill opacity=0.5] (E) -- (I) -- (G) -- cycle;
 		\draw[thick] (I) -- (G) -- (H) -- cycle;
 		
 		\foreach \x in {(E), (F), (G),(H),(I)} \fill[red] \x circle (2pt);
 		
 		\coordinate (A1) at (6,0,0);
 		\coordinate (T) at (7,1.732,0);
 		\coordinate (B1) at (8,0,0);
 		\coordinate (C1) at (7,-1,-1);
 		\coordinate (D1) at (6,-1,-1);
 		\draw[thick, fill=gray!50,fill opacity=0.5] (A1) -- (B1) -- (T) -- cycle;
 		\draw[thick, fill=gray!30,fill opacity=0.5] (B1) -- (C1) -- (T) -- cycle;
 		\draw[thick, fill=red!20,fill opacity=0.5] (D1) -- (C1) -- (T) -- cycle;
 		\draw[thick] (A1) -- (D1) -- (T) -- cycle;
 		\draw[thick, fill=green!20,fill opacity=0.5] (A1) -- (B1) -- (C1)--(D1) -- cycle;
 		
 		\foreach \x in {(A1), (B1), (C1),(D1),(T)} \fill[red] \x circle (2pt);
 		
 		\coordinate (A2) at (10,0,0);
 		
 		\coordinate (B2) at (12,0,0);
 		\coordinate (C2) at (12,2,0);
 		\coordinate (D2) at (10,2,0);
 		\coordinate (A3) at (10,0,2);
 		
 		\coordinate (B3) at (12,0,2);
 		\coordinate (C3) at (12,2,2);
 		\coordinate (D3) at (10,2,2);
 		\draw[thick, fill=gray!50,fill opacity=0.5] (A2) -- (B2) -- (C2) --(D2) --cycle;
 		\draw[thick, fill=gray!30,fill opacity=0.5] (A3) -- (B3) -- (C3) --(D3) --cycle;
 		
 		\draw[thick, fill=gray!20,fill opacity=0.5] (A2) -- (B2) -- (B3) --(A3) --cycle;
 		
 		\draw[thick, fill=gray!10,fill opacity=0.5] (C3) -- (B3) -- (B2) --(C2) --cycle;
 		
 		\draw[thick, fill=gray!20,fill opacity=0.5] (A3) -- (D3) -- (D2) --(A2) --cycle;
 		\draw[thick, fill=gray!30,fill opacity=0.5] (C3) -- (D3) -- (D2) --(C2) --cycle;

 		\foreach \x in {(A2), (B2), (C2),(D2),(A3), (B3), (C3),(D3)} \fill[red] \x circle (2pt);
 		
 	\end{tikzpicture}
 	\caption{We have tried several regular cell complex of $S^2$, and computations suggest $T_0(X,G)\simeq H_4(K(G,2),\ZZ)$ and $T_1(X,G)\simeq H_4(G,\ZZ)$. Note that $H_4(K(G,2),\ZZ)$ is the familiar anyon statistics, and the 1-skeleton of the first graph coincide with that of the T-junction process.}
 	\label{figLoopFusion}
 \end{figure*}

\begin{table*}[thb]
	\centering
	\renewcommand{\arraystretch}{1.5}
	\begin{tabular}{|c|c|c|c|}
		\hline
		& $G$-particles with $G= \prod_i \mathbb{Z}_{N_i}$ & $G$-loops with $G= \prod_i \mathbb{Z}_{N_i}$ & $G$-membranes with $G= \prod_i \mathbb{Z}_{N_i}$\\
		\hline
		(1+1)D & 
		$\begin{aligned}
			& H^3(G, \operatorname{U}(1)) \\
			=& \textstyle\prod_i \mathbb{Z}_{N_i} \textstyle\prod_{i<j} \mathbb{Z}_{(N_i, N_j)} \\
			&\textstyle\prod_{i<j<k} \mathbb{Z}_{(N_i, N_j, N_k)} 
		\end{aligned}$
		& &\\ 
		\hline
		(2+1)D & 
		$\begin{aligned}
			&H^4(K(G,2), \operatorname{U}(1)) \\
			=& \textstyle\prod_{i} \mathbb{Z}_{(N_i, 2) \times N_i} \textstyle\prod_{i<j} \mathbb{Z}_{(N_i, N_j)}
		\end{aligned}$
		& 
		$\begin{aligned}
			&H^4(G, \operatorname{U}(1)) \\
			=& \textstyle\prod_{i<j} \mathbb{Z}_{(N_i,N_j)}^2 \textstyle\prod_{i<j<k} \mathbb{Z}_{(N_i,N_j,N_k)}^2 \\
			&\textstyle\prod_{i<j<k<l} \mathbb{Z}_{(N_i,N_j,N_k,N_l)}
		\end{aligned}$
		&\\
		\hline
		(3+1)D & 
		$\begin{aligned}
			&H^5(K(G,3), \operatorname{U}(1)) \\
			=& \textstyle\prod_{i} \mathbb{Z}_{(N_i, 2)}
		\end{aligned}$
		& 
		$\begin{aligned}
			&H^5(K(G,2), \operatorname{U}(1)) \\
			=& \textstyle\prod_{i} \mathbb{Z}_{(N_i, 2)} \textstyle\prod_{i<j} \mathbb{Z}_{(N_i,N_j)}
		\end{aligned}$
		& 
		$\begin{aligned}
			&H^5(G, \operatorname{U}(1)) \\
			=& \textstyle\prod_{i} \mathbb{Z}_{N_i} \textstyle\prod_{i<j} \mathbb{Z}^2_{(N_i, N_j)} \\
			&\textstyle\prod_{i<j<k} \mathbb{Z}^4_{(N_i, N_j, N_k)}  \\
			&\textstyle\prod_{i<j<k<l} \mathbb{Z}^3_{(N_i,N_j,N_k,N_l)} \\
			&\textstyle\prod_{i<j<k<l<m} \mathbb{Z}_{(N_i,N_j,N_k,N_l,N_m)}
		\end{aligned}$ \\
		\hline
	\end{tabular}
	\caption{(Table I in \cite{Previous}) The cohomology of the Eilenberg–MacLane space $ K(G, n)$ for the finite Abelian group $G = \prod_i \mathbb{Z}_{N_i}$. The notation $(N_i, N_j, \cdots)$ denotes the greatest common divisor among the integers. 
	}
	\label{tab: particle loop membrane}
\end{table*}

Theorem~\ref{conjectureCohomology} is initially found by comparing the computational results like Fig.~\ref{figFSymbol} and \ref{figLoopFusion} and the Table~\ref{tab: particle loop membrane} of Eilenberg-MacLane spaces. More surprisingly, we find that for different simplicial complexes or even some polyhedrons $X$, as long as the underlying topological space $|X|$ is a manifold without boundary, then $T(m_p(X,G))$ depends only on $|X|$ but not on the triangulation. On the other hand, $T(m_p(X,G))$ can have very bad behavior if $|X|$ is not a manifold. Therefore, we have the following conjecture:
 
 \begin{conjecture}\label{conjectureDef}
 	If $X_1,X_2$ are two triangulations of the manifold $M$, then $T(m_p(X_1,G))\simeq T(m_p(X_2,G))$. In particular, if $X$ is a triangulation\footnote{We mean simplicial complexes but not $\Delta$-complexes; see Appendix~\ref{appendixSimplicialComplex}.} of $S^d$, then $T(m_p(X,G))\simeq H_{d+2}(K(G,d-p),\ZZ)$ and $T^*(m_p(X,G))\simeq H^{d+2}(K(G,d-p),\RR/\ZZ)$.
 \end{conjecture}

\subsection{Some discussion}

Here we show a theoretical reasoning of Conjecture~\ref{conjectureDef} based on cohomology operations.
In $m_p(X,G)$, the configuration group $A=B_p(X,G)$ is the boundary group but not the coboundary group. To use cohomology operations, we replace our excitation patterns $m_p(X,G)$ (Definition~\ref{expSimplicialComplex}) by a cochain version. This is less geometric intuitive, and we should specify the dimension $d$.

\begin{definition}\label{expDualCell}
	Let $X$ be a $d$-dimensional combinatorial manifold. Then, the excitation pattern $m^q(X,G)$ is constructed using the following data:
	\begin{enumerate}
		\item \(A = B^q(X, G)\).
		\item \(S = G_0\times X_{q-1}\). 
		\item  \(\operatorname{supp}(s)\) is the set of all $d$-cells containing $s$, and \(\partial: S\rightarrow A\) is the differential map.
	\end{enumerate}
	In other words, $\supp(s_1)\cap\cdots\cap\supp(s_k)=\emptyset$ if and only if these $(q-1)$-cells do not share (i.e., on the boundary of) any common $d$-cell.
\end{definition}

This definition directly implies
\begin{equation}
	m^q(X,G)\simeq m_{d-q}(X^\#,G),
\end{equation}
where $X^\#$ is the dual cell complex of $X$. Note that $\partial\Delta^{d+1}$ is the dual cell complex of it self, so we have
\begin{equation}
	m^q(\partial\Delta^{d+1},G)\simeq m_{d-q}(\partial\Delta^{d+1},G).
\end{equation}

Therefore, Theorem~\ref{conjectureCohomology} is equivalent to
\begin{equation}
	T^*(m^q(\partial\Delta^{d+1},G))\simeq H^{d+2}(K(G,q),\RR/\ZZ),
\end{equation}
and Conjecture~\ref{conjectureDef} for $|X|=S^d$ says that
\begin{equation}
	T^*(m^q(X,G))\simeq H^{d+2}(K(G,q),\RR/\ZZ).
\end{equation}

Here is a method to construct realizations of $m^q(X,G)$ from cohomology classes in $ H^{d+2}(K(G,q),\RR/\ZZ)$.

\begin{theorem}\label{thmEvaluation}
	We use $Z^n(\cdot,G)$ to denote the functor that gives any simplicial complex $X$ the Abelian group $Z^n(X,G)$. For any natural transformation
	 $\nu:Z^q(\cdot,G)\rightarrow Z^{d+2}(\cdot,\RR/\ZZ)$, there exist a natural map $\Theta: C^{q-1}(\cdot,G)\times Z^q(\cdot,G)\rightarrow C^{d+1}(\cdot,\RR/\ZZ)$ such that
	\begin{equation}
		d\Theta(\alpha,\beta)=\nu(d\alpha+\beta)-\nu(\beta).
	\end{equation}
	then 
	\begin{equation}
		\theta(s,a)=\int_{\Delta^{d+1}}\Theta(s,a)-\Theta(0,a)
	\end{equation}
	 is a realization of $m^q(\partial\Delta^{d+1},G)$, and its equivalence class in $T^*(m^q(\partial\Delta^{d+1},G))\simeq H^{d+2}(K(G,q),\RR/\ZZ)$ is identified with the cohomology operation $[\nu]:H^q(\cdot,G)\rightarrow H^{d+2}(\cdot,\RR/\ZZ)$.
\end{theorem}

Similar construction seems can be use to construct
realizations of $m^q(X,G)$ when $X$ is any triangulation of $S^d$. Using the cone construction, we treat $X\simeq S^d$ as the boundary of the disk $D^{d+1}$. For $s\in S=G\times X_{q-1}\subset C^{q-1}(X,G)$ and $a\in A=B^{q}(S^d,G)$, we extend $s,a$ to $D^{d+1}$, and define

\begin{equation}
	\theta(s,a)=2\pi\int_{D^{d+1}} \Theta(s,a)-\Theta(0,a).
\end{equation}
Extending $a$ from $S^d$ to $D^{d+1}$ is actually subtle. We think that the locality axiom still holds in this case but fails when $|X|$ is not a sphere. Although we do not have a rigorous statement, our intuition is that

\begin{conjecture}
	\textbf{	Any proposal claiming that certain physical theories are classified by $H^n(K(G,q),\RR/\ZZ)$ } (i.e., in contrast to the generalized cohomology theories like cobordism)\textbf{, is essentially working on $S^d$.}
\end{conjecture}

We have another independent idea about what is the statistics in a manifold $M$. Instead of choosing a triangulation of $M$, we try all simplicial complexes $X$ together with an embedding into $M$, and then these $T(m_p(X,G))$ should characterize some properties of $M$. Since $T(m_p(X,G))$ only depends on the $(p+1)$-skeleton of $X$, we assume $X$ is $(p+1)$-dimensional. Then, whether some $(p+1)$-dimensional simplicial complex $X$ can be embedded into the manifold $M$ becomes the crucial question. 

\begin{figure}
	\centering
	\begin{tikzpicture}
		\coordinate (A) at (0,0,0);
		\coordinate (B) at (1,1.732,0);
		\coordinate (C) at (2,0,0);
		\coordinate (D) at (1,-1,-1);
		\draw[thick] (A) -- (B) -- (C) -- cycle;
		\draw[thick] (D) -- (A);
		\draw[thick] (D) -- (B);
		\draw[thick] (D) -- (C);
		\fill[red] (D) circle (2pt);
		\coordinate (Centroid) at (1, 0.333, -0.25);
		\fill[red] (Centroid) circle (2pt);
		\foreach \x in {(A), (B), (C)} \fill[red] \x circle (2pt);
		\draw[thick, dashed] (Centroid) -- (A);
		\draw[thick, dashed] (Centroid) -- (B);
		\draw[thick, dashed] (Centroid) -- (C);
		\draw[thick, dashed] (Centroid) -- (D);
		
		\coordinate (A1) at (3,1.5);    
		\coordinate (A2) at (3,0.5);    
		\coordinate (A3) at (3,-0.5);   
		\coordinate (B1) at (5,1.5);    
		\coordinate (B2) at (5,0.5);    
		\coordinate (B3) at (5,-0.5);   
		\draw[thick] (A1) -- (B1);
		\draw[thick] (A1) -- (B2);
		\draw[thick] (A1) -- (B3);
		\draw[thick] (A2) -- (B1);
		\draw[thick] (A2) -- (B2);
		\draw[thick] (A2) -- (B3);
		\draw[thick] (A3) -- (B1);
		\draw[thick] (A3) -- (B2);
		\draw[thick] (A3) -- (B3);
		\fill[red] (A1) circle (2pt);
		\fill[red] (A2) circle (2pt);
		\fill[red] (A3) circle (2pt);
		\fill[red] (B1) circle (2pt);
		\fill[red] (B2) circle (2pt);
		\fill[red] (B3) circle (2pt);
	\end{tikzpicture}
	\caption{Non-planar graphs corresponds to particles in 3 dimensions. For $G=\oplus_i\ZZ_{N_i}$, we have $T=\oplus_i\ZZ_{(N_i,2)}$: exchange phase can only be $0$ (bosons) or $\pi$ (fermions) and braiding between different particles is trivial.}
	\label{figFermion}
\end{figure}

For example, for particle excitations, we consider the existence of embedding from graphs to manifolds. A triangle or a square, as in Fig.~\ref{figFSymbol}, can be embedded in any close manifolds; \hbox{\raisebox{-2ex}{\includegraphics[width=1.2cm]{triangle0123}}} can be embedded in two dimensions or higher, but it cannot be embedded in $S^1$; non-planar graphs such as $K_{3,3}$ and $K_5$ cannot be embedded in $S^2$, but they can be embedded in $S^3$; see Fig.~\ref{figFermion}. By computing the corresponding $T_p(X,G)$, we find that different manifolds $M$ have different statistics "because" whether a $(p+1)$-dimensional simplicial complexes $X$ can be embedded in $M$ depends on the topology of $M$. For example, the statistics in $S^3$ is characterized by non-planar graphs. An interesting theorem in point-set topology \cite{MunkresTopology} says that every finite $(p+1)$-dimensional simplicial complex (with Lebesgue covering dimension $p+1$) can be embedded in $\mathbb{R}^{2p+3}$, so we conjecture that when one increase the spatial dimension, the classification of statistics will be stable since $d=2p+3$. This aligns with Theorem~\ref{conjectureCohomology} and the stability of Eilenberg-MacLane space \cite{miller}. The really interesting thing is that although we do not have a theory for non-Abelian and non-invertible excitations yet, our prediction from the perspective of embeddings aligns with the stabilities of $E_n$ monoidal categories \cite{ncatlabPeriodicTable,Baez1995HigherTQFT}; according to \cite{kong2024highercondensationtheory}, all topological excitations in $d$-dimensional space form an $E_{d-p}$ multi-fusion $(p+1)$-category.
 
 Finally, we note that excitation patterns that do not come from simplicial complex may also admit a physical interpretation; see Fig.~\ref{figFSymbolAndBraiding}. In \cite{Previous}, loop-membrane statistics and particle-membrane statistics are also found in this way. If one compute them using the excitation pattern described in Remark~\ref{remarkMixDimension}, the computational complexity will be too large, and the output process will also be complicated.
 
 \begin{figure}[htb]
 	\centering
 	\begin{subfigure}[b]{0.45\textwidth}
 		\begin{tikzpicture}
 			\coordinate (A) at (0,0);    
 			\coordinate (B) at (1,0);    
 			\coordinate (C) at (2,0);    
 			\coordinate (D) at (3,0);    
 			\draw[thick] (A) to[out=60, in=120] (C);   
 			\draw[thick] (A) to[out=-60, in=-120] (C); 
 			\draw[thick] (B) to[out=60, in=120] (D);   
 			\draw[thick] (B) to[out=-60, in=-120] (D); 
 			\fill[red] (A) circle (2pt) node[below] {0};
 			\fill[red] (B) circle (2pt) node[below] {1};
 			\fill[red] (C) circle (2pt) node[below] {2};
 			\fill[red] (D) circle (2pt) node[below] {3};
 		\end{tikzpicture}
 		\caption{$T=\ZZ_2$ describes the braiding process of two anyons.}
 		\label{fig:sub1}
 	\end{subfigure}
 	\begin{subfigure}[b]{0.45\textwidth}
 		\begin{tikzpicture}
 			\coordinate (A1) at (0,1);    
 			\coordinate (A2) at (0,-1);   
 			\coordinate (B) at (1,0);     
 			\coordinate (C) at (2,0);     
 			\coordinate (D1) at (3,1); 
 			\coordinate (D2) at (3,-1);
 			\draw[thick] (A1) -- (C);   
 			\draw[thick] (A2) -- (C);   
 			\draw[thick] (A1) -- (A2);  
 			\draw[thick] (D1) -- (B);   
 			\draw[thick] (D2) -- (B);   
 			\draw[thick] (D1) -- (D2);
 			\fill[red] (A1) circle (2pt); 
 			\fill[red] (A2) circle (2pt) ;
 			\fill[red] (B) circle (2pt) ;
 			\fill[red] (C) circle (2pt) ;
 			\fill[red] (D1) circle (2pt) ;
 			\fill[red] (D2) circle (2pt) ;
 		\end{tikzpicture}
 		\caption{$T=\ZZ_2^3$ describes the braiding process of two anyons together with the F-symbol of both anyons}
 	\end{subfigure}
 	\caption{Two excitation patterns of $\ZZ_2$ anyons, while their underlying graphs are not simplicial complexes.}
 	\label{figFSymbolAndBraiding}
 \end{figure}
 
 \subsection{Constructing statistical processes}
 
Recall that the statistics of a realization $U\in R(m)$ is identified with the equivalence class $[U]\in T^*(m)$, but in general, even though we already know $T^*(m)$, it is difficult to decide which equivalence class it belongs to. That is why we must study the pairing
\begin{equation*}
	\text{realizations }\times \text{ statistical processes }\rightarrow \RR/\ZZ
\end{equation*}
or
\begin{equation}
	R(m)\times E_{\inv}(m)\rightarrow \RR/\ZZ.
\end{equation}
From a theoretical point of view, the cleanest object is $T=E_\inv/E_\id$, but its elements are also equivalence classes which are difficult to write down explicitly. In practice, the best way is still to construct an explicit statistical process $g\in \operatorname{F}(S)$. Next, for any realization $U$, we compute the unitary operator $U(g)$, which should be a pure phase factor due to Theorem~\ref{thmInitialStateIndependence}. This phase factor tells the statistics of $U$.

To do this construction, we should first choose a representative $e\in E_\inv$ of an element $x\in T=E_\inv/E_\id$ and then write $e$ as $(g,a)$. The Smith decomposition algorithm in the computation of $T$ automatically produce a representative $e$, but it is in general very complicated. We want to select the simplest expression in the class $e+E_\id$. For any expression $e=\sum c(s,a)(s,a)$, we define the square-norm $|e|^2=\sum |c(s,a)|^2$. Thus, our goal is to minimize the square-norm among $e+E_\id$.
 
In practice, we do not really need to make the norm strictly smallest; thus, we can find a relatively-small one in the following way. Initially, we begin with an expression $e\in E_\inv$; next, we randomly pick a basic locality identity $\{e_i\in E_{\operatorname{id}}$; if $2|e\cdot e_i|> |e_i|^2$, we can replace $e$ by $e\pm e_i$ such that the norm decreases. We repeat it until $2|e\cdot e_i|\le  |e_i|^2$ for all basic locality identities, and then we obtain a local minimum. Repeating this procedure many times with different random choices, we select the best result found. Then, we may draw the corresponding $1$-cycle in $\operatorname{G}(m)$ on a computer.

\begin{figure*}
 	\centering
 	\includegraphics[width=0.9\linewidth]{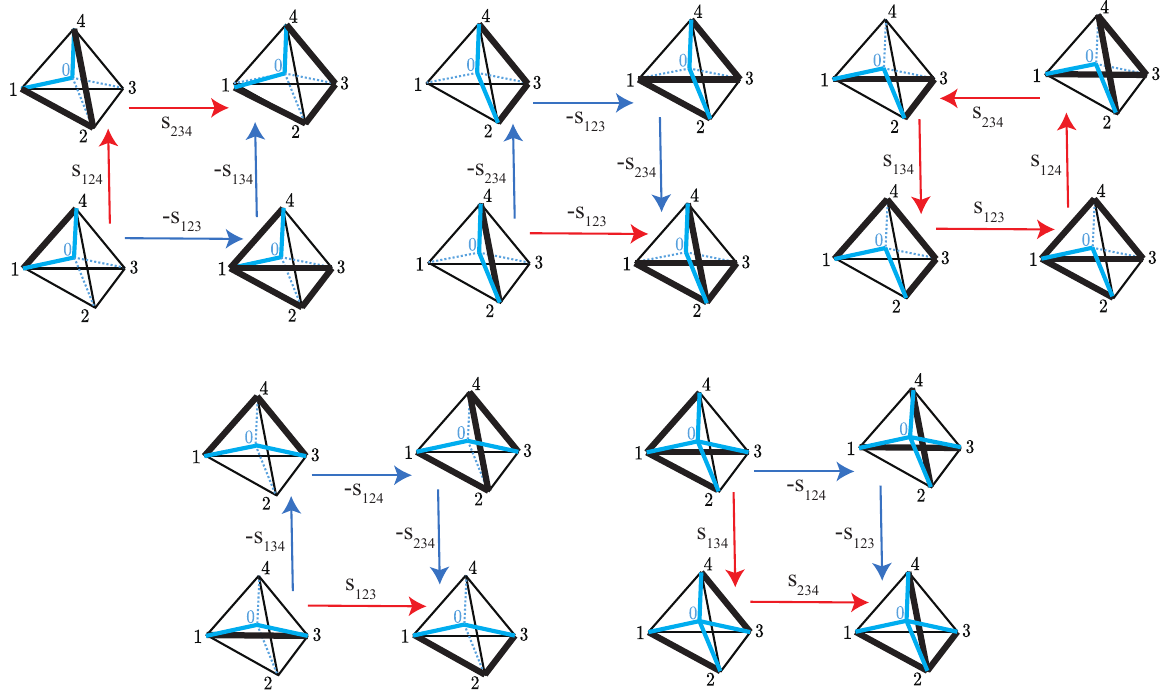}
 	\caption{The figure shows a statistical expression $e$ of fermionic loop statistics consisting $20$ of $(s,a)$-terms, drawn as a $1$-cycle in the configuration graph $\operatorname{G}(m)$. Compared to the $24$-step process in Fig.~\ref{fig: 24 step process}, this expression has many features. First, it is not connected; second, these configurations are not single-loop configurations, and the geometric intuition of reversing the orientation is no longer meaningful; third, only four outer membranes $s_{123},s_{134},s_{124},s_{234}$ are involved, sharply different from the $24$-step process, where only $s_{0ij}$ are involved. Since this expression is equivalent to the $24$-step process, one can conclude that the statistics is independent to any excitation operator. One can always construct statistical expressions in such forms, i.e., avoiding or always containing the vertex $\overline{0}$; see Theorem~\ref{thmReplacable1} and Theorem~\ref{thmStar}.
 		}
 	
 	\label{figFermionicloop20step}
 \end{figure*}

This algorithm was used to find the 24-step process for fermionic loop statistics \cite{Previous}, while the value $24$ is the minimum among numerous computations. We actually find some expressions with $|e|^2=20$, but all of them are not connected, so converting them into processes will involve more steps; see Fig.~\ref{figFermionicloop20step}.
 
 We apply this algorithm to the particle-fusion statistics and find the following process: 
  when the fusion group is $G=\ZZ_n$, we have $T=\ZZ_n$ generated by the statistical process $[s_3,s_2^n]$, where $s_1,s_2,s_3$ correspond to three edges of the triangle. 
 This process provides a new physical interpretation of particle-fusion statistics in one dimension. Traditionally, this statistics is called the F-symbol and interpreted as the associator in the fusion of three quasi-particles. Our theory gives another interpretation: \textbf{ the $\ZZ_n$ F-symbol is the obstruction to $ U(s)^n=1$.} If a realization $U$ has nontrivial statistics, then $U([s_3,s_2^n])\ne1$; although $U(s_2)^n|a\rangle=e^{i\theta(s_2^n,a)}|a\rangle$, the phase factor $e^{i\theta(s_2^n,a)}$ must have dependence on $a\in A$.
 
 This means that statistics imposes general restrictions on excitation operators. For instance, a special class of lattice models is Pauli stabilizer models. In a $\ZZ_n$ qubit spanned by $\{|0\rangle,\cdots,|n-1\rangle\}$, the Pauli $X$ and Pauli $Z$ operators are defined by $X|k\rangle=|k+1 \text{ mod n}\rangle$ and $Z|k\rangle=e^{\frac{2k\pi i}{n}}|k\rangle$, satisfying $ZX = e^{\frac{2\pi i}{n}} XZ$. We say that a realization $U$ is a $\ZZ_n$ Pauli stabilizer model, if all excitation operators $ U(s)$ are products of Pauli $X$ and Pauli $Z$, possibly on many different qubits. In this case, we have $ U([s,s'])\propto1$ and $ U(s^n)\propto1$, so phases correspond to $[s'',[s',s]]$ and $[s',s^n]$ must vanish.
 
 Therefore, topological orders with $\mathbb{Z}_n$ anyons cannot be described by $\mathbb{Z}_n$ Pauli stabilizer models unless it has trivial F-symbol. For example, the $\mathbb{Z}_2$ toric code is a $\ZZ_2$ Pauli stabilizer model, and all its anyons have trivial F-symbol\footnote{Although it is two-dimensional, we can treat it as an one-dimensional system by focusing on a specific triangle.}. In contrast, the F-symbol of a semion is nontrivial in $\ZZ_2$. This implies that the double-semion model cannot be realized as a $\ZZ_2$ Pauli stabilizer model. Nonetheless, it can be realized in a $\mathbb{Z}_4$ Pauli stabilizer model \cite{Ellison_2022}.
 
 A similar discussion applies to membrane-fusion statistics. For $\ZZ_n$-membranes,  the statistical expression generating $T\simeq \ZZ_n$ can be written as a sum of $\pm (s_1^n, a)$ over $4n$ different configurations $a$; more specifically, the statistical process is $g = (s_4 s_3)^{-n} (s_4s_3 [ s_3, [s_2, s_1^n]])^n$ \cite{Previous}. It is a bit surprising that only $s_1^n$ is involved in the expression $e$, but the existence of such a form is not a coincidence; see Theorem~\ref{thmReplacable1}. Similar to the case of F-symbol, \textbf{the fusion statistics of $\ZZ_n$-membranes is the obstruction to $U(s)^n=1$}.
 
 \begin{remark}
 	The discussion of $\mathbb{Z}_n$ F-symbol also applies to mutual statistics. When the fusion group is $G=\ZZ_2\times \ZZ_2$, the statistics is $T=\ZZ_2^3$, which can be viewed as the self-statistics of three particle types $a$, $b$, and $ab$. Actually, $T$ is generated by three statistical processes $[s_{3a}, s_{2a}^n]$, $[s_{3b}, s_{2b}^n]$, and $[s_{3a}s_{3b}, (s_{2a}s_{2b})^n] $. In contrast to particles, the statistics of $\ZZ_2\times \ZZ_2$-membranes is $T=\ZZ_2^4$, so it cannot be interpret as the self-statistics of three different types of membranes.
 \end{remark}
 
The method of random simplification does not always produce an enlightening expression, especially when $\dim E$ is too large. The perspective of obstructions sometimes provides better insights.

Let us begin with a simple example. We have used the statistical process $[s_3,s_2^n]$ to show that the $\ZZ_n$ F-symbol is the obstruction to $U(s)^n = 1$; now, we reverse this logic, assuming we do not know nothing in priori. To find the statistical process, we try to construct $F = \operatorname{span}\{(s_2^n, a) | a \in A\} \subset E$. Taking any $e\in E_{\operatorname{inv}}$ (not in its simplest form) generating $T$, we can use row reduction to check if $e\in F + E_{\operatorname{id}}$, which is true in this case. Thus, we know the generator of $T$ has a representative of the form $\sum_i c_i (s_2^n, a_i)$. This gives us information about the potential form of representatives, and it exactly means that the F-symbol is the obstruction to $U(s_2^n)=1$. 
 
 In general, one may select several processes $\{g_1, \cdots, g_k\}$, construct $F=\operatorname{span}\{(g_i,a)|1\le i\le k,a\in A\}$ and check if $E_{\operatorname{inv}}\subset F + E_{\operatorname{id}}$. If that is the case, every statistics $t\in T(m)$ can be represented by expressions in $F$, and then we say the statistics $T(m)$ is the obstruction to $\{ U(g_1)=1, \cdots,  U(g_k)=1\}$.
 
 This method is used to study the fusion statistics of $\ZZ_n\times \ZZ_n$ loops. Note that the self-statistics of $\ZZ_n$ loops is trivial because $ H_4(\mathbb{Z}_n, \ZZ) = 0$. We intend to find the expressions generating $T_1(\partial \Delta^3,\ZZ_n\times \ZZ_n) = H_4(\mathbb{Z}_n \times \mathbb{Z}_n, \ZZ) = \mathbb{Z}_n^2$.
 
 One way is to try obstructions like $ U(s)^n=1$, similar to the particle-fusion statistics. When $n = 2$, there are three nontrivial loops: $a$, $b$, and their fusion $ab$. Let $U_a,U_b$ be two membrane operators for $a$ and $b$ on a specific $2$-simplex. According to the computation, \textbf{the $\ZZ_2\times \ZZ_2$-loop-fusion statistics is the obstruction to $ U(s_a)^2=1, U(s_b)^2=1$, and $ U(s_as_b)^2=1$}. Choosing only two of them is not enough.
 
 Another way is to try double-commutators. We generate $F$ by $([s'', [s', s]], a)$ for all $s'', s', s \in S,a\in A$; our computer program shows $E_{\operatorname{inv}}\subset F + E_{\operatorname{id}}$. Therefore, \textbf{the statistics is the sum of higher commutators terms, which is always trivial in any Pauli stablizer models}. 
 
 When $n>2$, another formula works better. We have verified that  \textbf{the $\ZZ_n\times \ZZ_n$-loop-fusion statistics is the obstruction to $ U(s_{1a} s_{2a} s_{3a} s_{4a})=1$ and $ U(s_{1b} s_{2b} s_{3b} s_{4b})=1$}. $s_{1a} s_{2a} s_{3a} s_{4a}$ means swiping the $a$-loop on the whole sphere, which can be viewed as a symmetry that creates no excitation. This obstruction implies that when the fusion statistics is nontrivial, it is impossible to truncate the global symmetry such that the patch symmetry operator commutes with the global symmetry operator. We have further found an explicit statistical process generating $T$: one is $(s_{4a} s_{3a})^{-n} (s_{4a} s_{3a} [s_{3a}, [s_{2a}, s_{1b} s_{2b} s_{3b} s_{4b}]])^n$, and the other is obtained by exchanging $a,b$.

\section{Relations Across Different Dimensions and Fusion Groups}\label{secFunctorial}

In algebraic topology, there are canonical maps between Eilenberg-MacLane spaces, and they can also be interpreted from the side of statistics. These correspondences are exact because our construction in the proof of Theorem~\ref{conjectureCohomology} is quite natural. 

First, any map $f:G'\rightarrow G$ induces a map $K(G',d-p)\rightarrow K(G,d-p)$ and  $H^{d+2}(K(G,d-p),\RR/\ZZ)\rightarrow H^{d+2}(K(G',d-p),\RR/\ZZ)$. This corresponds to the map from $R(m_p(\Delta^{d+1},G))\rightarrow R(m_p(\Delta^{d+1},G'))$. Let $U\in R(m_p(\Delta^{d+1},G))$; then, its image $U'$ is defined by $U'(s)=U(f(s))$, where $s\in G'\times \Delta^{d+1}_{p+1}$ and $f$ acts on the $G'$ part. 

To see the picture more clearly, we consider two cases, the embedding $f:\ZZ_2\rightarrow \ZZ_4$ and the projection $g: \ZZ\rightarrow\ZZ_2$, from the viewpoint of anyon statistics. On the algebraic side, $f$ induces a map

\begin{equation}
	H^4(K(\ZZ_4,2),\RR/\ZZ)\simeq \ZZ_8\xrightarrow{1\mapsto 2} \ZZ_4\simeq H^4(K(\ZZ_2,2),\RR/\ZZ).
\end{equation}
From the physical side, initially, we have an anyon type $\alpha$ satisfying $4\alpha=0$; then, this map determines the topological spin of $2\alpha$ from the topological spin of $\alpha$. This map is not surjective, which indicates that $2\alpha$ can never be semions or anti-semions.

From the algebraic side, $g$ induces the map
\begin{equation}
	H^4(K(\ZZ_2,2),\RR/\ZZ)\simeq \ZZ_4\xrightarrow{1\mapsto \frac{1}{4}} \RR/\ZZ\simeq H^4(K(\ZZ,2),\RR/\ZZ).
\end{equation}
From the physical side, we have a $\ZZ_2$-anyon $\alpha$ which can be boson, fermion, semion, or anti-semion. If we do not require $2\alpha=0$, then the topological spin of $\alpha$ could be any phase factor. From the perspective of statistical processes, the induced map is
\begin{equation}
	H_4(K(\ZZ,2),\ZZ)\simeq \ZZ\xrightarrow{1\mapsto1}\ZZ_4\simeq H_4(K(\ZZ_2,2),\ZZ).
\end{equation}
The surjectivity of this map indicates that regardless of the fusion group, the topological spin is universally measured by the T-junction process. The membrane statistics in $4$ dimensions also has this property \cite{feng2025anyonicmembranespontryaginstatistics}\footnote{The statistics when $G=\ZZ$ cannot be computed directly using our computer program; there are some technical details to find this process for $G=\ZZ$.}. Other statistics do not have this property: for example, the F-symbol of $\ZZ_n$-particle is measured by $[s_3,s_2^n]$, and there is no universal process for all $\ZZ_n$ or $\ZZ$; also, the fermionic loop statistics is a unique behavior of $\ZZ_{2n}$-loops, and it does not exist when the fusion group is $\ZZ$. 

Second, the homotopy equivalence
\begin{equation}
	K(G,d-p-1)\simeq \Omega K(G,d-p)
\end{equation}
induces a map 
\begin{equation}
	\Sigma K(G,d-p-1)\rightarrow K(G,d-p)
\end{equation}
and further induces a canonical map 
\begin{equation}\label{eqSuspensionMap}
	H^{d+2}(K(G,d-p),\RR/\ZZ)\rightarrow H^{d+1}(K(G,d-p-1),\RR/\ZZ).
\end{equation}

From the physical side\footnote{We believe this can be rigorously proved using Kan suspension \cite{goerss-2009}.}, it maps a realization of $m_p(\Delta^{d+1},G)$ to a realization of $m_p(\Delta^d,G)$; the idea is to simply focus on a sub-manifold $S^{d-1}$. For example, let us consider a $\ZZ_2$ quasi-particle $\alpha$ in $3\rightarrow2\rightarrow1$ dimensions. The corresponding maps of cohomology groups are
\begin{equation}
	\ZZ_2\xrightarrow{1\mapsto2}\ZZ_4\xrightarrow{1\mapsto1}\ZZ_2.
\end{equation}
 In $3$ dimensions, $\alpha$ can be a boson or a fermion; a $\ZZ_2$ anyon in $2$ dimensions can also be a semion or an anti-semion, but these possibilities are forbidden if the anyon is mobile in $3$ dimension. For a generic $\ZZ_2$-anyon in $2$ dimensions, if we focus on a circle, then we will get its F-symbol. Semions and anti-semions has the F-symbol $1\in \ZZ_2$, while the F-symbol for bosons and fermions is trivial. So, the particle statistics in $1$, $2$, and $3$ are very different. Nothing new happens in $d>3$, and in general, the statistics of $p$-dimensional excitation stabilizes since $d\ge 2p+3$. This is because Eq.~\eqref{eqSuspensionMap} is an isomorphism when $d>2p+3$; see, for example, Proposition 70.2 of \cite{miller}.
 
 \begin{remark}\label{remarkThreeLoop}
 	Our current theory can not describe three-loop braiding statistics \cite{Wang2014Braiding}. For simplicity, we take $G=\ZZ_3\times \ZZ_3$. In our theory, the loop-fusion statistics in $2$ dimensions is classified by $H^4(G,\RR/\ZZ)\simeq \ZZ_3^2$, while its statistics in $3$ dimensions is trivial. This contradicts with the study of three-loop braiding statistics, which asserts a classification of $H^4(G,\RR/\ZZ)$ in $3$ dimensions. We believe that the three-loop braiding statistics is not the statistics of loops but a non-Abelian statistics between loops and quasi-particles, which is beyond our current theory.
 \end{remark}

\section{Local Statistics and Local Topology}\label{secUniversal}

In this section, we continue our theoretical analysis of statistics for excitation patterns, particularly focusing on the statistics of $m=m_p(X,G)$. We have already defined
\begin{equation*}
	T(m)=\frac{\cap_{x}E_\id(m|_x)}{E_\id(m)},
\end{equation*}
where the localization $m|_x$ at a vertex $x$ plays a central role. We may view $m$ as a "global" object and $m|_x$ as an "on-site" object; between the two extremes, what will happen if we do localization in a subspace smaller than the whole space $X$ but larger than a point $x$? For example, what can we say about the "local statistics" $T(m|_s)$ for some $s\in S$? This idea turns out to be the critical bridge between local topology and the global topology of $X$, which leads to a deeper understanding of statistics. We have already shown that the phase factor measured by the T-junction process has many striking properties related to its stability: it is quantized, independent to the initial state, and independent to the choice of excitation operators. In this section, we will link these properties to the manifold structure of $X$: although we can define $T(m_p(X,G))$ on any simplicial complex, the statistics enjoys these properties only when $X$ has good local topology. The surprising relationship between the manifold structure of $X$ and these properties makes us believe that both of them are profound. In order to bridge local statistics and global statistics and the operator independence, we develop two tools: quantum cellular automata and condensation. We borrow these names from physics but only for our own definitions. We note that most proofs in this section are very technical, so readers may only read the statement of these theorems. We also write a short sentence for every theorem to conclude its physical meaning.

\subsubsection{Quantum cellular automata}\label{subsecQCA}

\begin{definition}\label{defQCA}
	Let $m=(A,S,\partial,\supp)$ be an excitation model. A quantum cellular automata (QCA) of $m$ consists of a Hilbert space $\mathcal{H}$, a collection of \textit{configuration states} $\{|a\rangle|a\in A\}$ such that $|a\rangle$ and $|a'\rangle$ are either orthogonal or collinear, and an unitary operator $\mathcal{U}$ satisfying the following two axioms.
	\begin{enumerate}
		\item \textbf{The configuration axiom:} For any $a\in A$, the equation
		\begin{equation}\label{axiomConfigurationQCA}
			\mathcal{U}|a\rangle=e^{i\theta(a)}|a\rangle
		\end{equation}
		holds for some $\theta(a)\in \RR/2\pi\ZZ$.
		\item \textbf{The locality axiom:} \(\forall s_1, s_2, \dots, s_k \in S\) satisfying \(\operatorname{supp}(s_1) \cap \operatorname{supp}(s_2) \cap \cdots \cap \operatorname{supp}(s_k) = \emptyset\), the equation
		\begin{align}\label{axiomLocalityIdentityQCA}
			\sum_{c_1,\cdots,c_k\in\{0,1\}}(-1)^{c_1+\cdots+c_k}\theta(a+\sum_{i=1}^kc_i\partial s_i)=0\in \RR/2\pi\ZZ
		\end{align}
		holds.
	\end{enumerate}
	
\end{definition}

In parallel to realizations, we use the symbol $\mathcal{U}$ to label a QCA, and we denote all QCAs of an excitation pattern $m$ by $Q(m)$, which is the solution space of Eq.~\eqref{axiomLocalityIdentityQCA}. We similarly define the expression group 
\begin{equation}
	D(m)=\oplus_{a\in A}\left(a\right)\simeq \ZZ[A],
\end{equation} 
locality identities $D_\id(m)$ generated by
\begin{equation}\label{eqGeneratorOfDid}
	\sum_{c_1,\cdots,c_k\in\{0,1\}}(-1)^{c_1+\cdots+c_k}\left(a+\sum_{i=1}^kc_i\partial s_i\right)
\end{equation}
for \(\operatorname{supp}(s_1) \cap \operatorname{supp}(s_2) \cap \cdots \cap \operatorname{supp}(s_k) = \emptyset\). The statistics is 
\begin{equation}
	\tau^*(m)=Q(m)/\sum_{x}Q(m|_x),
\end{equation}
and its dual group is
\begin{equation}
	\tau(m)=D_\inv(m)/D_\id(m),
\end{equation}
where 
\begin{equation}
	D_\inv(m)=\cap_xD_\id(m|_x).
\end{equation}

These constructions are actually simpler than the statistics we have studied before. While $E(m)\simeq \ZZ[S\times A]$ corresponds to $1$-chains in $\operatorname{G}(m)$, $D(m)\simeq \ZZ[A]$ corresponds to $0$-chains. In Appendix \ref{appendixSimplicialSet}, we prove that
\begin{theorem}\label{thmTau}
	\begin{equation}
		\tau^*(m_p(\partial\Delta^{d+1},G))\simeq H^{d+1}(K(G,d-p),\RR/\ZZ);
	\end{equation}
	\begin{equation}
		\tau(m_p(\partial\Delta^{d+1},G))\simeq H^{d+1}(K(G,d-p),\ZZ).
	\end{equation}
\end{theorem}
It is of course interesting to explore its physical meaning \cite{CarolynZhangSPTEntangler}, but in this paper we only use it as a technical tool to study the operator independence as follows. Let $m=(A,S,\partial,\supp)$ be an excitation pattern and $U\in R(m)$. Next, imagine we replace a specific unitary operator $U(s)$ by some $U'(s)$ having the same support and also satisfying the configuration axiom. Introducing $U'(s)$ is equivalent to introducing $\mathcal{U}=U'(s)U(s)^{-1}$, which satisfies Eq.~\eqref{axiomConfigurationQCA}. Then, for any $s_1,\cdots,s_k$ such that $\supp(s)\cap (\cap_i\supp(s_i))=\emptyset$, we have 
\begin{equation}
	[U(s_k),\cdots,[U(s_1),\mathcal{U}]]=1,
\end{equation}
which is exactly in the form of Eq.~\eqref{axiomLocalityIdentityQCA}; this implies $\mathcal{U}\in Q(m|_s)$. If $\tau^*(m|_s)=0$, then $\mathcal{U}$ decomposes into on-site terms, and therefore replacing $U(s)$ by $U'(s)$ do not change statistics, i.e., they are in the same equivalence class of $T^*(m)$. Although $\tau^*(m)$ is nonzero in general, we usually have $\tau^*(m|_s)=0$, and in this case the strong operator independence holds.

We can make similar arguments in terms of expression groups. Let $m=(A,S,\partial,\supp)$ be an excitation pattern and $s\in S$; we construct another excitation pattern $m'=(A,S\cup \{\sigma\},\partial,\supp)$ by introducing an additional element $\sigma$ such that $\partial\sigma=0$ and $\supp(\sigma)=\supp(s)$. Then, we have

\begin{equation}\label{eqDecomposeD}
	E(m')\simeq D(m|_s)\oplus E(m)
\end{equation}
 by identifying $(\sigma,a)\in E(m')$ with $(a)\in D(m|_s)$. Moreover, generators of $E_{\operatorname{id}}(m')$ split into two classes: the first class, with the form $([s_k,[\cdots,[s_2,s_1]]],a)$, generates $E_{\operatorname{id}}(m)$; the second class, with the form $([s_k,[\cdots,[s_1,\sigma]]],a)$, exactly generates $D_\id(m|_s)$. We have
 \begin{equation}
 	E_\id(m')\simeq D_\id(m|_s)\oplus E_\id(m)
 \end{equation}
 and
\begin{equation}
	E_\inv(m')\simeq D_\inv(m|_s)\oplus E_\inv(m).
\end{equation}
Thus, we have proved that
\begin{theorem}\label{thmSymmetryOperator}
	Consider two excitation patterns in $M$: $m=(A,S,\partial,\supp)$ and $m'=(A,S\cup \{\sigma\},\partial,\supp)$ , satisfying $\partial\sigma=0$ and $\supp(\sigma)=N$. Then we have
	\begin{equation}
		T(m')\simeq T(m)\oplus\tau(m|_N).
	\end{equation}
	
\end{theorem}

\begin{remark}\label{remarkPlaceHolder}
	It is convenient to write Eq.~\eqref{eqGeneratorOfDid} as $([s_k,[\cdots,[s_1,\sigma]]],a)$, where $\sigma$ is nothing but a placeholder with $\partial\sigma=0$. To recover Eq.~\eqref{eqGeneratorOfDid}, one needs to expand it using Eq.~\eqref{eqTheta(g,a)} and substitute all $(\sigma,a)$ by $(a)$.
\end{remark}

\subsubsection{Condensation}\label{subsecCondensation}

\begin{definition}\label{defCondensation}
	Let $m=(A,S,\partial,\supp)$ be an excitation pattern and $A_0\subset A$ is a subgroup. We define the corresponding \textit{condensed pattern} $m/A_0$ as $(A/A_0,S,q\circ \partial,\supp)$, where $q:A\rightarrow A/A_0$ is the quotient map.\footnote{Because condensation and localization deal with configurations and supports independently, $(m|_N)/A_0$ and $(m/A_0)|_N$ are the same.}
\end{definition}

A realization of the condensed pattern $m/A_0$ is automatically a realization of $m$; they share the same Hilbert space and operators, while $|a\rangle $ for $a\in A$ is identified to $|q(a)\rangle$. Different $a\in A$ having the same image in $A/A_0$ correspond to the same configuration state $|q(a)\rangle$, and that is why in Definition~\ref{defRealization}, we allow different configuration states to be identical. This construction of condensing $A_0$ also induces maps $D(m)\rightarrow D(m/A_0)$ defined by $(a)\mapsto (q(a))$ and $E(m)\rightarrow E(m/A_0)$ defined by $(s,a)\mapsto (s,q(a))$. They further induce maps for $D_\id,D_\inv,\tau,E_\id,E_\inv$, and $T$, and we denote these maps by the same notation $q_*$. 

A common situation is to condense $A_V=\{\partial s|s\in V\}$ for a given subset $V\subset S$, and then we write $m/A_V=(A/A_V,S,q\circ \partial,\supp)$ as $m/V$. Because $q\circ \partial(s)=0,\forall s\in V$, similar to Eq.~\eqref{eqDecomposeD}, there is a canonical decomposition
\begin{equation}\label{eqDecomposeE}
	E(m/V)=\big(\oplus_{s\in V}D(m|_s/V)\big)\bigoplus E(m'/V),
\end{equation}
where $m'/V=(A/A_V,V^c,q\circ \partial,\supp)$ is obtained by replacing $S$ by the complement $V^c=S-V$. Similarly, we have 
\begin{equation}\label{eqDecomposeEid}
	E_\id(m/V)=\big(\oplus_{s\in V}D_\id(m|_s/V)\big)\bigoplus E_\id(m'/V)
\end{equation}
and
\begin{equation}
	T(m/V)=\big(\oplus_{s\in V}\tau(m|_s/V)\big)\bigoplus T(m'/V).
\end{equation}

The map $T(m)\rightarrow T(m/V)$ induced by condensation is neither injective or surjective in general. On the other hand, this map is an isomorphism if all operators in $V$ have empty support, which usually appears in localization.

\begin{lemma}\label{lemmaQuotientLifting}
	Let $m=(A,S,\partial,\supp)$ be an excitation pattern, and $V$ is a subset of $S$ satisfying $\supp(s)=\emptyset, \forall s\in V$. Let $q_*:E(m)\rightarrow E(m/V)$ be the map induced by condensing $A_V$; then, we have
	\begin{equation}\label{eqQuotientExpression1}
		E_\id(m/V)=q_*[E_\id(m)],
	\end{equation}
	\begin{equation}\label{eqQuotientExpression2}
		E_\id(m)=q_*^{-1}[E_\id(m/V)]\cap E_\id(m|_\emptyset),
	\end{equation}
	\begin{equation}\label{eqQuotientExpression3}
		E_\inv(m/V)=q_*[E_\inv(m)],
	\end{equation}
	and
	\begin{equation}\label{eqQuotientExpression4}
		E_\inv(m)=q_*^{-1}[E_\inv(m/V)]\cap E_\id(m|_\emptyset).
	\end{equation}
	We omit the parallel equations for $D(m)$, as they are always simpler.
\end{lemma}
\begin{proof}
	$E_\id(m/V)$ is generated by $([s_k,\cdots,[s_2,s_1]],q(a))$ for $\cap_i\supp(s_i)=\emptyset$, which is the image of $([s_k,\cdots,[s_2,s_1]],a)\in E_\id(m)$. Thus we have Eq.~\eqref{eqQuotientExpression1}.
	
	The "$\subset$" part of Eq.~\eqref{eqQuotientExpression2} is trivial; conversely, assume $e\in E_\id(m|_\emptyset)$ and $q_*(e)\in E_\id(m/V)$, we want to prove $e\in E_\id(m)$. Using Eq.~\eqref{eqQuotientExpression1}, there exists $e_0\in E_\id(m)$ such that $q_*(e_0)=q_*(e)$. Thus, $e-e_0$ can be decomposed in to terms of the form $(s,a+\delta)-(s,a)$, with $\delta\in A_V$. For any $\delta\in A_V$, we fix $g_\delta\in \operatorname{F}(V)$ such that $\partial g_\delta=\delta$. Then, we expand $([g_\delta,s],a)\in E_\id(m)$ as
	\begin{equation}
		(s,a)-(g_\delta,a+\partial s)-(s,a+\delta)-(g_\delta,a+\partial s).
	\end{equation} 
	Adding these expressions to $e-e_0$, we obtain an expression $e'\in E_\id(m|_\emptyset)$ only involving $(s,a)$ terms for $s\in V$.  Since $e'$ is closed, we can write $e'=\sum_i (g_i,a_i)$ for $g_i\in \operatorname{F}(V),\partial g_i=0, a_i\in A$. Because $([s,g_i],a)\in E_{\operatorname{id}}(m),\forall s\in S, a\in A$, $e'$ is equivalent to $\sum_i (g_i,0)=(g_1g_2\cdots g_n,0)$, and then it obviously belongs to $E_\id(m)$.
	
	To prove Eq.~\eqref{eqQuotientExpression3}, $q_*[E_\inv(m)]\subset E_\inv(m/V)$ is obvious; conversely, assuming $e\in E_\inv(m/V)$, we construct a pre-image $e_0\in E_\id(m|_\emptyset)$ such that $q_*(e_0)=e$. Using Eq.~\eqref{eqQuotientExpression2} for every $m|_x$, we find $e_0\in E_\id(m|_x),\forall x$. Thus, $e\in E_\inv(m)$.
	
	Eq.~\eqref{eqQuotientExpression4} follows directly from $E_\inv(m)=\cap_x E_\id(m|_x)$.
\end{proof}

\begin{theorem}\label{thmQuotientEmptySupport} \textbf{Operators with empty support do not contribute to the statistics.}
	
	Let $m=(A,S,\partial,\supp)$ be an excitation pattern, and $V$ is a subset of $S$ satisfying $\supp(s)=\emptyset, \forall s\in V$.  Then, $q_*$ induces isomorphisms $T(m)\simeq T(m/V)$ and $\tau(m)\simeq \tau(m/V)$.
\end{theorem}
\begin{proof}
	For the homomorphism $q_*:T(m)\rightarrow T(m/V)$, the injectivity follows Eq.~\eqref{eqQuotientExpression2}, and the surjectivity follows Eq.~\eqref{eqQuotientExpression3}. The case for $\tau(m)\simeq \tau(m/V)$ is similar.
\end{proof}

As an application, we give a characterization of statistical expressions $e\in E_\inv(m)=\cap_xE_\id(m|_x)$. First, a necessary condition is $e\in E_\id(m|_\emptyset)$; on top of that, we want to check $e\in E_\id(m|_x)$. Let
\begin{equation}\label{eqVx}
	V(x)=\{s\in S|x\in \supp(s)\}
\end{equation}
 be all operators containing $x$, and its complement is $V(x)^c=S-V(x)$. We use $q_x$ to denote the quotient map $A\rightarrow A/A_{V(x)^c}$.
 
 Using Eq.~\eqref{eqQuotientExpression2}, we have
\begin{equation}
	e\in E_\id(m|_x)\iff q_{x*}(e)\in E_\id\big(m|_x/V(x)^c\big),
\end{equation}
where $q_{x*}$ maps $(s,a)$ to $(s,q_x(a))$. In the decomposition Eq.~\eqref{eqDecomposeE}, components in $D(m|_s/V)$ are in $D_\id$ automatically because $e\in E_\id(m|_\emptyset)$. The $E(m'/V)$ component is obtained by further map $(s,q_x(a))$ to $0$ when $s\in V(x)^c$, and we denote the total map by $\tilde{q}_{x*}$. All operators in $m'/V$ have supports containing $x$, so $E_\id(m'/V)=0$. Thus, we have proved Theorem~\ref{thmCheckByHand}:
\begin{equation}
	e\in E_\id(m|_x)\iff \tilde{q}_{x*}(e)=0.
\end{equation}

\begin{remark}
	Theorem~\ref{thmCheckByHand} applies to $D_\inv$ similarly. We say $d=\sum_{a\in A} c(a)(a)$ is zero-weight if $\sum c(a)=0$. This is equivalent to $d\in D_\id(m|_\emptyset)$. When $d$ is zero-weight, we have
	\begin{equation}
		d\in D_\id(m|_x)\iff q_{x*}(d)=0
	\end{equation}
	and
	\begin{equation}\label{eqDinv}
		d\in D_\inv\iff q_{x*}(d)=0,\forall x,
	\end{equation}
	where $q_{x*}: \ZZ[A]\rightarrow \ZZ[A/A_{V(x)^c}]$ is induced by the quotient map $A\rightarrow A/A_{V(x)^c}$.
\end{remark}

\subsection{Strong operator independence}

In Section~\ref{subsecQCA}, we have already discuss an explanation of strong operator independence, and now we discuss them more delicately. To begin with, we introduce some notations.

\begin{definition}
	Let $m=(A,S,\partial,\supp)$ be an excitation model. For any $s\in S$,
	
	 We define the "picking-$s$ homomorphism" $\sigma_s: E(m)\rightarrow D(m)$, mapping $(s,a)$ to $(a)$ and $(t,a)$ to $0$ for $t\ne s$. 

	Fr any $b\in A$, we introduce the "translate-and-compare homomorphism" $\Delta_b:D(m)\rightarrow D(m)$ by
	\begin{equation}
		\Delta_b(a)=(a+b)-(a).
	\end{equation}
	We use the same notation $\Delta_b$ for the homomorphism $E(m)\rightarrow E(m)$ defined by
	\begin{equation}\label{eqTranslationDifference}
		\Delta_b(s,a)=(s,a+b)-(s,a).
	\end{equation}
\end{definition}

\begin{lemma}\label{lemmaBoundary}
	For any $s\in S$,
	\begin{enumerate}
		\item $e\in E_{\operatorname{id}}(m)\implies\sigma_{s}(e)\in D_{\operatorname{id}}(m|_s)$.
		\item $e$ satisfies Eq.~(\ref{eqEinv}) $\implies \sigma_s(e)$ satisfies Eq.~\eqref{eqDinv}.
	\end{enumerate}
\end{lemma}
\begin{proof}
	Let $e=([s_n,\cdots,[s_1,s]],a)$ be a basic locality identity in $ E_{\operatorname{id}}(m)$, then $\sigma_s(e)=([s_n,\cdots,[s_1,\sigma]],a)\in D_{\operatorname{id}}(m|_s)$ in the sense of Remark~\ref{remarkPlaceHolder}.
	
	The second assertion is obvious.
\end{proof}

The map in Eq.~\eqref{eqTranslationDifference} can bring $D_{\operatorname{id}}(m|_s)$ to $D_{\operatorname{id}}(m)$. 

\begin{lemma}\label{lemmaTranslation}
	Let $m=(A,S,\partial,\supp)$ be an excitation pattern and $t\in S$ satisfies $\tau(m|_t)=0$. Then, 
	\begin{itemize}
		\item If $d\in D(m)$ satisfies $q_{x*}(d)=0,\forall x\in\supp(t)$, then $\Delta_{\partial t}(d)\in D_\id(m)$.
		\item If $e\in E(m)$ is closed and satisfies $q_{x*}(e)=0,\forall x\in\supp(t)$, then $\Delta_{\partial t}(e)\in E_\id(m)$.
	\end{itemize}
\end{lemma}
\begin{proof}
	We only prove the assertion for $E$, and the assertion for $D$ is similar. 
	
	 Geometrically, $\Delta_{\partial t}(e)$ are the difference of two parallel loops in $\operatorname{G}(m)$ differed by the translation along $t$. Following this picture, writing $e=\sum_{s\in S,a\in A}c(s,a)(s,a)$, we have
	\begin{equation}
		\Delta_{\partial t}(e)=\sum_{s\in S,a\in A}c(s,a)([s,t],a).
	\end{equation}
	Our destination is to prove that for any $s\in S$,
	\begin{equation}
		\Delta_{\partial t}(e)_s=\sum_{a\in A}c(s,a)([s,t],a)\in E_\id(m).
	\end{equation}
	When $\supp(s)\cap\supp(t)=\emptyset$, then this equation holds automatically; when $\supp(s)\cap\supp(t)\ne\emptyset$,
	$q_{x*}(e)=0,\forall x\in \supp(s)$ implies that $\sigma_s(e)=\sum_{a\in A}c(s,a)(a)$ is zero-weight. Using Lemma \ref{lemmaBoundary} and Eq.~\eqref{eqDinv}, we have $\sigma_s(e)\in D_\inv(m|_s)=D_\id(m|_s)$.	
	We expand $\sigma_s(e)$ as the sum of basic locality identities $([s_k,[\cdots,[s_1,\sigma]]],a)\in D_\id(m|_s)$, and $\Delta_{\partial t}(e)_s$ corresponds to replacing them by $([s_k,[\cdots,[s_1,[s,t]]]],a)$. The result is in $ E_\id(m)$ obviously. 
\end{proof}
Using this lemma, we immediately get the following theorem.
\begin{theorem}\label{thmInitialStateIndependence}
	\textbf{The phase factor evaluated by a statistical process is initial-state independent.}
	
	Let $m=(A,S,\partial,\supp)$ be an excitation pattern satisfying $\tau(m|_s)=0, \forall s\in S$. Then, we have
	\begin{itemize}
		\item $d\in D_\inv(m)\implies \Delta_a(d)\in D_\id(m),\;\;\forall a\in A$. 
		\item $e\in E_\inv(m)\implies \Delta_a(e)\in E_\id(m),\;\;\forall a\in A$. 
		\item If $g$ is a statistical process, then $[(g,a)]\in T(m)$ does not depend on $a\in A$.
	\end{itemize}
\end{theorem}

\begin{example}
	Theorem~\ref{thmInitialStateIndependence} may not hold if $\tau(m|_s)\ne0$. An example is like\hbox{\raisebox{-2ex}{\begin{tikzpicture}
			\draw[thick] (0,0) -- (2,0); 
			
			\foreach \x in {(0,0), (2,0)} \fill[red] \x circle (2pt);
			\node[red, below] at (0,0) {$s_1$};
			\node[red, below] at (2,0) {$s_2$};
			\node[black, below] at (1,0) {$s_3$};
	\end{tikzpicture}}} , which is a modification of $m_{-1}(S^0,\ZZ_2)$ discussed in Section~\ref{subsecSymmetry}. Two red points correspond to two excitation operators $s_1,s_2$, and the black line corresponds to $s_3$. We define $A=\ZZ_2\oplus\ZZ_2,S=\{s_1,s_2,s_3\},\partial s_1=\partial s_2=(1,0),\partial s_3=(0,1)$. If we do not have $s_3$ and the additional $\ZZ_2$ part, we will have $\tau=\ZZ_2$, generated by $(1)-(0)\in D$. But with the additional $\ZZ_2$ part, we have $\tau=\ZZ_2\oplus\ZZ_2$, generated by $d=((1,0))-((0,0))$ and $\Delta_{\partial s_3}(d)\notin D_{\operatorname{id}}$. We can construct a similar example that $\Delta_a(e)\notin E_{\operatorname{id}}$ by modifying $m_{-1}(S^0,\ZZ_2\times \ZZ_2)$.

\end{example}

Next, we prove three theorems about the operator independence.

\begin{theorem}\label{thmReplacable1}
	\textbf{One can eliminate any operator in a  statistical expression.}
	
	Let $m=(A,S,\partial,\supp)$ be an excitation pattern in $M$. For any $x\in M$, if $\tau(m|_t)=0,\forall t\in V(x)$ (see Eq.~\eqref{eqVx}), then 
	\begin{equation}
		E_\inv(m)\subset E_\id(m)+\ZZ[V(x)^c\times A].
	\end{equation}
	In other words,	every element in $T(m)$ can be represented by an expression $e$ consisting only of terms $(s,a)$ that $s\in V(x)^c$.
\end{theorem}

\begin{proof}
	Taking a representative $e'\in E_{\operatorname{inv}}(m)$, we should annihilate all its terms involving operators in $V(x)$. Fixing $t\in V(x)$, we denote the corresponding terms by $\sum_ic_i(t,a_i)$. Using Lemma~\ref{lemmaBoundary} and the condition $\tau(m|_t)=0$, we have $\sigma_t(e')=\sum_ic_i(a_i)\in D_{\operatorname{inv}}(m|_t)=D_{\operatorname{id}}(m|_t)$. $\sigma_t(e')$ is generated by basic locality identities of the form $([s_p,\cdots,[s_1,\sigma]],a)$; they satisfy $(\bigcap_{i=1}^p \supp(s_i))\cap\supp(t)=\emptyset$, and without loss of generality, we assume $s_1\in V(x)^c$. 
	We substitute all terms like $([s_p,\cdots,[s_1,\sigma]],a)\in D_{\operatorname{id}}(m|_t)$ by $([s_p,\cdots,[s_1,t]],a)\in E_{\operatorname{id}}(m)$ and subtract them from $e'$. Because of Eq.~\eqref{eqBasicLocalityIdentityInEid}, we have eliminate all $t$-terms, while we also introduce some other terms $(s_1,a)$ for $s_1\in V(x)^c$. Repeat this method for all $t\in V(x)$, and then we are done.
\end{proof}

In Theorem~\ref{thmReplacable1}, we have proved the existence of a representative in $\ZZ[V\times A]$. Sometimes we want the stronger result that there is a representative in $\ZZ[V\times A_V]$, where $A_V$ is the subgroup generated by $\{\partial s|s\in V\}$. If this is true, then we can represent it as $(g,0)$, where $g\in\operatorname{F}(S)$.

\begin{theorem}\label{thmReplacable2}\textbf{One can eliminate any operator in a statistical process.}
	
	Let $m=(A,S,\partial,\supp)$ be an excitation pattern in $M$. $\alpha\subset M$ satisfies $\tau(m|_\alpha)=0$. Let $V$ be a subset of $S$ satisfying $\forall s\in V$, $\supp(s)=\alpha$, and then we have
	\begin{enumerate}
		\item $D_\inv(m|_\alpha)\subset D_\id(m)+\ZZ[A_{V^c}]$.
		\item $E_\inv(m|_\alpha)\subset E_\id(m)+\ZZ[V^c\times A_{V^c}]$.
	\end{enumerate}
	
\end{theorem}
\begin{proof}
	
	For any $d\in D_{\operatorname{inv}}(m)$, we have $q_{x*}(d)=0,\forall x\in \alpha$. We do the decomposition $d=\sum_{\alpha\in A/A_{V^c}}d_\lambda$, where $d_\lambda$ only contains $(a)$ terms such that $[a]=\lambda\in A/A_{V^c}$. Geometrically, $d$ is divided into several layers corresponding to the cosets of $A_{V^c}$. For each $\lambda$, there exists  $g\in \operatorname{F}(V)$ such that $[\partial g]=\alpha$, and then $d_\lambda+\Delta_{\partial g}(d_\lambda)\in \ZZ[A_{V^c}]$. Thus, we only need to prove $\Delta_{\partial g}(d_\lambda)\in D_\id(m)$, and by Lemma~\ref{lemmaTranslation}, it is enough to prove $q_{x*}(d_\lambda)=0$ for any $\lambda$ and $x\in \alpha$. This follows directly from $q_{x*}(d)=0,\forall x\in \alpha$ because the map $q_{x*}$ acts separately in different cosets.
	
	For any $e_0\in E_{\operatorname{inv}}(m|_\alpha)$, we construct an equivalent expression $e$ that consists only terms $(s,a)$ that $s\in {V^c}$. We do the decomposition $e=\sum_{\lambda\in A/A_{V^c}}e_\lambda$, where $e_\lambda$ only contains $(s,a)$ terms such that $s\in {V^c}, [a]=\lambda\in A/A_{V^c}$. Geometrically, $e$ is a $1$-cycle in $\operatorname{G}(m)$ containing no $t$-terms so it can be divided into several layers corresponding to the cosets of $A_{V^c}$, and every part is still a $1$-cycle. Similar to the proof of $D$, we do translation for each $e_\lambda$ and prove $\Delta_{\partial g}(e_\lambda)\in E_\id(m)$ using Lemma~\ref{lemmaTranslation}.
\end{proof}

Next, we prove a theorem that we have used implicitly everywhere. When we define $m_p(X,G)$ (Definition~\ref{expSimplicialComplex}), a generating set $G_0$ is taken as input data, but it is not reflected in the notation $m_p(X,G)$. This abuse of notation does not create problems only when $T(m_p(X,G))$ does not depend on $G_0$, which needs the additional assumption $\tau(m|_\alpha)=0$ for all $(p+1)$-simplex $\alpha$. Let $U$ be the excitation operators supported at $\alpha$, and let $V=U^c$. No matter how $G_0$ is chosen, $A_U=\operatorname{span}\{\partial s|s\in U\}$ is the same; we want to prove that $U$ only contribute to $T$ through $A_U$.

We need a linear algebra fact:
\begin{lemma}\label{lemmaLatticeIsomorphic}
	If $E'\subset E$ satisfies $E_{\operatorname{inv}}\subset E'+E_{\operatorname{id}}$, then $E_{\operatorname{inv}}/E_{\operatorname{id}}=(E_{\operatorname{inv}}\cap E')/(E_{\operatorname{id}}\cap E')$.
\end{lemma}

\begin{theorem}\label{thmG0independence}
	
	\textbf{$T(m_p(X,G))$) does not depend on $G_0$.}
	
	 Let $m=(A_V,V,\partial,\supp)$ be an excitation pattern. We consider two "enlarged" excitation patterns $m_1=(A,S=V\sqcup U_1,\partial,\supp)$ and $m_2=(A,S=V\sqcup U_2,\partial,\supp)$, satisfying $\supp(s)=\alpha$ if $s\in U_1$ or $s\in U_2$.
	 \begin{itemize}
	 	\item If $\operatorname{span}\{\partial s|s\in U_1\}=\operatorname{span}\{\partial s|s\in U_2\}$, then $\tau(m_1)=\tau(m_2)$.
	 	\item If we further have $\tau(m|_\alpha)=0$, then $T(m_1)=T(m_2)$.
	 \end{itemize}

\end{theorem}
\begin{proof}
	We only prove the case for $T(m)$, and for $\tau(m)$ it is similar. We define a group $E'$ generated by $\{(s,a)|s\in V,a\in A\}$. Using Theorem~\ref{thmReplacable2} and Lemma~\ref{lemmaLatticeIsomorphic}, it is sufficient to prove $E_{\operatorname{inv}}(m_1)\cap E'=E_{\operatorname{inv}}(m_2)\cap E'$ and $E_{\operatorname{id}}(m_1)\cap E'=E_{\operatorname{id}}(m_2)\cap E'$. Because of Definition~\ref{defStatisticsMain} and the symmetry between $m_1,m_2$, proving $E_{\operatorname{id}}(m_1)\cap E'\subset E_{\operatorname{id}}(m_2)\cap E'$ is enough. 
	
	Let $e\in E_{\operatorname{id}}(m_1)\cap E'$, then $e$ is the sum of terms $([s_n,\cdots,[s_2,s_1]],a)$, $s_i\in V\sqcup U_1$, $\cap_i\supp(s_i)=\emptyset$. By assumption, for any $u\in U_1$, there exists $g_u\in \operatorname{F}(U_2)$ such that $\partial g_u=\partial u$; thus, we construct $e'$ by substituting every $s_i\in U_1$ in $([s_n,\cdots,[s_2,s_1]],a)$ by $g_{s_i}$. On one hand, since $e\in E'$, this replacement is actually doing nothing, i.e., $e=e'$;
	on the other hand, we see $e'\in E_{\operatorname{id}}(m_2)$ by expanding $g_u$ in the sense of Eq.~\eqref{eqBilinearCommutator}. Therefore, $E_{\operatorname{id}}(m_1)\cap E'\subset E_{\operatorname{id}}(m_2)\cap E'$.

\end{proof}

	In these theorems, the local triviality of statistics such as $\tau(m|_s)=0,\forall s\in S$ turns out to be crucial, but this condition is not satisfied by $m_p(X,G)$ for any simplicial complex $X$, and this failure indicates that we should only focus on manifolds. Before we prove it in the next section, we give a simple counterexample when $X$ is not a manifold.

\begin{example}\label{expNotManifold}
	Consider $\ZZ_2$ particles in the graph below. Using the computer program, we compute $\tau(m)$ and $\tau(m|_\alpha)$ for every simplex. The only nontrivial group is $\tau(m|_{\overline{12}})=\ZZ_2$, generated by $(\overline{3}+\overline{4})-(0)$.  In this case, Theorem~\ref{thmG0independence} no longer holds: when choosing $G_0=\{1\}$, we have $T=\ZZ_4$, and the generator exactly corresponds to the T-junction process; when choosing $G_0=\{0,1\}$, we have $T=\ZZ_4\oplus \ZZ_2$. Theorem~\ref{thmReplacable1} and \ref{thmReplacable2} also do not work: $s_{\overline{12}}$ must appear in any expression of the order-$4$ statistics. This is in contrast to the case in Section~\ref{subsecSymmetry}, where we can construct a process without $s_{\overline{12}}$.

	~~~~~~~~~~~~~~~~\begin{tikzpicture}
		\centering
		\draw[thick] (0,0) -- (0,2) -- (1.732,1) -- cycle; 
		\draw[thick] (0,0) -- (0,2) -- (-1.732,1) -- cycle;
		\fill[red] (0,0) circle (2pt);
		\fill[red] (0,2) circle (2pt);
		\fill[red] (1.732,1) circle (2pt);
		\fill[red] (-1.732,1) circle (2pt);
		
		\node[red, right] at (0,0) {1};
		\node[red, right] at (0,2) {2};
		\node[red, right] at (1.732,1) {3};
		\node[red, right] at (-1.732,1) {4};
	\end{tikzpicture}

\end{example}

\subsection{Topological excitations and manifold structure}

In the previous section, equations of the form $\tau(m|_s)=0$ appear in the assumptions of almost every theorem, so we want to find some general conditions. Equivalently, we want to answer why in some situations $T(m)$ or $\tau(m)$ can be nontrivial. The physical intuition is that nontrivial statistics occurs when excitations are "topological". We have not found a very convenient definition for this notion, but at least intuitively, the excitation pattern $m_p(X,G)$ is "topological" because of the boundary map $\partial: G_0\times X_{p+1}\rightarrow B_p(X,G)$. 

In contrast, we may define a non-topological version as follows.

\begin{definition}
	We construct the excitation pattern $\tilde{m}_p(X,G)$ such that
	\begin{itemize}
		\item $A=C_{p+1}(X,G)$ is the $(p+1)$-chain group of simplicial complex $X$.
		\item $\partial:G_0\times X_{p+1} \rightarrow C_{p+1}(X,G)$ is the natural map.
	\end{itemize}
	Similarly, we modify Definition~\ref{expDualCell} to obtain
	$\tilde{m}^q(X,G)$ such that
	\begin{itemize}
		\item $A=C^{q-1}(X,G)$.
		\item $\partial:G_0\times X_{q-1} \rightarrow C^{q-1}(X,G)$ is the natural map.
	\end{itemize}
\end{definition}

 We do not yet have a general definition for non-topological excitation patterns, but the following definition and theorem capture some of the relevant intuition.
 
\begin{definition}\label{constructionSum}
	Let $m_1=(A_1,S_1,\partial_1,\supp_1)$, $m_2=(A_2,S_2,\partial_2,\supp_2)$ be two excitation patterns in $M$. $m_1\times m_2=(A,S,\partial,\supp)$ is defined by
	\begin{itemize}
		\item $A=A_1\oplus A_2$, $S=S_1\sqcup S_2$.
		\item For $s\in S_i$, $\partial s=\partial_is$ and $\supp(s)=\supp_i(s)$.
	\end{itemize}
	Note that the support of excitation operators in $S_1,S_2$ may have nonempty intersection.
\end{definition}

\begin{theorem}\label{thmDirectSum}
	\textbf{Non-topological excitations have trivial statistics.}
	
	Let $N_i,i\in\{1,\cdots,n\}$ be subspaces of $M$, $m_i=(A_i,S_i,\partial_i,\supp_i),i\in\{1,\cdots,n\}$ be excitation patterns satisfying $\supp_i(s_i)=N_i,\forall i$, and $m=m_1\times \cdots\times m_n$. Then we have $T(m)=\tau(m)=0$.
\end{theorem}
\begin{proof}
	We prove the theorem by induction on $n$. For $n=1$ the statement is obvious. Assuming the theorem is true for $m=m_1\times\cdots\times m_{n-1}$, we now prove the theorem for $m'=m\times m_{n}$.
	
	We first prove $\tau(m')=0$. For any element in $\tau(m')$, we use Theorem~\ref{thmReplacable2} to construct a representative $d\in D_{\operatorname{inv}}(m')\cap D(m)$. We find $d\in D_{\operatorname{inv}}(m)$ by checking Eq.~(\ref{eqDinv}) directly. By the induction hypothesis, $d\in D_{\operatorname{id}}(m)$. Therefore, we have $d\in D_{\operatorname{id}}(m')$ and $\tau(m')=0$.
	
	Secondly, we prove $T(m')=0$. For any element in $T(m')$, we use Theorem~\ref{thmReplacable2} to construct a representative $e\in E_{\operatorname{inv}}(m')\cap E(m)$. We find $e\in E_{\operatorname{inv}}(m)$ by checking Eq.~(\ref{eqEinv}) directly. By the induction hypothesis, $e\in E_{\operatorname{id}}(m)$. Therefore, we have $e\in E_{\operatorname{id}}(m')$ and $T(m')=0$.
\end{proof}

Thus, we have $T(\tilde{m}_p(X,G))=0$. Recall that the only difference between $\tilde{m}_p(X,G)$ and $m_p(X,G)$ is that their configuration groups are different, which corresponds to the boundary map $\partial:C_{p+1}(X,G)\rightarrow B_p(X,G)$. In the sense of Definition~\ref{defCondensation}, $m_p(X,G)$ is obtained from $\tilde{m}_p(X,G)$ by condensing all closed excitations:
\begin{equation}
	m_p(X,G)=\tilde{m}_p(X,G)/Z_{p+1}(X,G).
\end{equation}
When $p=0$, this gives the familiar picture of string-net condensation \cite{Levin_2005}: when closed strings are condensed, vertices of open strings become anyons and can have nontrivial statistics.

In general, every excitation pattern $m=(A,S,\partial,\supp)$ can be obtained by condensing a non-topological excitation pattern. For example, we can construct $m_{\operatorname{free}}=( \ZZ[S],S,\partial_0,\supp)$ by taking $\ZZ[S]=\oplus_{s\in S}\ZZ s$ and $\partial_0$ maps $s$ to the corresponding basis element of $\ZZ[S]$.

\begin{remark}\label{remarkAnotherDefOfEid}
	$E_\inv(m)$ has a good characterization from Theorem~\ref{thmCheckByHand}, 
	while generating $E_\id(m)$ by higher commutators seems weird. We can give $E_\id(m)$ a better characterization as follows. We write $m$ as the condensation from a non-topological pattern $\tilde{m}$, such as $m_{\operatorname{free}}$. For $m=m^q(X,G)$, we may choose $\tilde{m}=\tilde{m}^q(X,G)$, and then
	\begin{equation}
		m=\tilde{m}/Z^{q-1}(X,G).
	\end{equation}
	Here, the condensation $\pi: C^{q-1}(X,G)\rightarrow B^q(X,G)$ is exactly the differential. Because $T(m)=0$, we have $E_\id(\tilde{m})=E_\inv(\tilde{m})$. Also, we have $E_\id(m)=\pi_*[E_\id(\tilde{m})]$ which directly follows the definition of $E_\id$. Thus, we have
	Thus, we have
	\begin{equation}
		E_{\operatorname{id}}(m)=\pi_*[E_{\operatorname{inv}}(\tilde{m})].
	\end{equation}
	Similarly, we have
	\begin{equation}
		D_{\operatorname{id}}(m)=\pi_*[D_{\operatorname{inv}}(\tilde{m})].
	\end{equation}
\end{remark}

\begin{theorem}\label{thmFiniteOrder}
	\textbf{Statistics is discrete.}
	
	Let $m=(A,S,\partial,\supp)$ be an excitation pattern that $A,S$ are finite. Then $T(m)=T_f(m)$.
\end{theorem}
\begin{proof}
	Let $e\in E(m)$. $[e]\in T_f(m)$ means $\exists n>1$ such that $ne\in E_{\operatorname{id}}$, so $ne$ satisfies the conditions of Theorem~\ref{thmCheckByHand}. Because these conditions are "divisable", $e$ also satisfies them, so $e\in E_{\operatorname{inv}}(m)$ and $T_f(m)\subset T(m)$.
	
	The proof of the inverse direction $T(m)\subset T_f(m)$ involves a construction similar to $m_{\operatorname{free}}$. For any $s\in S$, let $n(s)\in \ZZ_{>0}$ be the order of $\partial s$, and we construct another excitation pattern $m_0=(A_0,S,\partial_0,\supp)$, where $A_0=\oplus_{s\in S}\ZZ_{n(s)}$, and $\partial_0$ maps $s$ to the corresponding generator of $\ZZ_{n(s)}$.  There is a natural map $q:A_0\rightarrow A$ that $q(\sum_ic_is_i)=\sum_ic_i\partial s_i$, and the corresponding map $f:E(m_0)\rightarrow E(m)$ is $(s,a_0)\mapsto(s,q(a_0))$. Conversely, we define a map $g: E(m)\rightarrow E(m_0)$ by
	\begin{equation}
		g(s,a)=\sum_{a_0\in A_0,q(a_0)=a}(s,a_0).
	\end{equation}
	 Both $f,g$ map $E_{\operatorname{id}}$ to $E_{\operatorname{id}}$ and $E_{\operatorname{inv}}$ to $E_{\operatorname{inv}}$. The critical observation is that $f\circ g=N\operatorname{id}: E(m)\rightarrow E(m)$ where $N=\frac{|A_0|}{|A|}$ is the number of elements of $\operatorname{ker}q$. If $e\in E_{\operatorname{inv}}(m)$, then $g(e)\in E_{\operatorname{inv}}(m_0)$, so $g(e)\in E_{\operatorname{id}}(m_0)$ by Theorem~\ref{thmDirectSum}. Finally, $Ne=f(g(e))\in E_{\operatorname{id}}(m)$.
	
\end{proof}

The next theorem is a more general interpretation that non-topological excitation patterns have trivial statistics. The idea can be illustrated in the following example: consider the excitation pattern \hbox{\raisebox{-2ex}{\begin{tikzpicture}
			\draw[thick] (0,0) -- (2,0); 
			
			\foreach \x in {(0,0), (2,0)} \fill[red] \x circle (2pt);
			\node[red, below] at (0,0) {$s_1$};
			\node[red, below] at (2,0) {$s_2$};
			\node[black, below] at (1,0) {$s_3$};
\end{tikzpicture}}}
where $A=\ZZ_2\oplus\ZZ_2$, $\partial s_1=(1,0)$, $\partial s_2=(0,1)$, and $\partial s_3=(1,1)$. With only $s_1$ and $s_2$, the statistics is obviously trivial; we want to show that after introducing $s_3$, the statistics is still trivial.

\begin{theorem}\label{thmTrivialStatisticsInduction}
	\textbf{Statistics is trivial if excitations can be annihilated locally.}
	
	Let $m=(A,S=V\sqcup U,\partial,\supp)$ satisfying $\forall s\in U$, $\supp(s)=\alpha$ and $A_U\cap A_V\subset \operatorname{span}\{\partial s|s\in V,\supp(s)\subset \alpha\}$. We write $n=(A_V,V,\partial,\supp)$.
	\begin{enumerate}
		\item If $\tau(n|_\alpha)=\tau(n)=0$, then $\tau(m)=0$.
		\item If $\tau(n|_\alpha)=T(n)=0$, then $T(m)=0$.
	\end{enumerate}
\end{theorem}
\begin{proof}
	The proof is a slight modification to the last theorem, so we only present the proof for $T(m)=0$.
	
	For any element in $T(m)$, we use Theorem~\ref{thmReplacable2} to construct a representative $e\in E_{\operatorname{inv}}(m)\cap E(n)$. Now, we compare the difference of Eq.~(\ref{eqEinv}) between $E_{\operatorname{inv}}(m)$ and $E_{\operatorname{inv}}(n)$. For $x\in \alpha$, Eq.~(\ref{eqEinv}) is the same for $m$ and $n$ automatically. For $x\notin \alpha$, in the Eq.~(\ref{eqEinv}) for $E_{\operatorname{inv}}(m)$, the additional effect of $q_{x*}$ is that $(s,a)-(s,a')$ should be mapped to $0$ if $a-a'\in A_U$. However, since $a,a'\in A_V$, we have $a-a'\in A_U\cap A_V$, which is already mapped to zero in the Eq.~(\ref{eqEinv}) for $E_{\operatorname{inv}}(n)$ because $A_U\cap A_V\subset \operatorname{span}\{\partial s|s\in V,\supp(s)\subset \alpha\}$. Therefore, we have $e\in E_{\operatorname{inv}}(n)=E_{\operatorname{id}}(n)$, so $e\in E_{\operatorname{id}}(m)$.
\end{proof}

Now, we come to the central theorem in this section, where we prove local statistics is trivial if $X$ is a combinatorial manifold (Definition~\ref{defCombinatorialManifold}).

\begin{theorem}\label{thmLocalTopologyCondition}
	\textbf{Statistics have good behavior in manifolds.}
	
	Let $m=m_{p_1}(X,G_1)\times\cdots\times m_{p_n}(X,G_n)$. If $X$ is a combinatorial manifold, then for any simplex $c\in X$, we have $T(m|_c)=\tau(m|_c)=0$.
\end{theorem}
\begin{proof}
	For notation simplicity, we assume $m=m_p(X,G)$, where the configuration group is $B_p(X,G)$ and the set of excitation operators is $G\times X_{p+1}$. For any $s\in S$, we always reserve the notation $\supp(s)$ for the corresponding $(p+1)$-simplex in $X$ (instead of $\supp(s)\cap c$). Also, we may implicitly view it as a special $(p+1)$-chain and do addition.
	
	 Let the $\Sigma$ be the set of vertices of $c$, then, for any $s\in S$, $\supp(s)\cap c$ corresponds to a subset of of $\Sigma$. We divide $S$ into different classes according to the intersection with $c$:
	\begin{equation}
		S=\bigsqcup_{\alpha\subset \Sigma}S_\alpha,
	\end{equation}
	where
	\begin{equation}
		S_\alpha=\{s\in S|\supp(s)\cap c=\alpha\}.
	\end{equation}
	
	These subsets of $\Sigma$ have a partial order $\subset $ corresponding to the inclusion. We choose an arbitrary total order $\le$ on these subsets comparable with the inclusion. In other words, we write subsets of $\Sigma$ in a sequence:
	\begin{equation}
		\emptyset=\alpha_1<\alpha_2<\alpha_3\cdots<\alpha_N=\sigma,
	\end{equation}
	such that
	\begin{equation}
		\alpha_i\subset \alpha_j\implies i\le j.
	\end{equation}
	For example, if $\Sigma=\{1,2,3\}$, one can choose these $\alpha_i$ to be
	\begin{equation}
		\emptyset<\{1\}<\{2\}<\{3\}<\{1,2\}<\{1,3\},\{2,3\}<\{1,2,3\}.
	\end{equation}

	To prove $T(m|_c)=\tau(m|_c)=0$, we maintain a variable excitation pattern
	\begin{equation}
		n=(A,S,\partial,\supp).
	\end{equation}
	Initially, we take $A=0$ and $S=\emptyset$; in the $i$-th step, we enlarge $S$ by adding $S_{\alpha_i}$ to $S$, and replace $A$ by $A+\operatorname{span}\{\partial s|s\in S_{\alpha_i}\}$; finally, we add all $S_{\alpha_N}$ to $n$ and arrive at $n=m|_c$. We will use Theorem~\ref{thmTrivialStatisticsInduction} to prove that $T(n|_c)=\tau(n|_c)=0$ remains true in every step.
	
	In the first step, we only add excitation operators whose supports do not intersect with $c$. Then $T(n|_c)=\tau(n|_c)=0$ by Theorem~\ref{thmQuotientEmptySupport}.
	
	Now, assume $n=(A,S,\partial,\supp)$ satisfies $T(n|_c)=\tau(n|_c)=0$ after the $(k-1)$-th step. In the $k$-th step, we have (we write $\alpha_k$=$\alpha$ for short, and both $A,A_{S_{\alpha}}$ are subgroups of the configuration group of $m$)
	\begin{equation}
		n'=(A+A_{S_{\alpha}},S\sqcup S_\alpha,\partial,\supp),
	\end{equation}
	and we want to prove $T(n'|_c)=\tau(n'|_c)=0$.
	In order to use Theorem~\ref{thmTrivialStatisticsInduction}, we should prove 
	\begin{equation}
		A_{S_\alpha}\cap A\subset \operatorname{span}\{\partial s|s\in S,\supp(s)\cap c\subset \alpha\}.
	\end{equation}
	More concretely, if $g_1,\cdots,g_h\in G$ and $s_1,\cdots,s_h\in S_\alpha$ satisfies 
	$\lambda=\partial\sum g_is_i\in A$, then by definition $\lambda$ is generated by all $S_{\alpha_j}$ such that $\alpha_j<\alpha$; we want to prove that $\lambda$ is generated by all $S_{\alpha_j}$ such that $\alpha_j\subset \alpha$, which is a stronger condition.
	
	In this step, we need to seriously use the combinatorial topology of simplicial complex, reviewed in Appendix~\ref{appendixSimplicialComplex}.
	We write $c$ as the join $\alpha*c_0$. 
	 We also write any $s_i$ as  $s_i=\alpha*\beta_i$, where $\beta_i$ is a group element in $G$ times a simplex in $\operatorname{Lk}(\alpha)$. Because $\supp(s_i)\cap c=\alpha$, the underlying simplex of $\beta_i$ does not intersect with $c_0$. Using Eq.~\eqref{eqBoundaryJoin}, we have
	 
	 \begin{equation}
	 	\begin{aligned}
	 		\lambda&=\partial(\sum g_j s_j)\\&=\partial(\sum g_j\alpha*\beta_j)\\&=(\partial\alpha)*\sum g_j\beta_j+(-1)^{|\alpha|-1}\alpha*\sum g_j(\partial\beta_j).
	 	\end{aligned}
	 \end{equation}
	 
	 Because $\lambda$ is generated by all $S_{\alpha_j}$ such that $\alpha_j<\alpha$, the second term must be zero. So, we have
	 \begin{equation}
	 	\lambda=(\partial\alpha)*\sum g_j\beta_j
	 \end{equation}
	 and 
	 \begin{equation}
	 	\partial(\sum g_j\beta_j)=0.
	 \end{equation}
	 Because $X$ is a combinatorial manifold (Definition~\ref{defCombinatorialManifold}), $\operatorname{Lk}(\alpha)$ has the topology of a sphere, so $\operatorname{Lk}(\alpha)-c_0$ is contractible. Because $\sum g_j\beta_j$ is a closed chain in $\operatorname{Lk}(\alpha)-c_0$, it is also a boundary chain in $\operatorname{Lk}(\alpha)-c_0$:
	 \begin{equation}
	 	\sum g_j\beta_j=\partial \gamma.
	 \end{equation}
	 Thus, we have
	 \begin{equation}
	 	\lambda=(-1)^{|\alpha|}\partial (\partial\alpha*\gamma),
	 \end{equation}
	and $(-1)^{|\alpha|}(\partial\alpha)*\gamma$ is generated by $S_{\alpha_j}$ such that $\alpha_j\subset \alpha$.
\end{proof}

\begin{remark}
	The induction in the proof fails when $X$ is not a manifold. In Example~\ref{expNotManifold}, we choose $m=m_p(X,\ZZ_2)$, where $X$ is the graph
	\begin{tikzpicture}
		\centering
		\draw[thick] (0,0) -- (0,2) -- (1.732,1) -- cycle; 
		\draw[thick] (0,0) -- (0,2) -- (-1.732,1) -- cycle;
		\draw[blue, ultra thick] (0,0) -- node[right]{$c$} (0,2);
		\fill[red] (0,0) circle (2pt);
		\fill[red] (0,2) circle (2pt);
		\fill[red] (1.732,1) circle (2pt);
		\fill[red] (-1.732,1) circle (2pt);
		
		\node[red, right] at (0,0) {1};
		\node[red, right] at (0,2) {2};
		\node[red, right] at (1.732,1) {3};
		\node[red, right] at (-1.732,1) {4};
	\end{tikzpicture}
	and we have $\tau(m|_c)=\ZZ_2$. Similar to the previous proof, we have $S_\emptyset=\emptyset,\;S_{\{\overline{1}\}}=\{s_{\overline{13}},s_{\overline{14}}\},\;S_{\overline{2}}=\{s_{\overline{13}},s_{\overline{14}}\},$ and $S_{\{\overline{1},\overline{2}\}}=\{s_{\overline{12}}\}$. When we add $S_{\overline{2}}$ to $S=S_\emptyset\sqcup S_{\{\overline{1}\}}$, we have 
	\begin{equation}
		\partial s_{\overline{23}}+\partial s_{\overline{24}}=\overline{3}+\overline{4}. 
	\end{equation}
	This element can be generated by $S_\emptyset\sqcup S_{\{\overline{1}\}}$ as $\partial s_{\overline{13}}+\partial s_{\overline{14}}$, but touching the vertex $\overline{1}$ is inevitable. $\operatorname{Lk}(\overline{2})$ is generated by $\overline{13}$ and $\overline{14}$; it has the topology of an edge instead of a circle, and removing the vertex $\overline{1}$ makes it disconnected. In contrast, if one introduce an additional edge linking vertices $\overline{3}$ and $\overline{4}$ as in the T-junction process, $\operatorname{Lk}(\overline{2})$ becomes a circle and $\overline{3}+\overline{4}$ is generated by $s_{\overline{34}}\in S_\emptyset$.
\end{remark}

Finally, we give a theorem to explain why all membrane operators in Fig.~\ref{fig: 24 step process} are of the form $U_{0ij}$.

\begin{theorem}\label{thmStar}\textbf{In any statistical process, we can let all operators intersect at a point.}
	
	For the excitation pattern $m=m_p(\partial \Delta^{d+1},G)$, we define $V(\overline{0})=\{s\in S|\overline{0}\in \supp(s)\}$. For any $t\in T(m_p(\partial\Delta^{d+1},G))$, there exists $g\in \operatorname{F}(V(\overline{0}))$ such that $t=[(g,0)]$.
\end{theorem}
\begin{proof}
	We first choose a representative $e$ of $t$, and then annihilate $(s,a)$ terms for all $s\notin V(\overline{0})$. Fixing $s\notin V(\overline{0})$ and then using Lemma~\ref{lemmaBoundary} and Theorem~\ref{thmLocalTopologyCondition}, we find that $\sigma_s(e)\in D_{\operatorname{id}}(m|_s)$. We write $\sigma_s(e)$ as the sum of terms like $([s_p,\cdots,[s_1,\sigma]],a)$ with $(\bigcap_{i=1}^p \supp(s_i))\cap\supp(s)=\emptyset$. Now, for any $s_i\notin V(\overline{0})$, we consider the cone that the base and the apex are $\supp(s)$ and $\overline{0}$, respectively. It is obvious that $\partial s_i$ can be generated by excitation operators supported on the lateral surface of the cone. In other words, there is a process $g_i$ such that $\partial g_i=\partial s_i$ and only the excitation operators supported at the lateral surface are involved. In $([s_p,\cdots,[s_1,\sigma]],a)$, we replace $\sigma$ by $s$ and such $s_i$ by $g_i$, and the resulting expression is in $E_{\operatorname{id}}(m)$. Doing similarly for all terms of $\sigma_s(e)$ and subtracting the resulting expressions from $e$, we cancel all $(s,a)$ terms, while terms correspond to other operators in $V(\overline{0})^c$ are not involved. Therefore, we do the same construction for all $s\in V(\overline{0})^c$, and in the end, we get a representative $e'$ of $t$ that only involves operators in $V(\overline{0})$. Since $V(\overline{0})$ generates $A$, there exists $g\in \operatorname{F}(V(\overline{0}))$ such that $e'=(g,0)$.
\end{proof}

\section{Future Directions}\label{secFuture}

Topological excitations, and maybe also generalized symmetries, are interesting topics in theoretical physics. This work is rather independent of other studies and has its own advantages and disadvantages. On one hand, our theory is rigorous and self-contained, building a microscopic theory of Abelian excitations only from several axioms. On the other hand, there are many physical notions we do not include in our scenario. First, our theory does not describe non-Abelian and non-invertible excitations, which limits the applicability to describe three-loop braiding statistics \cite{Wang2014Braiding}; see Remark~\ref{remarkMixDimension} and \ref{remarkThreeLoop}. Second, there is a charge-decoration argument in \cite{FHH21} does not fit into our framework, and we still do not have a good understanding. 
Third, we do not discuss generalized symmetries, SPT phases, anomalies, and boundary-bulk relations; we have introduced quantum cellular automata and condensation but only as technical tools to study the strong independence. We feel that it is worth exploring these notions using our axiom system.

Our paper has an important unsolved problem, Conjecture~\ref{conjectureDef}, which claims that the statistics does not depend on the triangulation. If it is proved, then we will have a well-defined statistics \(T_p(M, G)\) for any manifold $M$, and then our theory will be more physically meaningful.
We believe the proof requires deeper tools in combinatorial topology and a more complete theory of condensation. Also, we mainly focus on simplicial complex, but their Poincar\'e duals are only cell complexes in general, so we may also need some stronger theorems.

Admitting this conjecture, it is also worth exploring how $T_p(M,G)$ is different from $T_p(S^d,G)\simeq H_{d+2}(K(G,d-p),\ZZ)$. We have found from computation that $T_p(M,G)$ can have more statistics and can also have less statistics. In our current understanding, the reason for $T_p(M,G)\subset T_p(S^d,G)$ is that nontrivial global topology lifts the ground-state degeneracy, so that our configuration axiom is violated. For example, we have computed the statistics of $\ZZ_2$ particles by taking $X$ as triangulations of torus, $\RR P^2$, and Klein bottle. The result is $T\simeq \ZZ_2$, corresponding to bosons and fermions. If a lattice model with a semion is placed in these surfaces, then the braiding of two semions along two non-contractible loops produce a phase factor $-1$, so the ground state degeneracy must $>1$, violating Definition~\ref{defRealization}.

Some other reasons may imply $T_p(S^d,G)\subset T_p(M,G)$. For example, the statistics of $\ZZ_2$-loop on $S^2$ is trivial, while the statistics is $\ZZ_2\times \ZZ_2$ on a torus. These statistics come from dimensional reduction: one can view a loop intertwined on the torus (two independent choice of directions) and behaves like a particle; then, the statistics actually corresponds to the particle fusion statistics.

We do not know the structure of $T_p(M,G)$ and whether these pictures exhaust all possibilities. It is important that our theory focus on a single manifold, completely different from TQFT methods, and we do not know how to link them sysmatically.

\section{Acknowledgements}
I thank Ryohei Kobayashi, Yuyang Li, Po-Shen Hsin, Yu-An Chen for their collaboration on the paper \cite{Previous}. I have very enlighening discussions with Yu-An Chen and Yitao Feng. Yuyang Li wrote the initial version of the computer program used in this paper. I also thank Ansi Bai, Xie Chen, Guochuan Chiang, Shenghan Jiang, Zibo Jin, Anton Kapustin, Liang Kong, Zhenpeng Li, Guchuan Li, Zijian Liang, Qingrui Wang, Chenjie Wang, Xiaogang Wen, Jiangnan Xiong, Hongjian Yang, Jinfei Zhou, and Tingyu Zhu for discussions. I especially thank Tingyu Zhu for numerous encouragements. Hanyu Xue is supported by the National Natural Science Foundation of China (Grant No.~12474491). Conflict of interest statement: NO; The data are avaliable from the authors upon reasonable request.

\bibliography{bibliography}

\begin{thebibliography}{49}%
\makeatletter
\providecommand \@ifxundefined [1]{%
 \@ifx{#1\undefined}
}%
\providecommand \@ifnum [1]{%
 \ifnum #1\expandafter \@firstoftwo
 \else \expandafter \@secondoftwo
 \fi
}%
\providecommand \@ifx [1]{%
 \ifx #1\expandafter \@firstoftwo
 \else \expandafter \@secondoftwo
 \fi
}%
\providecommand \natexlab [1]{#1}%
\providecommand \enquote  [1]{``#1''}%
\providecommand \bibnamefont  [1]{#1}%
\providecommand \bibfnamefont [1]{#1}%
\providecommand \citenamefont [1]{#1}%
\providecommand \href@noop [0]{\@secondoftwo}%
\providecommand \href [0]{\begingroup \@sanitize@url \@href}%
\providecommand \@href[1]{\@@startlink{#1}\@@href}%
\providecommand \@@href[1]{\endgroup#1\@@endlink}%
\providecommand \@sanitize@url [0]{\catcode `\\12\catcode `\$12\catcode
  `\&12\catcode `\#12\catcode `\^12\catcode `\_12\catcode `\%12\relax}%
\providecommand \@@startlink[1]{}%
\providecommand \@@endlink[0]{}%
\providecommand \url  [0]{\begingroup\@sanitize@url \@url }%
\providecommand \@url [1]{\endgroup\@href {#1}{\urlprefix }}%
\providecommand \urlprefix  [0]{URL }%
\providecommand \Eprint [0]{\href }%
\providecommand \doibase [0]{https://doi.org/}%
\providecommand \selectlanguage [0]{\@gobble}%
\providecommand \bibinfo  [0]{\@secondoftwo}%
\providecommand \bibfield  [0]{\@secondoftwo}%
\providecommand \translation [1]{[#1]}%
\providecommand \BibitemOpen [0]{}%
\providecommand \bibitemStop [0]{}%
\providecommand \bibitemNoStop [0]{.\EOS\space}%
\providecommand \EOS [0]{\spacefactor3000\relax}%
\providecommand \BibitemShut  [1]{\csname bibitem#1\endcsname}%
\let\auto@bib@innerbib\@empty
\bibitem [{\citenamefont {Zeng}\ \emph {et~al.}(2019)\citenamefont {Zeng},
  \citenamefont {Chen}, \citenamefont {Zhou},\ and\ \citenamefont
  {Wen}}]{zeng-2019}%
  \BibitemOpen
  \bibfield  {author} {\bibinfo {author} {\bibfnamefont {B.}~\bibnamefont
  {Zeng}}, \bibinfo {author} {\bibfnamefont {X.}~\bibnamefont {Chen}}, \bibinfo
  {author} {\bibfnamefont {D.-L.}\ \bibnamefont {Zhou}},\ and\ \bibinfo
  {author} {\bibfnamefont {X.-G.}\ \bibnamefont {Wen}},\ }\href
  {https://doi.org/10.1007/978-1-4939-9084-9} {\emph {\bibinfo {title}
  {{Quantum information meets quantum matter}}}}\ (\bibinfo {year}
  {2019})\BibitemShut {NoStop}%
\bibitem [{\citenamefont {Wen}(2015)}]{Wen_2015}%
  \BibitemOpen
  \bibfield  {author} {\bibinfo {author} {\bibfnamefont {X.-G.}\ \bibnamefont
  {Wen}},\ }\bibfield  {title} {\bibinfo {title} {A theory of 2+1d bosonic
  topological orders},\ }\href {https://doi.org/10.1093/nsr/nwv077} {\bibfield
  {journal} {\bibinfo  {journal} {National Science Review}\ }\textbf {\bibinfo
  {volume} {3}},\ \bibinfo {pages} {68–106} (\bibinfo {year}
  {2015})}\BibitemShut {NoStop}%
\bibitem [{\citenamefont {Kong}\ and\ \citenamefont
  {Wen}(2014)}]{Kong:2014qka}%
  \BibitemOpen
  \bibfield  {author} {\bibinfo {author} {\bibfnamefont {L.}~\bibnamefont
  {Kong}}\ and\ \bibinfo {author} {\bibfnamefont {X.-G.}\ \bibnamefont {Wen}},\
  }\bibfield  {title} {\bibinfo {title} {{Braided fusion categories,
  gravitational anomalies, and the mathematical framework for topological
  orders in any dimensions}},\ }\href@noop {} {\  (\bibinfo {year} {2014})},\
  \Eprint {https://arxiv.org/abs/1405.5858} {arXiv:1405.5858 [cond-mat.str-el]}
  \BibitemShut {NoStop}%
\bibitem [{\citenamefont {Kong}\ \emph {et~al.}(2024)\citenamefont {Kong},
  \citenamefont {Zhang}, \citenamefont {Zhao},\ and\ \citenamefont
  {Zheng}}]{kong2024highercondensationtheory}%
  \BibitemOpen
  \bibfield  {author} {\bibinfo {author} {\bibfnamefont {L.}~\bibnamefont
  {Kong}}, \bibinfo {author} {\bibfnamefont {Z.-H.}\ \bibnamefont {Zhang}},
  \bibinfo {author} {\bibfnamefont {J.}~\bibnamefont {Zhao}},\ and\ \bibinfo
  {author} {\bibfnamefont {H.}~\bibnamefont {Zheng}},\ }\href
  {https://arxiv.org/abs/2403.07813} {\bibinfo {title} {Higher condensation
  theory}} (\bibinfo {year} {2024}),\ \Eprint
  {https://arxiv.org/abs/2403.07813} {arXiv:2403.07813 [cond-mat.str-el]}
  \BibitemShut {NoStop}%
\bibitem [{\citenamefont {Kapustin}\ and\ \citenamefont
  {Sopenko}(2022)}]{Kapustin_2022}%
  \BibitemOpen
  \bibfield  {author} {\bibinfo {author} {\bibfnamefont {A.}~\bibnamefont
  {Kapustin}}\ and\ \bibinfo {author} {\bibfnamefont {N.}~\bibnamefont
  {Sopenko}},\ }\bibfield  {title} {\bibinfo {title} {Local noether theorem for
  quantum lattice systems and topological invariants of gapped states},\
  }\bibfield  {journal} {\bibinfo  {journal} {Journal of Mathematical Physics}\
  }\textbf {\bibinfo {volume} {63}},\ \href {https://doi.org/10.1063/5.0085964}
  {10.1063/5.0085964} (\bibinfo {year} {2022})\BibitemShut {NoStop}%
\bibitem [{\citenamefont {NAAIJKENS}(2011)}]{NAAIJKENS_2011}%
  \BibitemOpen
  \bibfield  {author} {\bibinfo {author} {\bibfnamefont {P.}~\bibnamefont
  {NAAIJKENS}},\ }\bibfield  {title} {\bibinfo {title} {Localized endomorphisms
  in kitaev’s toric code on the plane},\ }\href
  {https://doi.org/10.1142/s0129055x1100431x} {\bibfield  {journal} {\bibinfo
  {journal} {Reviews in Mathematical Physics}\ }\textbf {\bibinfo {volume}
  {23}},\ \bibinfo {pages} {347–373} (\bibinfo {year} {2011})}\BibitemShut
  {NoStop}%
\bibitem [{\citenamefont {Kitaev}(2003{\natexlab{a}})}]{Kitaev:1997wr}%
  \BibitemOpen
  \bibfield  {author} {\bibinfo {author} {\bibfnamefont {A.~Y.}\ \bibnamefont
  {Kitaev}},\ }\bibfield  {title} {\bibinfo {title} {{Fault tolerant quantum
  computation by anyons}},\ }\href
  {https://doi.org/10.1016/S0003-4916(02)00018-0} {\bibfield  {journal}
  {\bibinfo  {journal} {Annals Phys.}\ }\textbf {\bibinfo {volume} {303}},\
  \bibinfo {pages} {2} (\bibinfo {year} {2003}{\natexlab{a}})},\ \Eprint
  {https://arxiv.org/abs/quant-ph/9707021} {arXiv:quant-ph/9707021}
  \BibitemShut {NoStop}%
\bibitem [{\citenamefont {Levin}\ and\ \citenamefont {Wen}(2005)}]{Levin_2005}%
  \BibitemOpen
  \bibfield  {author} {\bibinfo {author} {\bibfnamefont {M.~A.}\ \bibnamefont
  {Levin}}\ and\ \bibinfo {author} {\bibfnamefont {X.-G.}\ \bibnamefont
  {Wen}},\ }\bibfield  {title} {\bibinfo {title} {{String-net condensation: A
  physical mechanism for topological phases}},\ }\bibfield  {journal} {\bibinfo
   {journal} {Physical Review B}\ }\textbf {\bibinfo {volume} {71}},\ \href
  {https://doi.org/10.1103/physrevb.71.045110} {10.1103/physrevb.71.045110}
  (\bibinfo {year} {2005})\BibitemShut {NoStop}%
\bibitem [{\citenamefont {Hahn}\ and\ \citenamefont
  {Wolf}(2020)}]{generalizedStringNet}%
  \BibitemOpen
  \bibfield  {author} {\bibinfo {author} {\bibfnamefont {A.}~\bibnamefont
  {Hahn}}\ and\ \bibinfo {author} {\bibfnamefont {R.}~\bibnamefont {Wolf}},\
  }\bibfield  {title} {\bibinfo {title} {Generalized string-net model for
  unitary fusion categories without tetrahedral symmetry},\ }\href
  {https://doi.org/10.1103/PhysRevB.102.115154} {\bibfield  {journal} {\bibinfo
   {journal} {Phys. Rev. B}\ }\textbf {\bibinfo {volume} {102}},\ \bibinfo
  {pages} {115154} (\bibinfo {year} {2020})}\BibitemShut {NoStop}%
\bibitem [{\citenamefont {Walker}\ and\ \citenamefont
  {Wang}(2011)}]{walker201131tqftstopologicalinsulators}%
  \BibitemOpen
  \bibfield  {author} {\bibinfo {author} {\bibfnamefont {K.}~\bibnamefont
  {Walker}}\ and\ \bibinfo {author} {\bibfnamefont {Z.}~\bibnamefont {Wang}},\
  }\href {https://arxiv.org/abs/1104.2632} {\bibinfo {title} {(3+1)-tqfts and
  topological insulators}} (\bibinfo {year} {2011}),\ \Eprint
  {https://arxiv.org/abs/1104.2632} {arXiv:1104.2632 [cond-mat.str-el]}
  \BibitemShut {NoStop}%
\bibitem [{\citenamefont {Haah}(2013)}]{HaahThesis}%
  \BibitemOpen
  \bibfield  {author} {\bibinfo {author} {\bibfnamefont {J.}~\bibnamefont
  {Haah}},\ }\href {https://arxiv.org/abs/1305.6973} {\bibinfo {title} {Lattice
  quantum codes and exotic topological phases of matter}} (\bibinfo {year}
  {2013}),\ \Eprint {https://arxiv.org/abs/1305.6973} {arXiv:1305.6973
  [quant-ph]} \BibitemShut {NoStop}%
\bibitem [{\citenamefont {Liang}\ \emph {et~al.}(2024)\citenamefont {Liang},
  \citenamefont {Xu}, \citenamefont {Iosue},\ and\ \citenamefont
  {Chen}}]{Zijian1}%
  \BibitemOpen
  \bibfield  {author} {\bibinfo {author} {\bibfnamefont {Z.}~\bibnamefont
  {Liang}}, \bibinfo {author} {\bibfnamefont {Y.}~\bibnamefont {Xu}}, \bibinfo
  {author} {\bibfnamefont {J.~T.}\ \bibnamefont {Iosue}},\ and\ \bibinfo
  {author} {\bibfnamefont {Y.-A.}\ \bibnamefont {Chen}},\ }\bibfield  {title}
  {\bibinfo {title} {Extracting topological orders of generalized pauli
  stabilizer codes in two dimensions},\ }\href
  {https://doi.org/10.1103/PRXQuantum.5.030328} {\bibfield  {journal} {\bibinfo
   {journal} {PRX Quantum}\ }\textbf {\bibinfo {volume} {5}},\ \bibinfo {pages}
  {030328} (\bibinfo {year} {2024})}\BibitemShut {NoStop}%
\bibitem [{\citenamefont {Liang}\ \emph {et~al.}(2025)\citenamefont {Liang},
  \citenamefont {Yang}, \citenamefont {Iosue},\ and\ \citenamefont
  {Chen}}]{Zijian2}%
  \BibitemOpen
  \bibfield  {author} {\bibinfo {author} {\bibfnamefont {Z.}~\bibnamefont
  {Liang}}, \bibinfo {author} {\bibfnamefont {B.}~\bibnamefont {Yang}},
  \bibinfo {author} {\bibfnamefont {J.~T.}\ \bibnamefont {Iosue}},\ and\
  \bibinfo {author} {\bibfnamefont {Y.-A.}\ \bibnamefont {Chen}},\ }\href
  {https://arxiv.org/abs/2410.11942} {\bibinfo {title} {Operator algebra and
  algorithmic construction of boundaries and defects in (2+1)d topological
  pauli stabilizer codes}} (\bibinfo {year} {2025}),\ \Eprint
  {https://arxiv.org/abs/2410.11942} {arXiv:2410.11942 [quant-ph]} \BibitemShut
  {NoStop}%
\bibitem [{\citenamefont {Pretko}\ \emph {et~al.}(2020)\citenamefont {Pretko},
  \citenamefont {Chen},\ and\ \citenamefont {You}}]{Fractonphasesofmatter}%
  \BibitemOpen
  \bibfield  {author} {\bibinfo {author} {\bibfnamefont {M.}~\bibnamefont
  {Pretko}}, \bibinfo {author} {\bibfnamefont {X.}~\bibnamefont {Chen}},\ and\
  \bibinfo {author} {\bibfnamefont {Y.}~\bibnamefont {You}},\ }\bibfield
  {title} {\bibinfo {title} {Fracton phases of matter},\ }\href
  {https://doi.org/10.1142/s0217751x20300033} {\bibfield  {journal} {\bibinfo
  {journal} {International Journal of Modern Physics A}\ }\textbf {\bibinfo
  {volume} {35}},\ \bibinfo {pages} {2030003} (\bibinfo {year}
  {2020})}\BibitemShut {NoStop}%
\bibitem [{\citenamefont {Haah}(2011)}]{Haah_2011}%
  \BibitemOpen
  \bibfield  {author} {\bibinfo {author} {\bibfnamefont {J.}~\bibnamefont
  {Haah}},\ }\bibfield  {title} {\bibinfo {title} {Local stabilizer codes in
  three dimensions without string logical operators},\ }\bibfield  {journal}
  {\bibinfo  {journal} {Physical Review A}\ }\textbf {\bibinfo {volume} {83}},\
  \href {https://doi.org/10.1103/physreva.83.042330}
  {10.1103/physreva.83.042330} (\bibinfo {year} {2011})\BibitemShut {NoStop}%
\bibitem [{\citenamefont {Chen}\ \emph {et~al.}(2013)\citenamefont {Chen},
  \citenamefont {Gu}, \citenamefont {Liu},\ and\ \citenamefont
  {Wen}}]{Chen2013Symmetry}%
  \BibitemOpen
  \bibfield  {author} {\bibinfo {author} {\bibfnamefont {X.}~\bibnamefont
  {Chen}}, \bibinfo {author} {\bibfnamefont {Z.-C.}\ \bibnamefont {Gu}},
  \bibinfo {author} {\bibfnamefont {Z.-X.}\ \bibnamefont {Liu}},\ and\ \bibinfo
  {author} {\bibfnamefont {X.-G.}\ \bibnamefont {Wen}},\ }\bibfield  {title}
  {\bibinfo {title} {Symmetry protected topological orders and the group
  cohomology of their symmetry group},\ }\href
  {https://doi.org/10.1103/PhysRevB.87.155114} {\bibfield  {journal} {\bibinfo
  {journal} {Phys. Rev. B}\ }\textbf {\bibinfo {volume} {87}},\ \bibinfo
  {pages} {155114} (\bibinfo {year} {2013})}\BibitemShut {NoStop}%
\bibitem [{\citenamefont {Kapustin}\ and\ \citenamefont
  {Thorngren}(2013)}]{Kapustin:2013uxa}%
  \BibitemOpen
  \bibfield  {author} {\bibinfo {author} {\bibfnamefont {A.}~\bibnamefont
  {Kapustin}}\ and\ \bibinfo {author} {\bibfnamefont {R.}~\bibnamefont
  {Thorngren}},\ }\bibfield  {title} {\bibinfo {title} {{Higher symmetry and
  gapped phases of gauge theories}},\ }\href@noop {} {\  (\bibinfo {year}
  {2013})},\ \Eprint {https://arxiv.org/abs/1309.4721} {arXiv:1309.4721
  [hep-th]} \BibitemShut {NoStop}%
\bibitem [{\citenamefont {Feng}\ \emph
  {et~al.}(2025{\natexlab{a}})\citenamefont {Feng}, \citenamefont {Kobayashi},
  \citenamefont {Chen},\ and\ \citenamefont
  {Ryu}}]{feng2025higherformanomalieslattices}%
  \BibitemOpen
  \bibfield  {author} {\bibinfo {author} {\bibfnamefont {Y.}~\bibnamefont
  {Feng}}, \bibinfo {author} {\bibfnamefont {R.}~\bibnamefont {Kobayashi}},
  \bibinfo {author} {\bibfnamefont {Y.-A.}\ \bibnamefont {Chen}},\ and\
  \bibinfo {author} {\bibfnamefont {S.}~\bibnamefont {Ryu}},\ }\href
  {https://arxiv.org/abs/2509.12304} {\bibinfo {title} {Higher-form anomalies
  on lattices}} (\bibinfo {year} {2025}{\natexlab{a}}),\ \Eprint
  {https://arxiv.org/abs/2509.12304} {arXiv:2509.12304 [cond-mat.str-el]}
  \BibitemShut {NoStop}%
\bibitem [{\citenamefont {Laughlin}(1983)}]{Laughlin1983Anomalous}%
  \BibitemOpen
  \bibfield  {author} {\bibinfo {author} {\bibfnamefont {R.~B.}\ \bibnamefont
  {Laughlin}},\ }\bibfield  {title} {\bibinfo {title} {Anomalous quantum hall
  effect: An incompressible quantum fluid with fractionally charged
  excitations},\ }\href {https://doi.org/10.1103/PhysRevLett.50.1395}
  {\bibfield  {journal} {\bibinfo  {journal} {Phys. Rev. Lett.}\ }\textbf
  {\bibinfo {volume} {50}},\ \bibinfo {pages} {1395} (\bibinfo {year}
  {1983})}\BibitemShut {NoStop}%
\bibitem [{\citenamefont {Stormer}(1999)}]{RevModPhys.71.875}%
  \BibitemOpen
  \bibfield  {author} {\bibinfo {author} {\bibfnamefont {H.~L.}\ \bibnamefont
  {Stormer}},\ }\bibfield  {title} {\bibinfo {title} {Nobel lecture: The
  fractional quantum hall effect},\ }\href
  {https://doi.org/10.1103/RevModPhys.71.875} {\bibfield  {journal} {\bibinfo
  {journal} {Rev. Mod. Phys.}\ }\textbf {\bibinfo {volume} {71}},\ \bibinfo
  {pages} {875} (\bibinfo {year} {1999})}\BibitemShut {NoStop}%
\bibitem [{\citenamefont {Freedman}\ \emph {et~al.}(2003)\citenamefont
  {Freedman}, \citenamefont {Kitaev}, \citenamefont {Larsen},\ and\
  \citenamefont {Wang}}]{freedman2003topological}%
  \BibitemOpen
  \bibfield  {author} {\bibinfo {author} {\bibfnamefont {M.}~\bibnamefont
  {Freedman}}, \bibinfo {author} {\bibfnamefont {A.}~\bibnamefont {Kitaev}},
  \bibinfo {author} {\bibfnamefont {M.}~\bibnamefont {Larsen}},\ and\ \bibinfo
  {author} {\bibfnamefont {Z.}~\bibnamefont {Wang}},\ }\bibfield  {title}
  {\bibinfo {title} {Topological quantum computation},\ }\href@noop {}
  {\bibfield  {journal} {\bibinfo  {journal} {Bulletin of the American
  Mathematical Society}\ }\textbf {\bibinfo {volume} {40}},\ \bibinfo {pages}
  {31} (\bibinfo {year} {2003})}\BibitemShut {NoStop}%
\bibitem [{\citenamefont {Nayak}\ \emph {et~al.}(2008)\citenamefont {Nayak},
  \citenamefont {Simon}, \citenamefont {Stern}, \citenamefont {Freedman},\ and\
  \citenamefont {Das~Sarma}}]{Nayak2008NonAbelian}%
  \BibitemOpen
  \bibfield  {author} {\bibinfo {author} {\bibfnamefont {C.}~\bibnamefont
  {Nayak}}, \bibinfo {author} {\bibfnamefont {S.~H.}\ \bibnamefont {Simon}},
  \bibinfo {author} {\bibfnamefont {A.}~\bibnamefont {Stern}}, \bibinfo
  {author} {\bibfnamefont {M.}~\bibnamefont {Freedman}},\ and\ \bibinfo
  {author} {\bibfnamefont {S.}~\bibnamefont {Das~Sarma}},\ }\bibfield  {title}
  {\bibinfo {title} {Non-abelian anyons and topological quantum computation},\
  }\href {https://doi.org/10.1103/RevModPhys.80.1083} {\bibfield  {journal}
  {\bibinfo  {journal} {Rev. Mod. Phys.}\ }\textbf {\bibinfo {volume} {80}},\
  \bibinfo {pages} {1083} (\bibinfo {year} {2008})}\BibitemShut {NoStop}%
\bibitem [{\citenamefont {Kitaev}(2006)}]{Kitaev:2005hzj}%
  \BibitemOpen
  \bibfield  {author} {\bibinfo {author} {\bibfnamefont {A.}~\bibnamefont
  {Kitaev}},\ }\bibfield  {title} {\bibinfo {title} {{Anyons in an exactly
  solved model and beyond}},\ }\href
  {https://doi.org/10.1016/j.aop.2005.10.005} {\bibfield  {journal} {\bibinfo
  {journal} {Annals Phys.}\ }\textbf {\bibinfo {volume} {321}},\ \bibinfo
  {pages} {2} (\bibinfo {year} {2006})}\BibitemShut {NoStop}%
\bibitem [{\citenamefont {Kitaev}(2003{\natexlab{b}})}]{Kitaev_2003}%
  \BibitemOpen
  \bibfield  {author} {\bibinfo {author} {\bibfnamefont {A.}~\bibnamefont
  {Kitaev}},\ }\bibfield  {title} {\bibinfo {title} {Fault-tolerant quantum
  computation by anyons},\ }\href
  {https://doi.org/10.1016/s0003-4916(02)00018-0} {\bibfield  {journal}
  {\bibinfo  {journal} {Annals of Physics}\ }\textbf {\bibinfo {volume}
  {303}},\ \bibinfo {pages} {2–30} (\bibinfo {year}
  {2003}{\natexlab{b}})}\BibitemShut {NoStop}%
\bibitem [{\citenamefont {Chen}\ and\ \citenamefont {Hsin}(2021)}]{CH21}%
  \BibitemOpen
  \bibfield  {author} {\bibinfo {author} {\bibfnamefont {Y.-A.}\ \bibnamefont
  {Chen}}\ and\ \bibinfo {author} {\bibfnamefont {P.-S.}\ \bibnamefont
  {Hsin}},\ }\bibfield  {title} {\bibinfo {title} {Exactly solvable lattice
  hamiltonians and gravitational anomalies},\ }\href@noop {} {\bibfield
  {journal} {\bibinfo  {journal} {arXiv preprint arXiv:2110.14644}\ } (\bibinfo
  {year} {2021})}\BibitemShut {NoStop}%
\bibitem [{\citenamefont {Fidkowski}\ \emph {et~al.}(2022)\citenamefont
  {Fidkowski}, \citenamefont {Haah},\ and\ \citenamefont {Hastings}}]{FHH21}%
  \BibitemOpen
  \bibfield  {author} {\bibinfo {author} {\bibfnamefont {L.}~\bibnamefont
  {Fidkowski}}, \bibinfo {author} {\bibfnamefont {J.}~\bibnamefont {Haah}},\
  and\ \bibinfo {author} {\bibfnamefont {M.~B.}\ \bibnamefont {Hastings}},\
  }\bibfield  {title} {\bibinfo {title} {Gravitational anomaly of
  $(3+1)$-dimensional ${\mathbb{z}}_{2}$ toric code with fermionic charges and
  fermionic loop self-statistics},\ }\href
  {https://doi.org/10.1103/PhysRevB.106.165135} {\bibfield  {journal} {\bibinfo
   {journal} {Phys. Rev. B}\ }\textbf {\bibinfo {volume} {106}},\ \bibinfo
  {pages} {165135} (\bibinfo {year} {2022})}\BibitemShut {NoStop}%
\bibitem [{\citenamefont {{Ryohei Kobayashi, Yuyang Li, Hanyu Xue, Po-Shen Hsin
  and Yu-An Chen}}()}]{Previous}%
  \BibitemOpen
  \bibfield  {author} {\bibinfo {author} {\bibnamefont {{Ryohei Kobayashi,
  Yuyang Li, Hanyu Xue, Po-Shen Hsin and Yu-An Chen}}},\ }\bibfield  {title}
  {\bibinfo {title} {{Universal microscopic descriptions for statistics of
  particles and extended excitations}}\ }\href {https://doi.org/10.48550}
  {10.48550},\ \Eprint {https://arxiv.org/abs/2412.01886} {arXiv:2412.01886
  [quant-ph]} \BibitemShut {NoStop}%
\bibitem [{\citenamefont {Thorngren}(2015)}]{Thorngren:2014pza}%
  \BibitemOpen
  \bibfield  {author} {\bibinfo {author} {\bibfnamefont {R.}~\bibnamefont
  {Thorngren}},\ }\bibfield  {title} {\bibinfo {title} {{Framed Wilson
  Operators, Fermionic Strings, and Gravitational Anomaly in 4d}},\ }\href
  {https://doi.org/10.1007/JHEP02(2015)152} {\bibfield  {journal} {\bibinfo
  {journal} {JHEP}\ }\textbf {\bibinfo {volume} {02}},\ \bibinfo {pages}
  {152}},\ \Eprint {https://arxiv.org/abs/1404.4385} {arXiv:1404.4385 [hep-th]}
  \BibitemShut {NoStop}%
\bibitem [{\citenamefont {Johnson-Freyd}\ and\ \citenamefont
  {Reutter}(2021)}]{johnsonfreyd2021minimal}%
  \BibitemOpen
  \bibfield  {author} {\bibinfo {author} {\bibfnamefont {T.}~\bibnamefont
  {Johnson-Freyd}}\ and\ \bibinfo {author} {\bibfnamefont {D.}~\bibnamefont
  {Reutter}},\ }\bibfield  {title} {\bibinfo {title} {{Minimal nondegenerate
  extensions}},\ }\href@noop {} {\  (\bibinfo {year} {2021})},\ \Eprint
  {https://arxiv.org/abs/2105.15167} {arXiv:2105.15167 [math.QA]} \BibitemShut
  {NoStop}%
\bibitem [{\citenamefont {Johnson-Freyd}(2020)}]{Johnson-Freyd:2020twl}%
  \BibitemOpen
  \bibfield  {author} {\bibinfo {author} {\bibfnamefont {T.}~\bibnamefont
  {Johnson-Freyd}},\ }\bibfield  {title} {\bibinfo {title} {{(3+1)D topological
  orders with only a $\mathbb{Z}_2$-charged particle}},\ }\href@noop {} {\
  (\bibinfo {year} {2020})},\ \Eprint {https://arxiv.org/abs/2011.11165}
  {arXiv:2011.11165 [math.QA]} \BibitemShut {NoStop}%
\bibitem [{\citenamefont {Wang}\ and\ \citenamefont
  {Levin}(2014)}]{Wang2014Braiding}%
  \BibitemOpen
  \bibfield  {author} {\bibinfo {author} {\bibfnamefont {C.}~\bibnamefont
  {Wang}}\ and\ \bibinfo {author} {\bibfnamefont {M.}~\bibnamefont {Levin}},\
  }\bibfield  {title} {\bibinfo {title} {Braiding statistics of loop
  excitations in three dimensions},\ }\href
  {https://doi.org/10.1103/PhysRevLett.113.080403} {\bibfield  {journal}
  {\bibinfo  {journal} {Phys. Rev. Lett.}\ }\textbf {\bibinfo {volume} {113}},\
  \bibinfo {pages} {080403} (\bibinfo {year} {2014})}\BibitemShut {NoStop}%
\bibitem [{\citenamefont {Gaiotto}\ and\ \citenamefont
  {Johnson-Freyd}(2019)}]{gaiotto2019condensationshighercategories}%
  \BibitemOpen
  \bibfield  {author} {\bibinfo {author} {\bibfnamefont {D.}~\bibnamefont
  {Gaiotto}}\ and\ \bibinfo {author} {\bibfnamefont {T.}~\bibnamefont
  {Johnson-Freyd}},\ }\href {https://arxiv.org/abs/1905.09566} {\bibinfo
  {title} {Condensations in higher categories}} (\bibinfo {year} {2019}),\
  \Eprint {https://arxiv.org/abs/1905.09566} {arXiv:1905.09566 [math.CT]}
  \BibitemShut {NoStop}%
\bibitem [{\citenamefont {Levin}\ and\ \citenamefont
  {Wen}(2003)}]{Levin2003Fermions}%
  \BibitemOpen
  \bibfield  {author} {\bibinfo {author} {\bibfnamefont {M.}~\bibnamefont
  {Levin}}\ and\ \bibinfo {author} {\bibfnamefont {X.-G.}\ \bibnamefont
  {Wen}},\ }\bibfield  {title} {\bibinfo {title} {Fermions, strings, and gauge
  fields in lattice spin models},\ }\href
  {https://doi.org/10.1103/PhysRevB.67.245316} {\bibfield  {journal} {\bibinfo
  {journal} {Phys. Rev. B}\ }\textbf {\bibinfo {volume} {67}},\ \bibinfo
  {pages} {245316} (\bibinfo {year} {2003})}\BibitemShut {NoStop}%
\bibitem [{\citenamefont {Kawagoe}\ and\ \citenamefont
  {Levin}(2020)}]{Kawagoe2020Microscopic}%
  \BibitemOpen
  \bibfield  {author} {\bibinfo {author} {\bibfnamefont {K.}~\bibnamefont
  {Kawagoe}}\ and\ \bibinfo {author} {\bibfnamefont {M.}~\bibnamefont
  {Levin}},\ }\bibfield  {title} {\bibinfo {title} {Microscopic definitions of
  anyon data},\ }\href {https://doi.org/10.1103/PhysRevB.101.115113} {\bibfield
   {journal} {\bibinfo  {journal} {Phys. Rev. B}\ }\textbf {\bibinfo {volume}
  {101}},\ \bibinfo {pages} {115113} (\bibinfo {year} {2020})}\BibitemShut
  {NoStop}%
\bibitem [{\citenamefont {Feng}\ \emph
  {et~al.}(2025{\natexlab{b}})\citenamefont {Feng}, \citenamefont {Xue},
  \citenamefont {Li}, \citenamefont {Cheng}, \citenamefont {Kobayashi},
  \citenamefont {Hsin},\ and\ \citenamefont
  {Chen}}]{feng2025anyonicmembranespontryaginstatistics}%
  \BibitemOpen
  \bibfield  {author} {\bibinfo {author} {\bibfnamefont {Y.}~\bibnamefont
  {Feng}}, \bibinfo {author} {\bibfnamefont {H.}~\bibnamefont {Xue}}, \bibinfo
  {author} {\bibfnamefont {Y.}~\bibnamefont {Li}}, \bibinfo {author}
  {\bibfnamefont {M.}~\bibnamefont {Cheng}}, \bibinfo {author} {\bibfnamefont
  {R.}~\bibnamefont {Kobayashi}}, \bibinfo {author} {\bibfnamefont {P.-S.}\
  \bibnamefont {Hsin}},\ and\ \bibinfo {author} {\bibfnamefont {Y.-A.}\
  \bibnamefont {Chen}},\ }\href {https://arxiv.org/abs/2509.14314} {\bibinfo
  {title} {Anyonic membranes and pontryagin statistics}} (\bibinfo {year}
  {2025}{\natexlab{b}}),\ \Eprint {https://arxiv.org/abs/2509.14314}
  {arXiv:2509.14314 [quant-ph]} \BibitemShut {NoStop}%
\bibitem [{\citenamefont {Chatterjee}\ and\ \citenamefont
  {Wen}(2023)}]{chatterjee-2023}%
  \BibitemOpen
  \bibfield  {author} {\bibinfo {author} {\bibfnamefont {A.}~\bibnamefont
  {Chatterjee}}\ and\ \bibinfo {author} {\bibfnamefont {X.-G.}\ \bibnamefont
  {Wen}},\ }\bibfield  {title} {\bibinfo {title} {{Symmetry as a shadow of
  topological order and a derivation of topological holographic principle}},\
  }\bibfield  {journal} {\bibinfo  {journal} {Physical review. B./Physical
  review. B}\ }\textbf {\bibinfo {volume} {107}},\ \href
  {https://doi.org/10.1103/physrevb.107.155136} {10.1103/physrevb.107.155136}
  (\bibinfo {year} {2023})\BibitemShut {NoStop}%
\bibitem [{\citenamefont {Mullhaupt}(2009)}]{sheafhom}%
  \BibitemOpen
  \bibfield  {author} {\bibinfo {author} {\bibfnamefont {P.}~\bibnamefont
  {Mullhaupt}},\ }\bibfield  {title} {\bibinfo {title} {Review of "sheafhom:
  software for sparse integer matrices", m. mcconnell, pure appl. math. q. 3
  (2007), no. 1, part 3, 307-322.},\ }\href@noop {} {\  (\bibinfo {year}
  {2009})}\BibitemShut {NoStop}%
\bibitem [{\citenamefont {Xue}(2024)}]{GitHubProject}%
  \BibitemOpen
  \bibfield  {author} {\bibinfo {author} {\bibfnamefont {H.}~\bibnamefont
  {Xue}},\ }\href@noop {} {\bibinfo {title} {Statistics calculation}},\
  \bibinfo {howpublished}
  {\url{https://github.com/HanyuXue2002/StatisticsCalculation}} (\bibinfo
  {year} {2024}),\ \bibinfo {note} {accessed: 2024-12-07}\BibitemShut {NoStop}%
\bibitem [{\citenamefont {Munkres}(1975)}]{MunkresTopology}%
  \BibitemOpen
  \bibfield  {author} {\bibinfo {author} {\bibfnamefont {J.~R.}\ \bibnamefont
  {Munkres}},\ }\href@noop {} {\emph {\bibinfo {title} {Topology: a First
  Course}}}\ (\bibinfo  {publisher} {Prentice-Hall, Inc.},\ \bibinfo {address}
  {Englewood Cliffs, NJ},\ \bibinfo {year} {1975})\ \bibinfo {note} {section
  7-9.}\BibitemShut {Stop}%
\bibitem [{\citenamefont {Miller}(2021)}]{miller}%
  \BibitemOpen
  \bibfield  {author} {\bibinfo {author} {\bibfnamefont {H.}~\bibnamefont
  {Miller}},\ }\href {https://doi.org/10.1142/12132} {\emph {\bibinfo {title}
  {Lectures on Algebraic Topology}}}\ (\bibinfo  {publisher} {WORLD
  SCIENTIFIC},\ \bibinfo {year} {2021})\ \Eprint
  {https://arxiv.org/abs/https://www.worldscientific.com/doi/pdf/10.1142/12132}
  {https://www.worldscientific.com/doi/pdf/10.1142/12132} \BibitemShut
  {NoStop}%
\bibitem [{\citenamefont {nLab}()}]{ncatlabPeriodicTable}%
  \BibitemOpen
  \bibfield  {author} {\bibinfo {author} {\bibnamefont {nLab}},\ }\href@noop {}
  {\bibinfo {title} {Periodic table}},\ \bibinfo {howpublished}
  {\url{https://ncatlab.org/nlab/show/periodic+table}}\BibitemShut {NoStop}%
\bibitem [{\citenamefont {Baez}\ and\ \citenamefont
  {Dolan}(1995)}]{Baez1995HigherTQFT}%
  \BibitemOpen
  \bibfield  {author} {\bibinfo {author} {\bibfnamefont {J.~C.}\ \bibnamefont
  {Baez}}\ and\ \bibinfo {author} {\bibfnamefont {J.}~\bibnamefont {Dolan}},\
  }\bibfield  {title} {\bibinfo {title} {Higher‐dimensional algebra and
  topological quantum field theory},\ }\href {https://doi.org/10.1063/1.531027}
  {\bibfield  {journal} {\bibinfo  {journal} {Journal of Mathematical Physics}\
  }\textbf {\bibinfo {volume} {36}},\ \bibinfo {pages} {6073} (\bibinfo {year}
  {1995})}\BibitemShut {NoStop}%
\bibitem [{\citenamefont {Ellison}\ \emph {et~al.}(2022)\citenamefont
  {Ellison}, \citenamefont {Chen}, \citenamefont {Dua}, \citenamefont
  {Shirley}, \citenamefont {Tantivasadakarn},\ and\ \citenamefont
  {Williamson}}]{Ellison_2022}%
  \BibitemOpen
  \bibfield  {author} {\bibinfo {author} {\bibfnamefont {T.~D.}\ \bibnamefont
  {Ellison}}, \bibinfo {author} {\bibfnamefont {Y.-A.}\ \bibnamefont {Chen}},
  \bibinfo {author} {\bibfnamefont {A.}~\bibnamefont {Dua}}, \bibinfo {author}
  {\bibfnamefont {W.}~\bibnamefont {Shirley}}, \bibinfo {author} {\bibfnamefont
  {N.}~\bibnamefont {Tantivasadakarn}},\ and\ \bibinfo {author} {\bibfnamefont
  {D.~J.}\ \bibnamefont {Williamson}},\ }\bibfield  {title} {\bibinfo {title}
  {Pauli stabilizer models of twisted quantum doubles},\ }\bibfield  {journal}
  {\bibinfo  {journal} {PRX Quantum}\ }\textbf {\bibinfo {volume} {3}},\ \href
  {https://doi.org/10.1103/prxquantum.3.010353} {10.1103/prxquantum.3.010353}
  (\bibinfo {year} {2022})\BibitemShut {NoStop}%
\bibitem [{\citenamefont {Goerss}\ and\ \citenamefont
  {Jardine}(2009)}]{goerss-2009}%
  \BibitemOpen
  \bibfield  {author} {\bibinfo {author} {\bibfnamefont {P.~G.}\ \bibnamefont
  {Goerss}}\ and\ \bibinfo {author} {\bibfnamefont {J.~F.}\ \bibnamefont
  {Jardine}},\ }\href {https://doi.org/10.1007/978-3-0346-0189-4} {\emph
  {\bibinfo {title} {{Simplicial homotopy Theory}}}}\ (\bibinfo {year}
  {2009})\BibitemShut {NoStop}%
\bibitem [{\citenamefont {Zhang}(2023)}]{CarolynZhangSPTEntangler}%
  \BibitemOpen
  \bibfield  {author} {\bibinfo {author} {\bibfnamefont {C.}~\bibnamefont
  {Zhang}},\ }\bibfield  {title} {\bibinfo {title} {Topological invariants for
  symmetry-protected topological phase entanglers},\ }\href
  {https://doi.org/10.1103/PhysRevB.107.235104} {\bibfield  {journal} {\bibinfo
   {journal} {Phys. Rev. B}\ }\textbf {\bibinfo {volume} {107}},\ \bibinfo
  {pages} {235104} (\bibinfo {year} {2023})}\BibitemShut {NoStop}%
\bibitem [{\citenamefont {Wan}\ \emph {et~al.}(2022)\citenamefont {Wan},
  \citenamefont {Wang},\ and\ \citenamefont {Wen}}]{Wan_2022}%
  \BibitemOpen
  \bibfield  {author} {\bibinfo {author} {\bibfnamefont {Z.}~\bibnamefont
  {Wan}}, \bibinfo {author} {\bibfnamefont {J.}~\bibnamefont {Wang}},\ and\
  \bibinfo {author} {\bibfnamefont {X.-G.}\ \bibnamefont {Wen}},\ }\bibfield
  {title} {\bibinfo {title} {invertible topological order for
  branch-independent bosonic systems},\ }\bibfield  {journal} {\bibinfo
  {journal} {Physical Review B}\ }\textbf {\bibinfo {volume} {106}},\ \href
  {https://doi.org/10.1103/physrevb.106.045127} {10.1103/physrevb.106.045127}
  (\bibinfo {year} {2022})\BibitemShut {NoStop}%
\bibitem [{\citenamefont {Bryant}(2001)}]{CombinationalManifold}%
  \BibitemOpen
  \bibfield  {author} {\bibinfo {author} {\bibfnamefont {J.~L.}\ \bibnamefont
  {Bryant}},\ }\bibfield  {title} {\bibinfo {title} {Piecewise linear
  topology}\ }(\bibinfo {year} {2001})\BibitemShut {NoStop}%
\bibitem [{\citenamefont {Lurie}(2018)}]{kerodon}%
  \BibitemOpen
  \bibfield  {author} {\bibinfo {author} {\bibfnamefont {J.}~\bibnamefont
  {Lurie}},\ }\href@noop {} {\bibinfo {title} {Kerodon}},\ \bibinfo
  {howpublished} {\url{https://kerodon.net}} (\bibinfo {year}
  {2018})\BibitemShut {NoStop}%
\bibitem [{\citenamefont {Weibel}(1994)}]{Weibel_1994}%
  \BibitemOpen
  \bibfield  {author} {\bibinfo {author} {\bibfnamefont {C.~A.}\ \bibnamefont
  {Weibel}},\ }\href@noop {} {\emph {\bibinfo {title} {An Introduction to
  Homological Algebra}}},\ Cambridge Studies in Advanced Mathematics\ (\bibinfo
   {publisher} {Cambridge University Press},\ \bibinfo {year}
  {1994})\BibitemShut {NoStop}%
\end{thebibliography}%
\appendix
\onecolumngrid   

\FloatBarrier

\section{Simplicial Complex and Combinatorial Topology}\label{appendixSimplicialComplex}

\subsubsection*{Simplicial complex and semi-simplicial set}
We briefly recall the notation for finite simplicial complexes that we use in this paper. 
The reader can think of these as combinatorial models of triangulated spaces: we glue points, line segments, triangles, tetrahedra, etc.\ along common faces to obtain a piecewise linear space. For example, the simplicial complex
\begin{equation}\label{eqSimplicialComplexExample}
	\begin{tikzpicture}
		\centering
		\draw[thick,fill=red,fill opacity=0.1] (0,0) -- (2,0) -- (1,1.732) -- cycle; 
		\draw [thick] (2,0)--(4,0);
		\fill[red] (0,0) circle (2pt);
		\fill[red] (2,0) circle (2pt);
		\fill[red] (4,0) circle (2pt);
		\fill[red] (1,1.732) circle (2pt);
		
		\node[red, right] at (0,0) {0};
		\node[red, right] at (2,0) {1};
		\node[red, right] at (1,1.732) {2};
		\node[red, right] at (4,0) {3};
	\end{tikzpicture}
\end{equation}
is obtained by gluing a $1$-simplex and a $2$-simplex. Including all faces, we list all its simplexes as follows.
\begin{itemize}
	\item $-1$-dimensional simplexes: $\emptyset$ (we always include one $(-1)$-simplex by convention);
	\item $0$-dimensional simplexes: $\{0\},\{1\},\{2\},\{3\}$;
	\item $1$-dimensional simplexes: $\{0,1\},\{0,2\},\{1,2\},\{1,3\}$;
	\item $2$-dimensional simplexes: $\{0,1,2\}$.
\end{itemize}

With this geometric intuition, it is cleaner to define simplicial complexes in a purely combinatorial way as follows.
\begin{definition}
	A  simplicial complex $X$ is a collection of finite subsets of $\NN$
	such that if $\sigma\in X$ and $\tau\subset\sigma$, then $\tau\in X$.  We always assume $X$ is finite. Any $\sigma\in X$ containing $(q+1)$ elements is called a $q$-simplex. We write $X_q$ for the set of all $q$-simplexes of $X$. We always assume $X$ contains a unique $-1$-simplex $\emptyset$.
\end{definition}
\begin{definition}
	Let $X$ be a simplicial complex. If $Y$ is a simplicial complex and $Y\subset X$, then we say $Y$ is a subcomplex of $X$. Let $S\subset X$ be a set of simplexes in $X$; we define the subcomplex $\overline{S}$ by including all faces of these simplexes and call it the subcomplex generated by $S$.
\end{definition}
	Note that in a simplicial complex $X$, the word "simplex" may have two different meanings: an element $\sigma \in X$ or the subcomplex generated by $\sigma$. We do not distinguish them in notation.

In this definition, vertices of $X$ have the standard total order induced from that of $\NN$; this further induces a total order on each simplex. For example, any $q$-simplex, which is a $(q+1)$-element set, can be written uniquely as
\[
\sigma = \{a_0,\dots,a_q\},\qquad 
 a_0 < a_1 < \cdots < a_q \le n-1
\]
We then denote it as $\sigma=[a_0,\cdots,a_q]$. In this notation, the simplicial complex in Eq.~\eqref{eqSimplicialComplexExample} can be written as
\begin{equation}
	X=\{[],[0],[1],[2],[3],[0,1],[0,2],[1,2],[1,3],[0,1,2]\}.
\end{equation}
This ordering is necessary to define a homology theory. In particular, we need to define the $i$-th face of a simplex:

\begin{equation}\label{eqPartiali}
	\partial_i [a_0,\cdots,a_q]=[a_0,\cdots,\hat{a}_i,\cdots,a_q],\qquad 0\le i\le q,
\end{equation}
where $[a_0,\cdots,\hat{a}_i,\cdots,a_q]$ means $[a_0,\cdots,a_{i-1},a_{i+1},\cdots,a_q]$.

There are several slightly different definitions about simplicial complex that do not assume the total ordering on vertices. Usually, it is enough to have a branching structure, which is a partial ordering of vertices that induces a total order on every simplex. Sometimes, people also use unordered simplicial complexes, where no ordering is assumed at all. The differences among these definitions may have some physical meaning \cite{Wan_2022}, but they are not important in our paper.

\begin{example}
	The standard $n$-simplex $\Delta^n$ is the simplicial complex containing all subsets of $\{0,\cdots,n\}$, and its boundary $\partial\Delta^n$ contains all subsets not equal to $\{0,\cdots,n\}$.
	
	For example, we have
	\begin{equation}
		\Delta^2=\{[],[0],[1],[2],[0,1],[0,2],[1,2],[0,1,2]\}.
	\end{equation}
	and
	\begin{equation}
		\partial\Delta^2=\{[],[0],[1],[2],[0,1],[0,2],[1,2]\}.
	\end{equation}
\end{example}

Sometimes simplicial complex is confused with a related but also different concept, called \emph{$\Delta$-complex} or \textit{semi-simplicial set}. In a simplicial complex, each $q$-simplex has $(q+1)$ distinct vertices and is uniquely determined by them; in contrast, in a semi-simplicial set, vertices of a simplex may coincide, and a simplex may not be uniquely determined by its vertices. Thus, to give the combinatorial data of a semi-simplicial set $X$, we should write down $X_q$, the set of all $q$-simplexes, directly. Every $q$-simplexes has $(q+1)$-faces, and giving the $i$-th face for all $q$-simplexes determines a map $\partial_i:X_q\rightarrow X_{q-1}$. Formally, a semi-simplicial set is defined as follows.
\begin{definition}\label{defSemiSimplicial}
	A semi-simplicial set $X$ is a sequence of sets $X_q$ for $q\ge 0$ together with maps $\partial_i: X_q\rightarrow X_{q-1}$ for each $q$ and $0\le i\le q$, satisfying that
	\begin{equation}
		\partial_i\partial_j=\partial_{j-1}\partial_i,\;\;\;\forall i<j.
	\end{equation}
	Additionally, we define $X_{-1}$ as the one-point set. Similarly, we define  semi-simplicial Abelian groups by requiring $X_q$ are groups and $\partial_i$ are homomorphisms.
\end{definition}

Using Eq.~\ref{eqPartiali}, we identify simplicial complexes (with a branching structure) as special semi-simplicial sets. Not all semi-simplicial sets are simplicial complexes, such as
\begin{equation}
	\vcenter{\hbox{\raisebox{-2.5ex}{
				\begin{tikzpicture}
					\coordinate (A) at (0,0);  
					\coordinate (C) at (2,0);   
					\draw[thick] (A) to[out=90, in=90] (C);   
					\draw[thick] (A) to[out=-90, in=-90] (C); 
					\fill[red] (A) circle (2pt) node[left] {0};
	\end{tikzpicture}}}} \;\;\; \text{and}
	\;\;\;
	\vcenter{\hbox{\raisebox{-2ex}{\begin{tikzpicture}
					\coordinate (A) at (0,0);  
					\coordinate (C) at (2,0);   
					\draw[thick] (A) to[out=60, in=120] (C);   
					\draw[thick] (A) to[out=-60, in=-120] (C); 
					\fill[red] (A) circle (2pt) node[left] {0};
					\fill[red] (C) circle (2pt) node[right] {1};
	\end{tikzpicture}}}}.
\end{equation}
 In our theory of particle-fusion statistics, although they have the topology of $S^1$ and Definition~\ref{expSimplicialComplex} also works for semi-simplicial set, the corresponding excitation patterns do not give expected statistic because they lack locality identities.

\subsubsection*{Simplicial chains and homology}

Now we define the homology theory of semi-simplicial sets, and that of simplicial complexes is treated as a special case. For any semi-simplicial set $X$, we first construct a semi-simplicial Abelian group $\ZZ[X]$, such that $\ZZ[X]_q=\ZZ[X_q]$ is the free Abelian group generated by the set $X_q$ of $q$-simplexes. The group homomorphism $\partial_i:\ZZ[X]_q\rightarrow \ZZ[X]_{q-1}$ is induced by $\partial_i:X_q\rightarrow X_{q-1}$. Next, we construct the chain complex $C_*(X,\ZZ)$ as follows. The chain group in dimension $q$ is
\[
C_q(X,\ZZ) = \ZZ[X]_q,
\]
and the boundary homomorphism
\[
\partial^q\colon C_q(X,\ZZ) \longrightarrow C_{q-1}(X,\ZZ)
\]
 is defined by
 \begin{equation}
 	\partial^q=\sum_{i=0}^q(-1)^i\partial_i.
 \end{equation}

In a simplicial complex, this formula says
\begin{equation}\label{eqChainBoundary}
	\partial^q [a_0,\dots,a_q]
	:= \sum_{i=0}^q (-1)^i [a_0,\dots,\widehat{a_i},\dots,a_q],
\end{equation}
where the hat $\widehat{a_i}$ means that the vertex $a_i$ is omitted.

Geometrically, $\partial^q$ sends an oriented $q$-simplex to the signed sum of its oriented $(q-1)$-dimensional faces.
One checks that $\partial^{q-1}\circ\partial^q = 0$ for all $q$, so $(C_*(X,\ZZ),\partial)$ is a chain complex.
The $q$-th simplicial homology group is
\[
H_q(X,\ZZ) := \ker(\partial^q)\big/\operatorname{im}(\partial^{q+1}) .
\]

More generally, for an abelian group $A$ we set
\[
C_q(X,G) := C_q(X,\ZZ)\otimes_\ZZ G
\]
and define $H_q(X,G)$ in the same way.  

Note that there are two slightly different versions of chain complexes. In the chain complex $C_*(X,\ZZ)$, we define $C_q(X,\ZZ)=0$ for all $q<0$; but in a \textit{augmented chain complex} $\tilde{C}_*(X,\ZZ)$, we define $\tilde{C}_{-1}(X,\ZZ)=\ZZ$, and $\partial^0$ maps any vertex to the generator of $\ZZ$. The homology differs only at the $0$ dimension:
\begin{equation}
	H_0(X,\ZZ)\simeq \tilde{H}_0(X,\ZZ)\oplus\ZZ.
\end{equation}

\subsubsection*{Simplicial cochains and cohomology}

Let $G$ be an abelian group of coefficients.  
The group of $q$-cochains is
\[
C^q(X,G) := \operatorname{Hom}(C_q(X,\ZZ),G),
\]
which can be identified with the group of functions $c\colon X_q\to G$.
Thus a cochain assigns an $G$-valued “discrete $q$-form” to each oriented $q$-simplex.

The coboundary homomorphism
\[
\delta^q \colon C^q(X,G) \longrightarrow C^{q+1}(X,G)
\]
is determined by
\[
(\delta^q c)([a_0,\dots,a_{q+1}])
=c(\partial[a_0,\dots,a_{q+1}]) =\sum_{i=0}^{q+1} (-1)^i\, c([a_0,\dots,\widehat{a_i},\dots,a_{q+1}]).
\]
Then $\delta^{q+1}\circ\delta^q = 0$ for all $q$, and the $q$-th simplicial cohomology group is
\[
H^q(X,G) := \ker(\delta^q)\big/\operatorname{im}(\delta^{q-1}).
\]

\subsubsection*{Join, star, and link}

The construction in this subsection is only for simplicial complexes but not for semi-simplicial sets. Let $X$ be a simplicial complex and $\sigma,\tau\in X$ are two of its simplexes. We say $\sigma$ and $\tau$ are \textit{joinable} if $\sigma\cap \tau=\emptyset$ and $\sigma\cup \tau\in X$. Then, we say $\sigma\cup\tau$ is their \textit{join}, denoted by $\sigma*\tau$. For example, in the simplicial complex of Eq.~\eqref{eqSimplicialComplexExample}, $[1]$ and $[0,2]$ are joinable, and their join is $[0,1,2]$. In general, the join of a $p$-simplex and a $q$-simplex is a $(p+q+1)$-simplex. It is obvious that if $\sigma$ and $\tau$ are joinable, then any pair of faces of $\sigma$ and $\tau$ are joinable.

\begin{definition}
	Let $\sigma\in X$ be a simplex. Then the \textit{link} of $\sigma$ is the subcomplex containing all simplexes joinable with $\sigma$. We denote it by $\operatorname{Lk}(\sigma)$:
	\begin{equation}
		\operatorname{Lk}(\sigma)=\{\tau|\tau\cap \sigma=\emptyset,\tau\cup\sigma\in X\}.
	\end{equation}
	The \textit{star} of $\sigma$, denoted by $\operatorname{St}(\sigma)$, is the subcomplex generated by all simplexes containing $\sigma$.
	\begin{equation}
		\operatorname{St}(\sigma)=\overline{\{\tau\in X|\sigma\subset \tau\}}.
	\end{equation}
	 Equivalently, it is
	generated by all joins between $\sigma$ and $\tau\in \operatorname{Lk}(\sigma)$:
	\begin{equation}
		\operatorname{St}(\sigma)=\overline{\{\tau*\sigma|\tau\in \operatorname{Lk}(\sigma)\}}.
	\end{equation}
	
\end{definition}
Geometrically, $\operatorname{Lk}(\sigma)$ consists of the simplexes that are disjoint from $\sigma$ but appear together with $\sigma$ as faces of some larger simplex in $X$.
If $X$ triangulates a manifold and $\sigma$ is a $q$-simplex, then $\operatorname{Lk}(\sigma)$ triangulates a small sphere in the normal directions to $\sigma$.

\begin{example}
	In Eq.~\eqref{eqSimplicialComplexExample}, $\operatorname{Lk}([1])$ is the subcomplex generated by $\{[0,2],[3]\}$, and  $\operatorname{St}([1])$ is the whole complex. $\operatorname{Lk}([0,1])=\{[],[2]\}=\overline{[2]}$ is the subcomplex generated by the vertex $[2]$, and $\operatorname{St}([0,1])=\overline{[0,1,2]}$ is the triangle.
\end{example}

\begin{example}
	In $\Delta^{d+1}$, for any $p$-simplex $\sigma$, $\operatorname{Lk}(\sigma)$ is isomorphic to $\Delta^{d-p}$.
	
	In $\partial\Delta^{d+1}\simeq S^d$, for any $p$-simplex $\sigma$, $\operatorname{Lk}(\sigma)$ is isomorphic to $\partial\Delta^{d-p}\simeq S^{d-p-1}$.
\end{example}

\begin{definition}\label{defCombinatorialManifold}
	A simplicial complex $X$ is a $d$-dimensional \textit{combinatorial manifold}, if for any $p$-simplex $\sigma$, $\operatorname{Lk}(\sigma)$ is homeomorphic to $S^{d-p-1}$  \cite{CombinationalManifold}.
\end{definition}
\subsubsection*{A boundary formula of joins}

Let $\sigma$ and $\tau$ are two joinable simplexes in $X$ of dimension $p$ and $q$, respectively. $\sigma*\tau$ is a $(p+q+1)$-simplex having $p+q+2$ codimension-one faces $\partial_i(\sigma*\tau)$, and they are divided into two classes: $p+1$ of them are of the form $(\partial_i \sigma)*\tau$, while other $q+1$ of them are of the form $\sigma*(\partial_i\tau)$.

We want to obtain a chain-level formula, so we now treat orientations more carefully. Recall that we denote a generic $p$-simplex by $\sigma=[a_0,\cdots,a_p]$, where $a_0<\cdots <a_p$; now, for any permutation $\mathcal{P}$ of $\{0,\cdots,p\}$, we define
\begin{equation}
	[a_{\mathcal{P}(0)},\cdots,a_{\mathcal{P}(p)}]=\operatorname{sgn}(\mathcal{P})\sigma
\end{equation}
In other words, we allow that the vertices of a simplex are not written in the standard order; but if the sequence of vertices is obtained from an odd permutation, we will multiply it by a minus sign. We say $[a_{\mathcal{P}(0)},\cdots,a_{\mathcal{P}(p)}]$ is a signed simplex, and its underlying simplex is $[a_0,\cdots,a_p]$.

The advantage of this notation is that it is comparable with the chain boundary map Eq.~\eqref{eqChainBoundary}.
 It is originally defined under the assumption $a_0<\cdots a_p$, but now we apply this definition for general cases without this assumption. To prove that these definitions are consistent, we should prove that 
\begin{equation}
	\partial 	[a_{\mathcal{P}(0)},\cdots,a_{\mathcal{P}(p)}]=\operatorname{sgn}(\mathcal{P})\partial  [a_0,\cdots,a_p].
\end{equation}
It is enough to prove that when $\mathcal{P}$ is the exchange of $a_I$ and $a_J$. We denote the original signed simplex by $\sigma=[a_0,\cdots,a_p]$. After the exchange, the new signed simplex is
\begin{equation}
	\sigma=\sigma'=-[a_0,\cdots,a_{I-1},a_J,a_{I+1},\cdots,a_{J-1},a_I,a_{J+1},\cdots, a_p].
\end{equation}
In the alternating sum of Eq.~\eqref{eqChainBoundary}, it is easy to see $\partial_i\sigma'=\partial_i \sigma$, while the remaining terms are

\begin{equation}
	\left\{
	\begin{aligned}
		(-1)^I\partial_I\sigma&= (-1)^I[a_0,\cdots,a_{I-1},a_{I+1},\cdots,a_{J-1},a_J,a_{J+1},\cdots, a_p]\\
		(-1)^J\partial_J\sigma&= (-1)^J[a_0,\cdots,a_{I-1},a_I,a_{I+1},\cdots,a_{J-1},a_{J+1},\cdots, a_p]\\
		(-1)^{I+1}\partial_I\sigma'&=(-1)^I[a_0,\cdots,a_{I-1},a_{I+1},\cdots,a_{J-1},a_I,a_{J+1},\cdots, a_p]\\
	(-1)^J\partial_{J+1}\sigma'&= (-1)^J[a_0,\cdots,a_{I-1},a_J,a_{I+1},\cdots,a_{J-1},a_{J+1},\cdots, a_p]
	\end{aligned}
	\right.
\end{equation}
Using the fact that the permutation $I\mapsto I+1,I+1\mapsto I+2,\cdots, J-2\mapsto J-1,J-1\mapsto I$ has the sign $(-1)^(J-I+1)$, we get $(-1)^J\partial_J\sigma=(-1)^I\partial_I\sigma'$, similarly, we get $	(-1)^I\partial_I\sigma=(-1)^J\partial_{J+1}\sigma'$, and we are done.

This notation is helpful to define the join of signed simplexes or chains.
Let $\sigma=[a_0,\cdots,a_p]$ and $\tau=[b_0,\cdots,b_q]$ be two signed simplex (we do not assume $a_0<\cdots a_p$ or $b_0<\cdots<b_q$) such that their underlying simplexes are joinable. Then, we define their join as
\begin{equation}
	\sigma*\tau=[a_0,\cdots,a_p,b_0,\cdots,b_q].
\end{equation}
If in two chains $\alpha$ and $\beta$, all pairs of simplexes are joinable, we define their join $\alpha*\beta$ linearly.

By the property of permutation, we have
\begin{equation}
	\sigma*\tau=(-1)^{(p-1)(q-1)}\tau*\sigma.
\end{equation}
We also obtain the boundary formula of joins:
\begin{equation}\label{eqBoundaryJoin}
	\begin{aligned}
		&\partial (\sigma*\tau)\\&=\sum_{i=0}^p(-1)^i[a_0,\cdots,\hat{a_i},a_p,b_0,\cdots,b_q]+\sum_{j=0}^q(-1)^{j+p+1}[a_0,\cdots,a_p,b_0,\cdots,\hat{b_j},\cdots,b_q]\\&=(\partial \sigma )*\tau+(-1)^{p+1}\sigma*(\partial\tau).
	\end{aligned}
\end{equation}

\section{Simplicial Homotopy Theory and Dold-Kan Correspondence}\label{appendixSimplicialSet}

In this section, we introduce the necessary mathematical tools to prove our theorem. In addition to topological spaces, we will introduce simplicial sets, simplicial Abelian groups, chain complexes, and discuss their relations. They are summarized in Fig.~\ref{figRelations}. 

\begin{figure}[t]
	\centering
\begin{adjustbox}{width=.8\linewidth}
	\begin{tikzcd}[column sep=3.8em, row sep=3.0em]
		\mathbf{Top}
		\arrow[r, shift left=1.2ex, "{\mathrm{Sing}}"] &
		\mathbf{sSet}
		\arrow[l, shift left=1.2ex, "{\lvert\,\cdot\,\rvert}"]
		\arrow[r, shift left=1.2ex, "{\mathbb{Z}[\,-\,]}"] &
		\mathbf{sAb}
		\arrow[l, shift left=1.2ex, "{U}"]
		\arrow[r, shift left=1.2ex, "{\mathrm{N}}"] &
		\mathbf{Ch}_{\ge 0}
		\arrow[l, shift left=1.2ex, "{\Gamma}"]
	\end{tikzcd}
\end{adjustbox}
	\caption{Relations of different concepts.}
	\label{figRelations}
\end{figure}

The standard $0$-simplex is a point; the standard $1$-simplex is an edge with two endpoints $0,1$; the standard $2$-simplex is a triangle with three vertices $0,1,2$. In general, the standard $n$-simplex $\Delta^n$ is a $n$-dimensional linear object with $(n+1)$ vertices labeled by $0,1,\cdots,n$. We focus on all linear maps that preserve the order of vertices, not necessarily injective. Let $f:\Delta^m\rightarrow \Delta^n$ be a such map, and then $f$ is determined by its action on vertices, a function from $\{0,\cdots,m\}$ to $\{0,\cdots,n\}$ such that $i\le j\implies f(i)\le f(j)$. These objects and maps define a category $\Delta^*$.

\begin{definition}
	A simplicial set is a contravariant functor from $\Delta^*$ to $\mathbf{Set}$; a simplicial Abelian group is a contravariant functor from $\Delta^*$ to $\mathbf{Ab}$. A simplicial map between two simplicial sets (or Abelian groups) is a natural transformation between the two functors. Simplicial sets (Abelian groups) and simplicial maps form the category $\mathbf{sSet}$ ($\mathbf{sAb}$). 
\end{definition}

This definition can be written more explicitly. Let  $X\in \mathbf{sSet}$. Then, for any $n\ge0$, there is a set $X_n$, called the set of all $n$-dimensional simplexes. Any (order-preserving) map $\Delta^m\rightarrow \Delta^n$ determines a structural map $X_n\rightarrow X_m$, which are generated by the special ones. For any $0\le i\le n$, we can embed $\Delta^{n-1}$ in $\Delta^n$ as the $i$-th face avoiding the vertex $i$, which determines a map $\partial_i:X_n\rightarrow X_{n-1}$, called "face map". For any $0\le i\le n$, there is a projection map $\Delta^{n+1}\rightarrow \Delta^n$ mapping $n,n+1$ to the same point $n$, which determines a map $s_i:X_n\rightarrow X_{n+1}$, called "degeneracy map". They satisfy the following equations:

\begin{equation}\label{eqSimplicialIdentity}
	\left\{
	\begin{aligned}
		\partial_i\,\partial_j &= \partial_{j-1}\,\partial_i && \text{if } i<j,\\
		\partial_i\, s_j &= s_{j-1}\, \partial_i && \text{if } i<j,\\
		\partial_j\, s_j &= 1 = \partial_{j+1}\, s_j,\\
		\partial_i\, s_j &= s_j\, \partial_{i-1} && \text{if } i>j+1,\\
		s_i\, s_j &= s_{j+1}\, s_i && \text{if } i\le j.
	\end{aligned}
	\right.
\end{equation}

Eq.~\eqref{eqSimplicialIdentity} exhaust all relations between $\partial_i$ and $s_i$. So, $\{X_n\}$ labeled by $n\ge0$ and maps $\partial_i,s_i$ satisfying Eq.~\eqref{eqSimplicialIdentity} can serve as a definition of simplicial set or simplicial Abelian group. A simplicial map $f:X\rightarrow Y$ consists of maps $f_n:X_n\rightarrow Y_n$ commute with all $\partial_i,s_i$.

\begin{remark}\label{remarkSemeSimplicial}
	The definition of simplicial set through Eq.~\eqref{eqSimplicialIdentity} is very similar to the definition of semi-simplicial set (Definition~\ref{defSemiSimplicial}). Actually, by forgetting degeneracy maps $s_i$, a simplicial set $X$ is automatically a semi-simplicial sets. We denoted this forgetful functor by $U$. As its adjoint, we can add degeneracy maps to a semi-simplicial set $Y$ directly to get a simplicial set $FY$. They have the adjoint relation:
	\begin{equation}
		\hom_{\mathbf{sSet}}(FY,X)\simeq \hom_{\mathbf{semisSet}}(Y,UX).
	\end{equation}
	 Actually, degeneracy maps are less intuitive than face maps, and they have no contribution in gluing simplexes at all. Usually we do not need to pay much attention to degeneracy maps, while their existence is still necessary to develop a homotopy theory. 
\end{remark}

Via the forgetful functor (also denoted by $U$) $\mathbf{sAb}\rightarrow\mathbf{sSet}$ that forgets the group structure, a simplicial Abelian group is automatically a simplicial set. Conversely, we can define the free Abelian group functor $\ZZ[\cdot]:\mathbf{sSet}\rightarrow\mathbf{sAb}$. For any simplicial set $X$, we define $\ZZ[X]_n=\ZZ[X_n]$ as the free Abelian group generated by the set $X_n$, whose elements are written as

\begin{equation}
	\sum_{x\in X_n}c(x)x,\;c(x)\in\ZZ.
\end{equation}

We can further compose them to get $\ZZ[U(\cdot)]:\mathbf{sAb}\rightarrow \mathbf{sAb}$. This construction will be heavily used, so we will denote $\ZZ[U(K)]$ by $\ZZ[K]$ for short. Elements in $\ZZ[K]_n$ are written as

\begin{equation}
	\sum_{\alpha\in K_n}c(\alpha)[\alpha],\;c(\alpha)\in \ZZ.
\end{equation}

Here, the notation $[\alpha]$ indicates that we are doing formal sum in $\ZZ[K]_n$, but not the addition in $K_n$.

Similar to simplicial complexes introduced in Appendix~\ref{appendixSimplicialComplex}, the purpose of simplicial sets is also to study topological spaces via triangulations. The building blocks are simplexes $X_n$ in different dimensions, and they are glued together using the $\partial_i$ maps. Via this gluing operation, one can always obtain a topological space $|X|$ from a simplicial set $X$. This map $|\cdot|$ is called the geometric realization functor.

As the adjoint of geometric realization functor, the singular functor $\operatorname{Sing}$ canonically constructs a simplicial set $\operatorname{Sing}(M)$ for any topological space $M$. We say a singular $n$-simplex of $M$ is a continuous map from $\Delta^n$ to $M$, and then $\operatorname{Sing}(M)_n$ is the set of all singular $n$-simplex. For any singular $n$-simplex $\sigma$, which is a map $\Delta^n\xrightarrow{\sigma}M$, and a map $\Delta^m\rightarrow \Delta^n$, the composite $\Delta^m\rightarrow\Delta^n\xrightarrow{\sigma}M$ defines a singular $m$-simplex, so we get a map from $\operatorname{Sing}(M)_n$ to $\operatorname{Sing}(M)_m$. For example,  $x\in \operatorname{Sing}(M)_0$ is a point in $M$, and $l\in \operatorname{Sing}(M)_1$ is a path in $M$. $\partial_0,\partial_1$ map a path to its two endpoints, and $s_0$ maps a point to the constant path at this point. These $X_n$ and maps between them determine the simplicial set $\operatorname{Sing}(M)$.

Note that in general, $\operatorname{Sing}(|X|)\ne X$ and $|\operatorname{Sing}(M)|\ne M$; however, they are equivalent up to homotopies. In fact,
parallel to the homotopy theory of topological spaces, there is a simplicial version of homotopy theory defined in a combinatorial way, and $\operatorname{Sing},|\cdot|$ passing one to the other. In the remaining part of this section, we do not distinguish simplicial sets and topological spaces. 

\subsubsection*{Homotopy and homology}

\begin{definition}
	A pointed simplicial set is a simplicial set $X$ with a set of specified base-points $*_n\in X_n$ such that any map $X_n\rightarrow X_m$ induced by $\Delta^m\rightarrow \Delta^n$ maps $*_n$ to $*_m$. 
	
	In a pointed simplicial set, $\sigma\in X_n$ is called spherical, if $\partial_i\sigma=*_{n-1}$ for any $0\le i\le n$.
	
	Two spherical $n$-simplexes $\sigma,\sigma'$ are called homotopic, if there is an $(n+1)$-simplex $\tau$ such that $\partial_{i}\tau=\sigma$, $\partial_{j}\tau=(-1)^{i-j}\sigma'$, and $\partial_k\tau=*_n$ for other $k$.
	
	$\pi_n(X)$ is defined as the equivalence class of spherical $n$-simplexes modulo homotopic relations.
	
\end{definition}

We have $\pi_n(X)\simeq \pi_n(|X|)$ only when $X$ is a Kan complex, which should additionally satisfy Kan's extension condition. Without this assumption, $\pi_n(X)$ usually has bad behavior. In particular, in a Kan complex, if $\sigma$ is spherical and $\sigma,\sigma'$ are homotopic, then $\sigma'$ is also homotopic.
We skip these details because $\operatorname{Sing}(M)$ and simplicial Abelian groups are all Kan complexes. 

A simplicial Abelian group $K$ is automatically a pointed simplicial set (by choosing $0\in K_n$ as the base-point $*_n$), and the group structure makes $\pi_n(K)$ simpler. More specifically, the group of spherical $n$-simplexes is
$
	\mathcal{Z}=\cap_{i=0}^n \ker \partial_i\subset K_n$, and the group of zero-homotopic spherical $n$-simplexes $\mathcal{B}$ is the image of $\cap_{i=0}^n \ker \partial_i\subset K_{n+1}$ under the map $\partial_{n+1}:K_{n+1}\rightarrow K_n$.
Then, the homotopy group is defined as their quotient:
\begin{equation}
	\pi_n(K)=\mathcal{Z}/\mathcal{B}.
\end{equation}

Next, we explain the relationship between simplicial Abelian groups and (non-negatively graded) chain complexes, known as the Dold-Kan correspondence.

A chain complex $A$ is a sequence of Abelian group $A_n$ for $n\ge 0$ and a sequence of homomorphism $\partial^n:A_n\rightarrow A_{n-1}$ satisfying $\partial^{n-1}\circ \partial^n=0$. The $n$-th homology group is defined by

\begin{equation}
	\operatorname{H}_n(A)=\frac{\ker \partial^n}{\Im \partial^{n-1}}.
\end{equation}

The Dold-Kan correspondence states that \textbf{the homotopy of simplicial Abelian groups is equivalent to the homology of chain complexes}. To see that, the quickest way is to construct a chain complex $\operatorname{N}(K)$ for any simplicial Abelian group $K$. We define $\operatorname{N}(K)_n=\cap_{i=0}^{n-1}\ker \partial_{i}\subset K_n$ and the boundary map $\partial^n:\operatorname{N}(K)_n\rightarrow\operatorname{N}(K)_{n-1}$ is exactly the $n$-th face map $\partial_n$. We immediately get

\begin{equation}\label{eqHomologyHomotopy1}
	\operatorname{H}_n(\operatorname{N}(K))= \pi_n(K).
\end{equation}

In fact, the functor $\operatorname{N}$ is a category equivalence, not only up to homotopy. We now define a functor $\Gamma$ mapping a chain complex to a simplicial Abelian group. The key is the object $C_*(\Delta^*,\ZZ)$. This object has two indices and plays double roles, and we may call it a "simplicial chain complex". When we fix both indices, $C_m(\Delta^n,\ZZ)$ is the familiar $m$-th simplicial chain group\footnote{Only non-degenerate sub-simplexes are involved here.} of the standard $n$-simplex. When $n$ is fixed, $C_*(\Delta^n,\ZZ)$ is the familiar chain complex on $\Delta^n$. Now, for any chain complex $A$, we define (see Construction 2.5.6.3 of \cite{kerodon})

\begin{equation}
	\Gamma(A)=\hom_{\mathbf{Ch}}(C_*(\Delta^*,\ZZ),A).
\end{equation}

While 

\begin{equation}
	\Gamma(A)_n=\hom_{\mathbf{Ch}}(C_*(\Delta^n,\ZZ),A)
\end{equation}
is an Abelian group, structure maps $\Gamma(A)_n\rightarrow \Gamma(A)_m$ are induced by the maps $\Delta^m\rightarrow \Delta^n$.

The central theorem of the Dold-Kan correspondence states that (for example, see Theorem 2.5.6.1 of \cite{kerodon})
\begin{equation}
	\Gamma\circ \operatorname{N}\simeq \operatorname{id} \text{ and } \operatorname{N}\circ \Gamma\simeq \operatorname{id}.
\end{equation}

In the literature, $\operatorname{N}(K)$ is called the \textit{normalized chain complex}. Although the definition of $\operatorname{N}(K)$ also makes sense when $K$ is a semi-simplicial Abelian group, $\operatorname{N}(K)$ has good properties only when $K$ is a simplicial Abelian group (Warning 2.5.6.22 of \cite{kerodon}). For semi-simplicial Abelian groups, one should \textit{alternating chain complex} $\operatorname{A}(K)$ as defined in Appendix \ref{appendixSimplicialSet}. Explicitly, we have $\operatorname{A}(K)_n=K_n$ and 
\begin{equation}
	\partial^n=\sum_{i=0}^n(-1)^i\partial_i.
\end{equation}
There are two relationships between chain complexes $\operatorname{A}$ and $\operatorname{N}$. First, they have the same homology for any simplicial Abelian group. This is known as the Eilenberg-MacLane normalization theorem (Theorem 2.4 in \cite{goerss-2009}).
\begin{equation}
	H_*(\operatorname{A}(K))\simeq H_*(\operatorname{N}(K)),\;\;\;\forall X\in \mathbf{sAb}.
\end{equation}
Second, the alternating chain complex of a semi-simplicial Abelian group $K$ is isomorphic to the normalized chain complex of the simplicial Abelian group $F(K)$, where $F(K)$ is obtained by adding degenerate elements (Remark~\ref{remarkSemeSimplicial}). This is mentioned in Chapter 8 of \cite{Weibel_1994}.
\begin{equation}
	\operatorname{A}(K)\simeq \operatorname{N}(FK),\;\;\;\forall K\in \mathbf{semisAb}.
\end{equation}

When we compute $H_*(X,\ZZ)$ of a topological space $X$, there are different choices of chain complexes (simplicial or semi-simplicial, normalized or alternating, and so on), but they produce the same homology. The standard choice is to identify $X$ with the singular simplicial set $\operatorname{Sing}(X)$ and us the normalized chain complex. Thus, we have

\begin{equation}\label{eqHomologyHomotopy2}
	H_n(X,\ZZ)\simeq \operatorname{H}_n(\operatorname{N}(\ZZ[X]))\simeq \pi_n\ZZ[X].
\end{equation}
In short, \textbf{the homology of topological space $X$ is isomorphic to the homotopy of simplicial Abelian group $\ZZ[X]$}.

Eq.~\eqref{eqHomologyHomotopy1}, \eqref{eqHomologyHomotopy2} show that homology and homotopy are intimately related, and this is important in our study of Eilenberg-MacLane space. Traditionally, for an Abelian group $G$ and $q\ge0$, $K(G,q)$ is defined as (the homotopy type of) a topological space such that $\pi_n(K(G,q))=G\delta_{nq}$; then we are interested in $H_*(K(G,q),\ZZ)$, which is closely related to cohomology operations. However, we may define $K(G,q)$ as a specific simplicial Abelian group satisfying $\pi_n(K(G,q))=G\delta_{nq}$, and the corresponding topological space is actually $|U(K(G,q))|$. In this sense, the "topological" homology groups $H_*(K(G,q),\ZZ)$ is isomorphic to $\pi_*(\ZZ[K(G,q)])$ in our new notation. This is indeed how $H_*(K(G,q),\ZZ)$ emerge in our theory of statistics.

We can construct $K(G,q)\in \mathbf{sAb}$ explicitly. Let $G[q]$ be a chain complex having the only nonzero component $G[q]_n=G\delta_{nq}$. We define $K(G,q)=\Gamma(G[q])$, and then we get $\pi_n(K(G,q))\simeq \operatorname{H}_n(G[q])=G\delta_{nq}$ by Dold-Kan correspondence. Expanding the definition, $K(G,q)_n=\hom_{\mathbf{Chain}}(C_*(\Delta^n,\ZZ),G[q])$ is determined by a map $f:C_q(\Delta^n,\ZZ)\rightarrow G$ such that $f\circ \partial=0$, where $\partial:C_{q+1}(\Delta^n,\ZZ)\rightarrow C_q(\Delta^n,\ZZ)$ is the boundary map. This means exactly that $f$ is a cocycle in $Z^q(\Delta^n,G)$. Thus, $K(G,q)\in \mathbf{sAb}$ is defined by
\begin{equation}
	K(G,q)_n=Z^q(\Delta^n,G),
\end{equation}
and, for example, the $i$-th face map $\partial_i$ is defined by restricting a cocycle to $\partial_i\Delta^n$.

There is a closely related simplicial Abelian group $L(G,q)$ for $q\ge 1$. Consider a chain complex $A$ whose only nonzero parts are
\begin{equation}\label{eqL(G,q)ChainChainComplex}
	0\xrightarrow{\partial^{q+1}} A_q\simeq G\xrightarrow{\partial^q=\operatorname{id}}G\simeq A_{q-1}\xrightarrow{\partial^{q-1}} 0,
\end{equation}
and we define $L(G,q)=\Gamma(A)$. By expanding the definition, we have

\begin{equation}
	L(G,q)_n=C^{q-1}(\Delta^n,G).
\end{equation}

$L(G,q)$ is contractible because $\operatorname{H}_*(A)=0$. There is a canonical simplicial map from $L(G,q)$ to $K(G,q)$, where for dimension $n$, the corresponding map is the differential $C^{q-1}(\Delta^n,G)\rightarrow Z^q(\Delta^n,G)$. Note that it is a surjection, and its kernel $Z^{q-1}(\Delta^n,G)$ is exactly $K(G,q-1)$. We say maps $K(G,q-1)\rightarrow L(G,q)\rightarrow K(G,q)$ form a path fibration sequence, in the following sense.

\begin{theorem}\label{thmFibration}
	(Remark 2.5.6.18 of \cite{kerodon}) Let $M\xrightarrow{i}L\xrightarrow{p}K$ be simplicial maps between simplicial Abelian groups. Then, $0\rightarrow M\xrightarrow{i}L\xrightarrow{p}K\rightarrow0$ is dimension-wise exact, i.e.
	\begin{equation}
		0\rightarrow M_n\xrightarrow{i_n} L_n\xrightarrow{p_n} K_n\rightarrow 0\;\;\;\text{is exact }\forall n\ge 0,
	\end{equation}
	 if and only if $0\rightarrow \operatorname{N}_*(M)\xrightarrow{i} \operatorname{N}_*(L)\xrightarrow{p} \operatorname{N}_*(K)\rightarrow 0$ is an exact sequence of chain complexes.
	
	 In this case, we say $M\xrightarrow{i}L\xrightarrow{p}K$ is a fibration sequence\footnote{In the standard definition, $L_0\rightarrow K_0$ are not required to be surjective. This surjection is necessary for the exactness of $\pi_{0}(L)\xrightarrow{p_*}\pi_{0}(K)\longrightarrow 0$ in the next theorem.}. If $L$ is contractible, i.e., all homotopy groups are zero, then we say $M\xrightarrow{i}L\xrightarrow{p}K$ is a path fibration sequence; $L$ is a path space of $K$ and $M$ is a loop space of $K$.
\end{theorem}

We will denote a loop space of $K$ by $\Omega K$, which is unique up to homotopy.

Note that under this definition, $K$ admits a loop space only when $\pi_0(K)=0$ (i.e., $K$ is connected). This follows the classical result of long exact sequences.

\begin{theorem}
	Let $M\xrightarrow{i}L\xrightarrow{p}K$ be a fibration sequence of simplicial Abelian groups. Then, there is a natural long exact sequence
	\begin{equation}
		\begin{aligned}
			&\cdots \longrightarrow
			\pi_{n}(M)\xrightarrow{i_*}\pi_{n}(L)\xrightarrow{p_*}\pi_{n}(K)
			\xrightarrow{\;\partial\;}\pi_{n-1}(M)\xrightarrow{i_*}\pi_{n-1}(L)\xrightarrow{p_*}\pi_{n-1}(K)
			\longrightarrow \cdots\\[2pt]
			&\cdots \xrightarrow{\;\partial\;}\pi_{1}(M)\xrightarrow{i_*}\pi_{1}(L)\xrightarrow{p_*}\pi_{1}(K)
			\xrightarrow{\;\partial\;}\pi_{0}(M)\xrightarrow{i_*}\pi_{0}(L)\xrightarrow{p_*}\pi_{0}(K)\longrightarrow 0.
		\end{aligned}
	\end{equation}
	
	The map $\pi_n(K)\rightarrow\pi_{n-1}(L)$ is induced by the map
	\begin{equation}
		\operatorname{N}_n(K)\xrightarrow{p_n^{-1}}\operatorname{N}_n(L)\xrightarrow{\partial_n}\operatorname{N}_{n-1}(L)\xrightarrow{i_{n-1}^{-1}}\operatorname{N}_{n-1}(M),
	\end{equation}
	which is well-defined, and the image in $\pi_{n-1}(M)$ is independent of the choice of $p_n^{-1}$.
\end{theorem}

\begin{corollary}
	Let $\Omega K\xrightarrow{i} L\xrightarrow{p} K$ be a path fibration sequence,	then there is a natural isomorphism	$\pi_n(\Omega K)\simeq \pi_{n+1}(K)$.
\end{corollary}

\begin{corollary}\label{corollaryZeroHomotopic}
	Let $ L\xrightarrow{p} K$ makes $L$ a path space of $K$, then a spherical $n$-simplex $\sigma \in K_n$ is zero-homotopic if and only if there is a spherical $n$-simplex $\tau\in L_n$ satisfying $p_n(\tau)=\sigma$.
\end{corollary}

\subsubsection*{Cohomology operations on the chain level}

It is well known in algebraic topology that the cohomology groups, as a functor $\mathbf{Top}\rightarrow \mathbf{Ab}$, is represented by Eilenberg-MacLane spaces: $H^q(\cdot,G)\simeq [\cdot,K(G,q)]$, where $[\cdot,K(G,q)]$ means the homotopy classes of continuous maps from any topological space to $K(G,q)$. When we work on (semi-)simplicial sets, we can write them at the level of cochain or cocycles, which is convenient for use in field theory. 

We use $U:\mathbf{sSet}\rightarrow \mathbf{semisSet}$ to denote the forgetful functor, and $F:\mathbf{semi-sSet}\rightarrow \mathbf{sSet}$ to denote its left adjoint; see Remark~\ref{remarkSemeSimplicial}.
\begin{theorem}
	For any simplicial set $X$, we have 
		\begin{enumerate}
		\item $Z_{\text{normalized}}^q(X,G)\simeq \hom_{\mathbf{sSet}}(X,K(G,q))$;
		\item $C^{q-1}_\text{normalized}(X,G)\simeq \hom_{\mathbf{sSet}}(X,L(G,q))$.
	\end{enumerate}
	For any semi-simplicial set $X$, we have 
	\begin{enumerate}
		\item $Z_{\text{alternating}}^q(X,G)\simeq \hom_{\mathbf{semisSet}}(X,UK(G,q))$;
		\item $C_{\text{alternating}}^{q-1}(X,G)\simeq \hom_{\mathbf{semisSet}}(X,UL(G,q))$.
	\end{enumerate}
\end{theorem}
\begin{proof}
	Using the Dold-Kan correspondence, we have 
	\begin{equation}
		Z_{\text{normalized}}^q(X,G)\simeq \hom_{\mathbf{Ch}} (C_*(X,\ZZ),G[q])\simeq  \hom_{\mathbf{sAb}}(\ZZ[X],K(G,q))\simeq  \hom_{\mathbf{sSet}}(X,K(G,q)).
	\end{equation}

	For any semi-simplicial set $X$, we have
	
	\begin{equation}
		Z_{\text{alternating}}^q(X,G)\simeq Z_{\text{normalized}}^q(FX,G)\simeq \hom_{\mathbf{sSet}}(FX,K(G,q))\simeq \hom_{\mathbf{semisSet}}(X,UK(G,q)).
	\end{equation}
	The case for $C^{q-1}(X,G)$ is similarly derived from Eq.~\eqref{eqL(G,q)ChainChainComplex}.
\end{proof}

The classical result of cohomological operations says that $\operatorname{Nat}(H^{q}(\cdot,G),H^{n}(\cdot,\RR/\ZZ))\simeq H^n(K(G,q),\RR/\ZZ)$. We also have a cochain version.
\begin{theorem}
	Treating $Z^q(\cdot,G)$ as the normalized cocycle functors from $\mathbf{sSet}$ to $\mathbf{Set}$, we have
	\begin{enumerate}
		\item $\operatorname{Nat}(Z^{q}(\cdot,G),Z^{n}(\cdot,\RR/\ZZ))\simeq Z^n(K(G,q),\RR/\ZZ);$
		\item $\operatorname{Nat}(C^{q-1}(\cdot,G),C^{n-1}(\cdot,\RR/\ZZ))\simeq C^{n-1}(L(G,q),\RR/\ZZ)$.
		
	\end{enumerate}
	Also, treating $Z^q(\cdot,G)$ as the alternating cocycle functors from $\mathbf{semisSet}$ to $\mathbf{Set}$, we have
	\begin{enumerate}
		\item $\operatorname{Nat}(Z^{q}(\cdot,G),Z^{n}(\cdot,\RR/\ZZ))\simeq Z^n(UK(G,q),\RR/\ZZ);$
		\item $\operatorname{Nat}(C^{q-1}(\cdot,G),C^{n-1}(\cdot,\RR/\ZZ))\simeq C^{n-1}(UL(G,q),\RR/\ZZ)$.
	\end{enumerate}
	Note that the right-hand side in two situations are \textit{different}. 
\end{theorem}
\begin{proof}
	Using the Yoneda Lemma,
	\begin{equation}
		\operatorname{Nat}(Z^{q}(\cdot,G),Z^{n}(\cdot,\RR/\ZZ))\simeq \operatorname{Nat}(\hom_{\mathbf{sSet}}(\cdot,K(G,q)),Z^{n}(\cdot,\RR/\ZZ))\simeq Z^{n}(K(G,q),\RR/\ZZ).
	\end{equation}
	For the semi-simplicial case,
	\begin{equation}
		\operatorname{Nat}(Z^{q}(\cdot,G),Z^{n}(\cdot,\RR/\ZZ))\simeq \operatorname{Nat}(\hom_{\mathbf{semisSet}}(\cdot,UK(G,q)),Z^{n}(\cdot,\RR/\ZZ))\simeq Z^{n}(UK(G,q),\RR/\ZZ).
	\end{equation}

	The case of $C^{q-1}$ is similar.
\end{proof}

	The semi-simplicial seems to be used more often.
	Using 
	\begin{equation}
		C^{n-1}(UL(G,q),\RR/\ZZ)\simeq \hom_{\mathbf{Set}}(C^{q-1}(\Delta^{n-1},G),\RR/\ZZ),
	\end{equation}
	we can write these correspondences can be written more explicitly.
	 For any natural transformation $\phi:C^{q-1}(\cdot,G)\rightarrow C^{n-1}(\cdot,\RR/\ZZ)$, we construct $f_\phi\in \hom_{\mathbf{Set}}(C^{q-1}(\Delta^{n-1},G),\RR/\ZZ)$  by $f_\phi(\alpha)=\int_{\Delta^{n-1}} f(\Delta^{n-1})(\alpha)$. Conversely, in order to reconstruct $\phi$ from $f_\phi$, we need to determine $\int_\sigma\phi_X(\alpha)$ for any semi-simplicial set $X$, cochain $\alpha\in C^{q-1}(X,G)$, and simplex $\sigma\in X_{n-1}$;
	 defining it as $f_\phi(\alpha|_\sigma)$ is enough. Similarly, a natural transformation $\nu:Z^{q}(\cdot,G)\rightarrow Z^{n}(\cdot,\RR/\ZZ)$ corresponds to a map $f_\nu:Z^q(\Delta^n,G)\rightarrow \RR/\ZZ$; because $\nu$ satisfies the equation
\begin{equation}
	\int_{\partial \Delta^{n+1}}\nu(\alpha)=0,\;\forall \alpha\in Z^q(\partial\Delta^{n+1},G),
\end{equation}
the function $f_\nu$ should satisfies
	\begin{equation}
		\sum_{i=0}^{n+1}(-1)^if_\nu(\partial^*_i\alpha)=0,\;\forall \alpha\in Z^q(\Delta^{n+1},G).
	\end{equation}
This is exactly the cocycle condition of the cochain complex $\hom(\operatorname{A}(\ZZ[K(G,q)]),\RR/\ZZ)$.

\subsection{Proof of main theorems}

\subsubsection*{Proof of Theorem~\ref{thmTau}}
A proof of Theorem~\ref{thmTau} immediately follows the mathematical tools we introduced in the previous section. Replacing the excitation pattern by a cochain version (Definition~\ref{expDualCell}), it is equivalent to prove that $\tau(m)\simeq H^{d+1}(K(G,q),\RR/\ZZ)$, where $m=m^q(\partial\Delta^{d+1},G)$. The expression group $D(m)$ is isomorphic to $\ZZ[K(G,q)]_{d+1}$, and the map  $q_{x*}$ is exactly the face map $\partial_i: \ZZ[K(G,q)]_{d+1}\rightarrow \ZZ[K(G,q)]_{d}$. Because we are using the dual cell, localizing at a point $x$ now corresponds to restricting at a face $\partial_i\Delta^{d+1}$. Thus, $D_\inv(m)=\cap_{i=0}^{d+1}\ker \partial_i$ is exactly the subgroup of spherical simplexes. On the other hand, we have $D_\id(m)=\pi_*[D_\inv(\tilde{m})]$ in the sense of Remark~\ref{remarkAnotherDefOfEid}. Similarly, $D(\tilde{m})\simeq \ZZ[L(G,q)]_{d+1}$ and $D_\inv (\tilde{m})$ corresponds to spherical simplexes of $\ZZ[L(G,q)]$. Using Corollary~\ref{corollaryZeroHomotopic}, $\pi_*[D_\inv(\tilde{m})]$ exactly corresponds to zero-homotopic simplexes of $\ZZ[L(G,q)]$. Thus, $\tau(m)=D_\inv(m)/D_\id(m)$ is identified with the homotopy group $\pi_{d+1}(\ZZ[K(G,q)])\simeq H_{d+1}(K(G,q),\ZZ)$, and we are done. It has a canonical pairing $H^{d+1}(K(G,q),\RR/\ZZ)\times H_{d+1}(K(G,q),\ZZ)\rightarrow \RR/\ZZ$: for any cohomology operation (either simplicial or semi-simplicial) $\nu:Z^q(\cdot,G)\rightarrow Z^{d+1}(\cdot,\RR/\ZZ)$ and 
\begin{equation}
	\sum_{a\in A\simeq Z^q(\Delta^{d+1},G)}c(a)(a)\in D_\inv,
\end{equation}
the pairing produces a number
\begin{equation}
	\sum_{a\in A} c(a)\int_{\Delta^{d+1}} \nu(a).
\end{equation}

\subsubsection*{Proof of Theorem~\ref{conjectureCohomology}}

Next, we prove that $T(m^q(\Delta^{d+1},G))\simeq H_{d+2}(K(G,q),\ZZ)$. Compared with the previous proof, there are two additional problems: first, we need to deal with $E=\ZZ[S\times A]$, where $S=G_0\times \Delta^{d+1}_{p-1}$ is not intuitive in algebraic topology; second, we need to explain the appearance of $d+2$. The answer to the first problem is that $E$ fits in a semi-simplicial subgroup, and to get $d+2$, we should use the path fibration sequence of $\ZZ[K(G,q)]$ and its long exact sequence to shift the dimension. However, $\pi_0(\ZZ[K(G,q)])=\ZZ$, so $\ZZ[K(G,q)]$ does not admit a path space in the sense of Theorem~\ref{thmFibration}. What we should do is to construct a modified simplicial Abelian group $\tZ[K(G,q)]$. The difference between $\tZ[X]$ and $\ZZ[X]$ corresponds to the augmented version of homology : $H_n(X,\ZZ)\simeq \widetilde{H}_n(X,\ZZ)$ for $n>1$, and $H_0(X,\ZZ)\simeq \widetilde{H}_0(X,\ZZ)\oplus\ZZ$. 

\begin{definition}
	Let $X$ be a simplicial set. Elements in $\ZZ[X]_n$ are written as $\lambda=\sum_{\alpha\in X_n}c(\alpha)[\alpha]$, where $c(\alpha)\in \ZZ$. We define $\tZ[X]_n\subset \ZZ[X]_n$ containing all elements $\sum_{\alpha\in X_n}c(\alpha)[\alpha]$ such that $\sum_\alpha c(\alpha)=0$. $\tZ[X]$ is automatically a simplicial Abelian group.
\end{definition}

Because $\tZ[X]\rightarrow \ZZ[X]\rightarrow\ZZ$ is a fibration sequence, we obtain from the long exact sequence that $\pi_n(\tZ[X])=\pi_n(\ZZ[X])$ for $n>0$ and $\pi_0(\ZZ[X])=\pi_0(\tZ[X])\oplus\ZZ$.

Next, we construct a simplicial Abelian group $L\subset \ZZ[L(G,q)\times K(G,q)]$ as a path space of $\tZ[K(G,q)]$. We take $L_n$ as the subgroup of $\ZZ[C^{q-1}(\Delta^n,G)\times Z^q(\Delta^n,G)]$, containing all elements

\begin{equation}
	\lambda=\sum_{\alpha\in C^{q-1}(\Delta^n,G),\beta\in Z^q(\Delta^n,G)}c(\alpha,\beta)[\alpha,\beta],\; c(\alpha,\beta)\in \ZZ
\end{equation}
such that 
\begin{equation}
	\sum_{\alpha\in C^{q-1}(\Delta^n,G),\beta\in Z^q(\Delta^n,G)}c(\alpha,\beta)[\beta]=0.
\end{equation}

We define two linear maps $d_0,d_1:\ZZ[L(G,q)\times K(G,q)]\rightarrow \ZZ[K(G,q)]$ by
\begin{equation}\emph{}
	\left\{
	\begin{aligned}
		d_0([\alpha,\beta]) &= [\alpha],\\
		d_1([\alpha,\beta]) &= [d\alpha+\beta].
	\end{aligned}
	\right.
\end{equation}
Thus, we have $L=\ker d_0$. It is easy to verify that $d_1$ is a surjective simplicial map $L\rightarrow \tZ[K(G,q)]$. In fact,  $L$ is a path space of $\tZ[K(G,q)]$, and we should prove $L$ is contractible. Before the proof, we prepare some elementary observations about cochains and cocycles on $\Delta^n$, which can be easily checked by drawing pictures.

\begin{lemma}\label{lemmaCochainTool}
	For any $0\le k\le n$, we define a linear map $\iota_k:C^{q-1}(\Delta^{n-1},G)\rightarrow C^{q-1}(\Delta^{n},G)$ by identifying $\Delta^{n-1}$ with $\partial_k\Delta^n$ and extending to $\Delta^n$ by zero. Then, we have
	\begin{enumerate}
		\item The composition of 
		\begin{equation}
			C^{q-1}(\Delta^{n-1},G)\xrightarrow{\iota_k}C^{q-1}(\Delta^{n},G)\xrightarrow{d}Z^{q}(\Delta^{n},G)
		\end{equation}
		is an isomorphism $C^{q-1}(\Delta^{n-1},G)\simeq Z^{q}(\Delta^{n},G)$.
		\item 
		\begin{equation}
			\partial_i\iota_k=\left\{
			\begin{aligned}
				\id\quad \quad & \text{if } i=k;\\
				\iota_{k-1}\partial_i\quad & \text{   if } i<k;\\
				\iota_{k}\partial_{i-1}\quad  & \text{   if } i>k.
			\end{aligned}
			\right.
		\end{equation}
		Here, $\partial_i$ is restricting a cochain on the $i$-th face.

	\end{enumerate}
\end{lemma}

\begin{lemma}\label{lemmaL(G,q)Contractible}
	$\tZ[L(G,q)]$ is contractible.
\end{lemma}
\begin{proof}
	$\pi_0(\tZ[L(G,q)])=0$ because $0$ is the only $0$-simplex of $\tZ[L(G,q)]$. To prove $\pi_n(\tZ[L(G,q)])=0$ for $n\ge 0$, for any $\lambda=\sum_{\alpha\in C^{q-1}(\Delta^n,G)} c(\alpha)[\alpha]$ such that $\partial_i\lambda=\sum_{\alpha\in C^{q-1}(\Delta^n,G)} c(\alpha)[\alpha|_{\partial_i\Delta^n}]=0$, we define
	\begin{equation}
		\mu=\sum_{\alpha\in C^{q-1}(\Delta^n,G)} c(\alpha)[\iota_{n+1}(\alpha)]
	\end{equation}
	Using Lemma~\ref{lemmaCochainTool}, we have
	\begin{equation}
		\partial_{n+1}\mu=\sum_{\alpha\in C^{q-1}(\Delta^n,G)} c(\alpha)[\partial_{n+1}\iota_{n+1}(\alpha)]=\lambda
	\end{equation}
	and for any $i<n+1$, we have
	\begin{equation}
		\begin{aligned}
			\partial_{i}\mu&=\sum_{\alpha\in C^{q-1}(\Delta^n,G)} c(\alpha)[\partial_{i}\iota_{n+1}(\alpha)]\\&=\sum_{\alpha\in C^{q-1}(\Delta^n,G)} c(\alpha)[\iota_{n}\partial_i(\alpha)]\\&=\iota_n\partial_i\lambda=0.
		\end{aligned}
	\end{equation}
	Here, we apply $\iota_n$ to $\partial_i\lambda\in \ZZ[C^{q-1}(\Delta^{n-1},G)]$ term-wisely.
\end{proof}

\begin{lemma}
	$L$ defined as $\ker d_0$ is contractible.
\end{lemma}

\begin{proof}
	Let 
	\begin{equation}
		\lambda=\sum_{\alpha\in C^{q-1}(\Delta^n,G),\beta\in Z^q(\Delta^n,G)}c(\alpha,\beta)[\alpha,\beta]
	\end{equation}
	satisfies $d_0\lambda=0$ and $\partial_i\lambda=0$ for any $0\le i\le n$.

	Similar to the proof of the previous lemma, the idea is to construct a lifting $f$ from $C^{q-1}(\Delta^n,G)\times Z^{q}(\Delta^n,G)$ to $C^{q-1}(\Delta^{n+1},G)\times Z^{q}(\Delta^{n+1},G)$ such that $\partial_{n+1}\circ f=\id$.
	 One can simply embed a cochain to a face of $\Delta^{n+1}$ and extend it by zero. However, when one is dealing with cocycles, the embedding will not closed in general. The best alternative is to use the degeneracy map $s_k: Z^q(\Delta^n,G)\rightarrow Z^q(\Delta^{n+1},G)$, which is induced by the projection $\Delta^{n+1}\rightarrow \Delta^n$ that maps $k,k+1$ to the same point. Next, we define
	
	\begin{equation}
		\mu=\sum_{\alpha\in C^{q-1}(\Delta^n,G),\beta\in Z^q(\Delta^n,G)}c(\alpha,\beta)[\iota_k(\alpha),s_k(\beta)].
	\end{equation}	
	
	Using Eq.~\eqref{eqSimplicialIdentity} and Lemma~\ref{lemmaCochainTool}, it is easy to prove $\partial_k\mu=\lambda$ and $\partial_i\mu=0$ if $i\ne k,k+1$. Then, we have
	\begin{equation}
		\begin{aligned}
			\partial_{k+1}\mu&=\sum_{\alpha\in C^{q-1}(\Delta^n,G),\beta\in Z^q(\Delta^n,G)}c(\alpha,\beta)[\partial_{k+1}\iota_k(\alpha),\partial_{k+1}s_k(\beta)]\\&=\sum_{\alpha\in C^{q-1}(\Delta^n,G),\beta\in Z^q(\Delta^n,G)}c(\alpha,\beta)[\iota_k\partial_{k}\alpha,\beta]\\
		\end{aligned}
	\end{equation}
	Although it is nonzero, $-\partial_{k+1}\mu$ is homotopic equivalent to $\lambda$ and simpler. Continuing to substitute $\lambda$ by $-\partial_{k+1}\mu$ for all $0\le k\le n$, we find that $\lambda$ is homotopic equivalent to
		\begin{equation}
		\pm\sum_{\alpha\in C^{q-1}(\Delta^n,G),\beta\in Z^q(\Delta^n,G)}c(\alpha,\beta)[0,\beta],
	\end{equation}
	which is zero because $d_0\lambda=0$.
\end{proof}

Now, we have a fibration sequence
\begin{equation}
	\Omega\tZ[K(G,q)]\rightarrow L\rightarrow \tZ[K(G,q)],
\end{equation}
where $\Omega\tZ[K(G,q)]=\ker d_0\cap \ker d_1$. More explicitly,
\begin{equation}
	\Omega\tZ[K(G,q)]_n=\left\{\sum_{\alpha\in C^{q-1}(\Delta^n,G),\beta\in Z^q(\Delta^n,G)}c(\alpha,\beta)[\alpha,\beta]|c(\alpha,\beta)\in \ZZ, \sum c(\alpha,\beta)[\beta]=0,\sum c(\alpha,\beta)[d\alpha+\beta]=0\right\}.
\end{equation}

We have
\begin{equation}
	\pi_{d+1}\Omega\tZ[K(G,q)]=H_{d+2}(K(G,q),\ZZ),
\end{equation}
which is exactly what we expect for $T(m^q(\partial\Delta^{d+1},G))$. To relate these two objects, we define a map 
\begin{equation}
	\begin{aligned}
		f:E&\longrightarrow L_{d+1},\\
		(s,a)&\mapsto [s,a]-[0,a].
	\end{aligned}
\end{equation}

By adding the term $-[0,a]$, the image of $f$ is automatically in $\ker d_0=L_{d+1}$. Although $d_1\circ f(s,a)=[ds+a]-[a]$ is nonzero in general, expressions in $E_\inv$ share a property "closed" that they have zero image under the map $(s,a)\mapsto [a+ds]-[a]$, which corresponds exactly to $d_1\circ f$. Thus, $f$ maps closed expressions, denoted by $E_c$, into $\Omega\tZ[K(G,q)]_{d+1}$. We take $G_0=G-\{0\}$ in the definition of $m^q(X,G)$ for simplicity; this does not affect the result by Theorem~\ref{thmG0independence}. Then, $f$ isomorphically maps $E_c$ to the subgroup of $\Omega\tZ[K(G,q)]_{d+1}$ containing only terms of the form $[0,\beta]$ or $[s,\beta]$ for $s\in G_0\times \Delta^{d+1}_{q-1}$. 
These subgroup in different dimensions form a \textbf{semi-simplicial subgroup} $K'\subset \Omega\tZ[K(G,q)]$. We note that $K'$ is closed under face maps but not closed under degeneracy maps: the restriction of $[s,\beta]$ on a face is either  $[s,\partial_i\beta]$ or $[0,\partial_i\beta]$, while a degeneracy map will generally bring $s$ to a $(p-1)$-cochain. Nonetheless, we can still define  $\mathcal{Z}(K')=\cap_{i=0}^{d+1}\ker \partial_i$ and $\mathcal{B}(K')=\partial_{d+2}[\cap_{i=0}^{d+1}\ker \partial_i]$. Similar to the previous proof of Theorem~\ref{thmTau}, we have $E_\inv\simeq \mathcal{Z}(K')$ and $E_\id\simeq \mathcal{B}(K')$; the first is derived from Theorem~\ref{thmCheckByHand}, and the second is derived from condensation (Remark~\ref{remarkAnotherDefOfEid}). Thus, we have
\begin{equation}
	T(m^q(\partial \Delta^{d+1},G))\simeq \mathcal{Z}(K')/\mathcal{B}(K'),
\end{equation} 
and the next step is to prove $\pi_{d+1}(K')\simeq \pi_{d+1}(\Omega\tZ[K(G,q)])$. More explicitly, we denote the spherical simplexes and zero-homotopic simplexes in $\Omega\tZ[K(G,q)]_{d+1}$ by $\mathcal{Z}=\cap_{i=0}^{d+1}\ker \partial_i$ and $\mathcal{B}=\partial_{d+2}[\cap_{i=0}^{d+1}\ker \partial_i]$, and we should prove that
\begin{equation}\label{eqB34}
	\left\{
	\begin{aligned}
		\mathcal{Z} &= \mathcal{Z}(K')+\mathcal{B}\\
		\mathcal{B}(K') &= \mathcal{Z}(K')\cap\mathcal{B}.
	\end{aligned}
	\right.
\end{equation}

To prove these equations, we will construct a semi-simplicial map $\mathcal{D}: \Omega\tZ[K(G,q)]\rightarrow K'$, which decomposes any cochain $C^{p-1}(\Delta^n,G)$ into basic elements in $G_0\times \Delta^n_{p-1}$ and satisfies that\footnote{$\iota$ is the embedding $K'\rightarrow \Omega\tZ[K(G,q)]$}
\begin{equation}
	\mathcal{D}\circ \iota=\id: K'_{d+1}\rightarrow K'_{d+1}
\end{equation}
and 
\begin{equation}
	\lambda\in\mathcal{Z}\implies\lambda-\mathcal{D}(\lambda)\in\mathcal{B}.
\end{equation}

We assign the lexicographic order to all subsets of $\NN$ containing $q$ elements, which will be used to construct a functorial decomposition of $(q-1)$-cochains into generators. Under this ordering, 
 we label the $(q-1)$-simplexes of $\Delta^{d+1}$ by $x_1<\cdots<x_N$. Any $\alpha\in C^{q-1}(\Delta^{d+1},G)$ can be written uniquely as $\alpha=\sum_{i=1}^N \alpha_ix_i,\;\alpha_i\in G$. Next, we define a decomposition map $\mathcal{D}$ on $\ZZ[C^{q-1}(\Delta^n,G)\times Z^q(\Delta^n,G)]$ for any $n\ge 0$ by
 \begin{equation}\label{eqB37}
 	\mathcal{D}([\sum_{i=1}^N \alpha_ix_i,\beta])=[0,\beta]+\sum_{i=1}^N\bigg([\alpha_ix_i,d(\sum_{j=1}^{i-1}\alpha_jx_j)+\beta]-[0,d(\sum_{j=1}^{i-1}\alpha_jx_j)+\beta]\bigg)
 \end{equation}
 We have $d_0\circ\mathcal{D}=d_0$ and $d_1\circ\mathcal{D}=d_1$ by a direct verification, so $\mathcal{D}$ is a dimension-wise group homomorphism $\Omega\tZ[K(G,q)]\rightarrow K'$. Because of the lexicographic order, the ordering of this decomposition is consistent in all faces. This makes $\mathcal{D}$ a semi-simplicial map:
 \begin{equation}
 	\mathcal{D}\circ \partial_i=\partial_i\circ \mathcal{D}.
 \end{equation}
 It is also direct to verify that
 \begin{equation}
 	\mathcal{D}\circ \iota=\id: K'\rightarrow K',
 \end{equation}
 where $\iota$ is the embedding $K'\rightarrow \Omega\tZ[K(G,q)]$.
 
 \begin{remark}\label{remarkDecomposition}
 	The intuition of $\mathcal{D}$ is similar to Eq.~\eqref{eqTheta(g,a)}, where we decompose the action of a process $(g,a)$ to elementary steps $(s_i,a_i)$. In this section, we view $[\sum_{i=1}^N\alpha_ix_i,\beta]$ as a successive action of $\alpha_1x_1,\alpha_2x_2,\cdots$; in the first step, the action of $\alpha_1x_1$ produce a term $[\alpha_1x_1,\beta]$, and the remain process corresponds to $[\sum_{i=2}^N\alpha_ix_i,d(\alpha_1x_1)+\beta]$. Thus, we get the decomposition
 	\begin{equation}
 		[\sum_{i=1}^N \alpha_ix_i,\beta]\mapsto [\alpha_1x_1,\beta]+[\sum_{i=2}^N\alpha_ix_i,d(\alpha_1x_1)+\beta]-[0,d(\alpha_1x_1)+\beta],
 	\end{equation}
 	where the last term is used to compensate $d_0$ and $d_1$. Keep doing similar decomposition, one will get Eq.~\eqref{eqB37} in the end.
 	
 	One may raise the question that why in Definition~\ref{expSimplicialComplex}, we use $S=G_0\times X_{p+1}$ but not $S=C_{p+1}(X,G)$ to define $m_p(X,G)$. The reason is from the locality axiom: a simplex naturally has a support, while the support of a chain is a bad notion. Actually, we have tried to use $S=C_{p+1}(X,G)$ as the set of excitation operators and to replace the locality axiom by some other conditions, but none of them is physically basic and produce Theorem~\ref{conjectureCohomology} at the same time.
 \end{remark}
 
 Now, we prove Eq.~\eqref{eqB34}. Both $\mathcal{Z}(K')+\mathcal{B}\subset \mathcal{Z}$ and $\mathcal{B}(K')\subset \mathcal{Z}(K')\cap\mathcal{B}$ are obvious. For any $\lambda\in\mathcal{Z}(K')\cap\mathcal{B}$, we have $\mathcal{D}\circ \iota(\lambda)=\lambda$ and $\lambda=\partial_{d+2}\mu$, where $\mu\in \operatorname{N}_{d+2}(\Omega\tZ[K(G,q)])$. Thus, 
 \begin{equation}
 	\lambda=\mathcal{D}\circ \iota \circ \partial_{d+2}\mu=\partial_{d+2}\circ \mathcal{D}\circ \iota( \mu)\in \mathcal{B}(K')
 \end{equation}
because $\mathcal{D}\circ \iota (\mu)\in \operatorname{N}_{d+2}(K')$. 

To prove $\mathcal{Z}\subset \mathcal{Z}(K')+\mathcal{B}$, it is enough to prove that
\begin{lemma}
	If $\lambda\in \mathcal{Z}$, then $\lambda-\mathcal{D}(\lambda)\in\mathcal{B}$.
\end{lemma}
\begin{proof}
	Following Remark~\ref{remarkDecomposition}, without loss of generality, we prove that for any
	\begin{equation}
		\lambda=\sum_{\alpha\in C^{q-1}(\Delta^{d+1},G),\beta\in Z^q(\Delta^{d+1},G)} c(\alpha,\beta)[\sum_{i=1}^N \alpha_ix_i,\beta]\in \mathcal{Z},
	\end{equation}
	we have
	\begin{equation}
		\mu=\sum c(\alpha,\beta)\bigg([\sum_{i=1}^N \alpha_ix_i,\beta]-[\alpha_1x_1,\beta]-[\sum_{i=2}^N \alpha_ix_i,\alpha_1dx_1+\beta]+[0,\alpha_1dx_1+\beta]\bigg)\in \mathcal{B}.
	\end{equation}
	The idea is to use Corollary~\ref{corollaryZeroHomotopic}. We have already constructed a path fibration sequence $\Omega\tZ[K(G,q)]\rightarrow L\rightarrow \tZ[K(G,q)]$; in parallel, we construct another path fibration sequence $\Omega\tZ[L(G,q)]\rightarrow L'\rightarrow \tZ[L(G,q)]$. More explicitly, the $\Omega\tZ[L(G,q)]$ is the subgroup $\ker d_0\cap \ker d_1$ of $\ZZ[L(G,q)\times L(G,q)]$, where 
	$d_0,d_1:\ZZ[L(G,q)\times K(G,q)]\rightarrow \ZZ[K(G,q)]$ are defined by
	\begin{equation}
		\left\{
		\begin{aligned}
			d_0([\alpha,\gamma]) &= [\alpha],\\
			d_1([\alpha,\gamma]) &= [\alpha+\gamma],
		\end{aligned}
		\right.
	\end{equation}
	where $\alpha,\gamma\in C^{q-1}(\Delta^n,G)$. There is a canonical simplicial map
	\begin{equation}
		\begin{aligned}
			\delta : \Omega\tZ[L(G,q)]&\rightarrow \Omega\tZ[K(G,q)]\\
			[\alpha,\gamma]&\mapsto[\alpha,d\gamma].
		\end{aligned}		
	\end{equation}
	
	We fix an integer $k_0$ such that $0\le k_0\le d+1$ and $x_1\nsubseteq \partial_{k_0}\Delta^{d+1}$. Using Lemma~\ref{lemmaCochainTool}, we have an isomorphism $d\circ \iota_{k_0}:C^{q-1}(\Delta^{d},G)\rightarrow Z^q(\Delta^{d+1},G)$. Replacing $\beta$ by $d \iota_{k_0}\gamma$ where $\gamma\in C^{q-1}(\Delta^{d},G)$, we have
	\begin{equation}
		\lambda=\sum_{\alpha,\gamma} c(\alpha,d \iota_{k_0}\gamma)[\sum_{i=1}^N \alpha_ix_i,d \iota_{k_0}\gamma]
	\end{equation}
	and
	\begin{equation}
		\mu=\sum_{\alpha,\gamma} c(\alpha,d \iota_{k_0}\gamma)\bigg([\sum_{i=1}^N \alpha_ix_i,d \iota_{k_0}\gamma]-[\alpha_1x_1,d \iota_{k_0}\gamma]-[\sum_{i=2}^N \alpha_ix_i,\alpha_1dx_1+d \iota_{k_0}\gamma]+[0,\alpha_1dx_1+d \iota_{k_0}\gamma]\bigg)
	\end{equation}
	Next, we define
	\begin{equation}\label{eqB54}
		\rho=\sum_{\alpha,\gamma} c(\alpha, d\iota_{k_0}\gamma)\bigg([\sum_{i=1}^N \alpha_ix_i, \iota_{k_0}\gamma]-[\alpha_1x_1, \iota_{k_0}\gamma]-[\sum_{i=2}^N \alpha_ix_i,\alpha_1x_1+ \iota_{k_0}\gamma]+[0,\alpha_1x_1+ \iota_{k_0}\gamma]\bigg)
	\end{equation}
	
	It is obvious that $d\rho=\mu$ and $d_0\rho=d_1\rho=0$. We now show that $\partial_k\rho=0$ for all $0\le k\le d+1$.

	If $x_1\nsubseteq \partial_{k}\Delta^{d+1}$, then in $\partial_k\mu$, all $\alpha_1x_1$ is replaced by zero, and the result is automatically zero.
	
	If $x_1\subseteq \partial_{k}\Delta^{d+1}$, we have $k\ne k_0$ by the selection of $k_0$. Because $\lambda\in\mathcal{Z}$, we have
	
	\begin{equation}
		\partial_k\lambda=\sum_{\alpha,\gamma} c(\alpha,d \iota_{k_0}\gamma)\bigg[\sum_{i=1}^N\partial_k \alpha_ix_i,d\partial_k\iota_{k_0}\gamma\bigg]=0.
	\end{equation}
	Using Lemma~\ref{lemmaCochainTool} (the case for $k<k_0$ and $k>k_0$ is similar, we assume $k>k_0$), we get
	\begin{equation}
		\sum_{\alpha,\gamma} c(\alpha,d \iota_{k_0}\gamma)\bigg[\sum_{i=1}^N\partial_k \alpha_ix_i,\partial_{k-1}\gamma\bigg]=0.
	\end{equation}
	
	This equation is strong enough to imply that all of the four terms in Eq.~\eqref{eqB54} is zero. In all, we have $\rho\in \cap_{i=0}^{d+1}\ker \partial_i$.
	
	Finally, we use $d\circ \iota_{d+2}$ to map $\rho$ into $\Omega\tZ[K(G,q)]_{d+2}$. Using Lemma~\ref{lemmaCochainTool}, we have $\partial_{d+2}\circ d\circ \iota_{d+2}(\rho)=\mu$ and $\partial_{i}\circ d\circ \iota_{d+2}(\rho)=0$ for $i<d+2$.
	This finishes the proof of this lemma and also Theorem~\ref{conjectureCohomology}.
\end{proof}
 
\subsubsection*{Proof of Theorem~\ref{thmEvaluation}}

Here we always use the cohomology operation on semi-simplicial sets. First, for any $\nu: Z^q(\cdot,G)\rightarrow Z^{d+2}(\cdot,\RR/\ZZ)$, we will construct 
\begin{equation}
	\Theta: C^{q-1}(\cdot,G)\times Z^q(\cdot,G)\rightarrow C^{d+1}(\cdot,\RR/\ZZ)
\end{equation}
 such that
\begin{equation}\label{eqB58}
	d\Theta(\alpha,\beta)=\nu(d\alpha+\beta)-\nu(\beta).
\end{equation}

Note that the natural transformation $\Theta$ is uniquely determined by $\Theta_{\Delta^{d+1}}: C^{q-1}(\Delta^{d+1},G)\times Z^q(\Delta^{d+1},G)\rightarrow C^{d+1}(\Delta^{d+1},\RR/\ZZ)$, and is further uniquely determined by the map 
\begin{equation}
	\tilde{\Theta}(\cdot,\cdot)=\int_{\Delta^{d+1}}\Theta_{\Delta^{d+1}}:C^{q-1}(\Delta^{d+1},G)\times Z^q(\Delta^{d+1},G)\rightarrow \RR/\ZZ.
\end{equation}
 Let $\alpha\in C^{q-1}(\Delta^{d+1},G)$ and $\beta\in Z^q(\Delta^{d+1},G)$; we construct a cocycle $\pi^*\beta+d\iota_*\alpha$ on $\Delta^{d+1}\times I$. Here, $\Delta^{d+1}\times I$ is the cylinder with the canonical triangulation, $\pi^*$ is the pullback of projection $\Delta^{d+1}\times I\rightarrow \Delta^{d+1}$, and $\iota_*$ embeds the cochain to the face $\Delta^{d+1}\times 1$. Next, we define
\begin{equation}
	\tilde{\Theta}(\alpha,\beta)=\int_{\Delta^{d+1}\times I}\nu(\pi^*\beta+d\iota_*\alpha).
\end{equation}

The corresponding natural transformation then satisfies
\begin{equation}
	\begin{aligned}
		&\int_{\Delta^{d+2}}d\Theta(\alpha,\beta)\\&=\int_{(\partial\Delta^{d+2})\times I}\nu(\pi^*\beta+d\iota_*\alpha)\\&=\int_{\Delta^{d+2}\times I}d\nu (\pi^*\beta+d\iota_*\alpha)+\int_{\Delta^{d+2}\times 1}\nu(\pi^*\beta+d\iota_*\alpha)-\int_{\Delta^{d+2}\times 0}\nu(\pi^*\beta+d\iota_*\alpha)\\&=\int_{\Delta^{d+2}}\nu(d\alpha+\beta)-\nu(\beta).
	\end{aligned}
\end{equation}

Next, we prove that
\begin{equation}
	\theta(s,a)=\tilde{\Theta}(s,a)-\tilde{\Theta}(0,a)
\end{equation}
satisfies the locality axiom. The axiom states that 
\begin{equation}
	\theta([s_k,[\cdots,[s_2,s_1]]],a)=0,
\end{equation}
if the underlying simplexes do not share any common $d$-face. We define $\tilde{\Theta}'(\alpha,\gamma)=\tilde{\Theta}(\alpha,d\gamma)$ for $\gamma\in C^{q-1}(\Delta^{d+1},G)$ and
\begin{equation}
	\theta'(s,b)=\tilde{\Theta}'(s,b)-\tilde{\Theta}'(0,b)
\end{equation}
 for $b\in C^{q-1}(\Delta^{d+1},G)$. Thus, it is enough to prove that
\begin{equation}
	\theta'([s_k,[\cdots,[s_2,s_1]]],b)=0,
\end{equation}
which actually corresponds to working on $\tilde{m}(\partial\Delta^{d+1},G)$.

We embed $\alpha,\beta$ into $\Delta^{d+2}$ using the map $\iota_{d+2}:C^{q-1}(\Delta^{d+1},G)\rightarrow C^{q-1}(\Delta^{d+2},G)$. Then, $\theta'(s,b)$ decompose into two parts: the first part corresponds to the integration of $\iota_{d+2}(s),\iota_{d+2}(b)$ on the boundary $\partial_i\Delta^{d+2}$ for $i\ne d+2$, and its contribution in $\theta'([s_k,[\cdots,[s_2,s_1]]],b)$ cancels due to locality; the second part corresponds to the volume integration of $\nu(ds+db)-\nu(db)$, whose contribution also cancels because $[s_k,[\cdots,[s_2,s_1]]]$ is closed.

Note that $\tilde{\Theta}:C^{q-1}(\Delta^{d+1},G)\times Z^q(\Delta^{d+1},G)\rightarrow\RR/\ZZ$ is actually a $(d+1)$-cochain of $\ZZ[L(G,q)\times K(G,q)]$.
 Here, we are referring to the alternating cochain complex $C^*(K)=\hom(\operatorname{A}(K),\RR/\ZZ)$; the  $n$-th cochain group is $C^n(K)=\hom (K_n,\RR/\ZZ)$, and the differential $d:C^n(K)\rightarrow C^{n+1}(K)$ is the dual map of $\sum_{i=0}^{n+1}(-1)^i\partial_i$. 

  We write down the $n$-th cochain group for $\Omega\tZ[K(G,q)], L$ and $\tZ[K(G,q)]$ explicitly. $C^n(\Omega\tZ[K(G,q)])$ is all functions $\tilde\Theta:C^{q-1}(\Delta^n,G)\times Z^q(\Delta^n,G)\rightarrow\RR/\ZZ$ modulo the relation
\begin{equation}
	\tilde\Theta(\alpha,\beta)\sim \tilde\Theta(\alpha,\beta)+\varphi_0(\beta)+\varphi_1(d\alpha+\beta),
\end{equation}
which is derived from $\Omega\tZ[K(G,q)]_n=\ker d_0\cap \ker d_1$. The $n$-th cochain group for $L$ is similar, but we do not modulo $\varphi_1(d\alpha+\beta)$. Finally, $C^n(\tZ[K(G,q)])$ is all functions $Z^q(\Delta^n,G)\rightarrow \RR/\ZZ$ modulo a constant.

We have already constructed a map from $\nu$ to $\Theta$, which represents an element in $\hom_{\mathbf{Set}}(\Omega\tZ[K(G,q)],\RR/\ZZ)$. Next, we construct another map in the opposite direction.

The fibration sequence $\Omega[\tZ[K(G,q)]]\rightarrow L\rightarrow\tZ[K(G,q)]$ induces an exact sequence of cochain complexes
\begin{equation}
	0	\rightarrow C^*(\tZ[K(G,q)])\xrightarrow{f} C^*(L)\xrightarrow{g} C^*(\Omega\tZ[K(G,q)])\rightarrow0.
\end{equation}
In the corresponding long exact sequence, the connecting homomorphism is an isomorphism $H^{d+1}(\Omega\tZ[K(G,q)])\rightarrow H^{d+2}(\tZ[K(G,q)])$. On the cocycle level, the corresponding map is $f^{-1}\circ d\circ g^{-1}$. For any cocycle $\lambda \in Z^{d+1}(\Omega\tZ[K(G,q)])$, its pre-image $g^{-1}(\lambda)$ is a map (still, only the equivalence class represented by)
\begin{equation}
	\tilde\Theta: C^{q-1}(\Delta^{d+1},G)\times Z^{q}(\Delta^{d+1},G)\rightarrow \RR/\ZZ,
\end{equation}
or equivalently, a natural transformation $\Theta:C^{q-1}(\cdot,G)\times Z^{q}(\cdot,G)\rightarrow C^{d+1}(\cdot, \RR/\ZZ)$.
 The cocycle condition of $\lambda$ means that there are two natural transformations $\varphi,\nu':Z^q(\cdot,G)\rightarrow Z^{d+2}(\cdot,\RR/\ZZ)$ such that
\begin{equation}\label{eqB68}
	d\Theta(\alpha,\beta)=\nu'(d\alpha+\beta)-\varphi(\beta).
\end{equation}
By choosing $\beta=0$, we find that $\nu$ is determined up to a constant. Thus, the composition $\nu\mapsto \Theta\mapsto \nu'$ is the identity up to a constant. Since the second part of the map induces an isomorphism on cohomology, the first part also induces an isomorphism on cohomology.

Note that Eq.~\eqref{eqB68} is slightly different from Eq.~\eqref{eqB58}. In the direction $\nu\rightarrow \Theta$, we choose an explicit construction of $\tilde{\Theta}$, but in the direction $\Theta\mapsto \nu'$, we construct $\tilde\Theta$ as a representative of $\lambda$, which has some freedom. This difference does not affect our arguments.

\end{document}